\documentclass[11pt]{article}

\usepackage{amsmath,amssymb,epsfig,cite,color,verbatim,array,bm,amsfonts,enumerate,enumitem}
\usepackage{graphics,float}
\usepackage[colorlinks=true,citecolor=DarkOrchid,urlcolor=black,linkcolor=Blue]{hyperref}
\hypersetup{ linktoc=page}
\usepackage{blindtext}
\usepackage{microtype}
\usepackage[latin9]{inputenc}
\usepackage{authblk}

\newcommand*{\approxident}{%
\mathrel{\vcenter{\offinterlineskip
\hbox{$\sim$}\vskip-.35ex\hbox{$\sim$}\vskip-.35ex\hbox{$\sim$}}}}
\numberwithin{equation}{section}
\allowdisplaybreaks

\usepackage[dvipsnames]{xcolor}

\title{A Lagrangian constraint analysis of first order classical field theories\\
with an application to gravity
}

\author[1]{Ver\'onica Errasti D\'iez\footnote{veroerdi@mppmu.mpg.de}}
\author[2]{Markus Maier\footnote{maier@mppmu.mpg.de}}
\author[3]{Julio A. M\'{e}ndez-Zavaleta\footnote{julioamz@mpp.mpg.de}}
\author[4]{Mojtaba Taslimi Tehrani\footnote{motaslimi@gmail.com}}

\affil[1,2,3,4]{ {\small {\it Max-Planck-Institut f\"ur Physik (Werner-Heisenberg-Institut),
F\"ohringer Ring 6, 80805 Munich, Germany}}}
\affil[2]{ {\small {\it Universit\"{a}ts-Sternwarte, Ludwig-Maximilians-Universit\"{a}t M\"{u}nchen,
Scheinerstr. 1, 81679 Munich, Germany} }}

\date{}

\begin{document}

%%%%%%%%%%%%%%%%%%%%%%%%%%%%%%%%%%%%%%%%%%%%%%%%%%%%%%%%%%%%%%%%%%%%%%%%%%%%%%%%%%%%%%%%%
\maketitle 
%%%%%%%%%%%%%%%%%%%%%%%%%%%%%%%%%%%%%%%%%%%%%%%%%%%%%%%%%%%%%%%%%%%%%%%%%%%%%%%%%%%%%%%%%

%%%%%%%%%%%%%%%%%%%%%%%%%%%%%%%%%%%%%%%%%%%%%%%%%%%%%%%%%%%%%%%%%%%%%%%%%%%%%%%%%%%%%%%%%
\begin{abstract}
We present a method that is optimized to explicitly obtain all the constraints
and thereby count the propagating degrees of freedom in (almost all) manifestly first order classical field theories.
Our proposal uses as its only inputs a Lagrangian density and the identification of the a priori independent field variables it depends on.
This coordinate-dependent, purely Lagrangian approach is complementary to and in perfect agreement with the related vast literature.
Besides, generally overlooked technical challenges and problems derived from an incomplete analysis are addressed in detail.
The theoretical framework is minutely illustrated in the Maxwell, Proca and Palatini theories for all finite $d\geq 2$ spacetime dimensions.  
Our novel analysis of Palatini gravity constitutes a noteworthy set of results on its own.  
In particular, its computational simplicity is visible, as compared to previous Hamiltonian studies.
We argue for the potential value of both the method and the given examples in the context of generalized Proca and their  coupling to gravity.
The possibilities of the method are not exhausted by this concrete proposal. 
\end{abstract}
%%%%%%%%%%%%%%%%%%%%%%%%%%%%%%%%%%%%%%%%%%%%%%%%%%%%%%%%%%%%%%%%%%%%%%%%%%%%%%%%%%%%%%%%%

%%%%%%%%%%%%%%%%%%%%%%%%%%%%%%%%%%%%%%%%%%%%%%%%%%%%%%%%%%%%%%%%%%%%%%%%%%%%%%%%%%%%%%%%%
\tableofcontents
%%%%%%%%%%%%%%%%%%%%%%%%%%%%%%%%%%%%%%%%%%%%%%%%%%%%%%%%%%%%%%%%%%%%%%%%%%%%%%%%%%%%%%%%%

%%%%%%%%%%%%%%%%%%%%%%%%%%%%%%%%%%%%%%%%%%%%%%%%%%%%%%%%%%%%%%%%%%%%%%%%%%%%%%%%%%%%%%%%%
\section{Introduction}
\label{sec:intro}
%%%%%%%%%%%%%%%%%%%%%%%%%%%%%%%%%%%%%%%%%%%%%%%%%%%%%%%%%%%%%%%%%%%%%%%%%%%%%%%%%%%%%%%%%

It is hard to overemphasize the importance of field theory in high energy physics.
Suffice it to recall that each and every of the fundamental interactions we are aware of as of yet
---the gravitational, electromagnetic, strong and weak interactions--- are described in terms of fields.
Correspondingly, their dynamics are studied by means of field theory.
Most often, this is done by writing a Lagrangian (or a Hamiltonian) density that is a real smooth function of the field components
(and their conjugate momenta) and that is then subjected to the principle of stationary action.
It is customary to encounter the situation where not all of the a priori independent quantities ---field components and/or conjugate momenta---
are conferred a dynamical evolution through the equations of motion.
In such a case, the field theory is said to be singular or constrained. 
For instance, it is well-known that all gauge theories are singular.
 
In this work, we focus on singular classical field theories that are manifestly first order
and analyze them employing exclusively the Lagrangian formalism.
Non-singular theories are also in (trivial) reach.
Throughout the paper, manifest first order shall stand for a Lagrangian that depends only on the field variables and their first derivatives.
This implies the equations of motion are guaranteed to be second order at most.
Within this framework, we present a systematic methodology that is optimized to determine
the number of field components that do propagate, which we denominate physical/propagating modes/degrees of freedom.
To do so, we explicitly obtain the constraints: specific functional relations among the field variables and their time derivatives
that avoid the propagation of the remaining field components.
Our approach is complementary to the similarly aimed procedures in~\cite{Kamimura:1981fe,Diaz:2014yua,Diaz:2017tmy}
and is markedly distinct from, yet equivalent to, that in~\cite{Heidari:2020cil}.

Apart from the intrinsic relevance of understanding and characterizing the constraint structure of those theories satisfying our postulates,
an ulterior motivation for this investigation is to pave the way towards a consistent theory building principle.
Indeed, theoretical physics is currently in need of new fundamental and effective field theories that are capable of accounting for experimental data;
the strong CP problem, neutrino masses and the nature of the dark sector, to mention but a few of the most relevant examples. 
A recurrent and challenging obstacle in the development of well-defined field theories consists in guaranteeing the correct number of physical modes.
In this context, most effort is devoted to avoiding the propagation of Ostrograski instabilities~\cite{Ostr} ---additional unphysical degrees of freedom,
which we shall denote ghosts for short.
The general problem is delineated in~\cite{Sbisa:2014pzo} and numerous realizations of this idea can be found, e.g.~\cite{GFtheories}.
However, it is equally important to ensure the theory is not overconstrained, i.e.~there are fewer than required propagating modes.
Our subsequent prescription provides a solid footing to this (double) end and is presented in a comprehensible and ready to be used manner,
with the goal of being useful to communities such as, but not limited to, theoretical cosmology and black hole physics.
We describe  how to convert the analytical tool here exposed into a constructive one,
but the concrete realization of this idea is postponed to future investigations. 

A specific materialization of the preceding general discussion (and the one we later on employ to ground our conversion proposal) is as follows.
We recall that an earlier version of the method here augmented and refined already allowed for the development of the most general
non-linear multi-vector field theory over four-dimensional flat spacetime: the Maxwell-Proca theory~\cite{ErrastiDiez:2019trb,ErrastiDiez:2019ttn}.
There, the inclusion of a dynamical gravitational field was beyond scope.
The present work provides a sound footing for the study of singular field theories defined over curved backgrounds.
Thus, it paves the way for the ghost-free coupling of Maxwell-Proca to gravity.

Bearing in mind the above future objective and in order to clarify the formal presentation of the method,
we (re)analyze the simplest spin one and two theories by means of our proposed procedure: Maxwell, Proca and Einstein's gravity. 
While the former two are manifestly first order, the latter is not.
Indeed, gravity, cast in the Einstein-Hilbert way, is a second order Lagrangian for the metric, up to a non-covariant boundary term. 
As such, it exceeds the domain of applicability of our approach.
Favorably, this property can be circumvented taking advantage of the deluge of reformulations available for the theory.
Among them, we single out the Palatini formalism ---see~\cite{Ferraris:1982} for a historical overview---,
which considers the metric and the affine connection as a priori independent fields. 

Our determination of the explicit constraints present in Palatini, while not yielding novel information about the theory, conforms a remarkable piece of work.
Not only it is carried out minutely and can be readily seen to be computationally easier and shorter than the previously performed Hamiltonian studies, e.g.~\cite{Ghalati:2007sv,Kiriushcheva:2005sk,Kiriushcheva:2006gp,McKeon:2010nf}.  It also provides the basis for a consistent inclusion of matter fields.
As such, we regard this comprehensive analysis as an intrinsically valuable result.

\vspace*{0.5cm}

\hspace*{-0.6cm}{\bf Organization of the paper.}\\
In the following section \ref{sec:method}, we introduce the Lagrangian methodology we shall use throughout the paper.
Our approach is complementary to the existing literature.
In particular, it is equivalent to the recent proposal in~\cite{Heidari:2020cil}, as argued and exemplified in section \ref{sec:comir}.

We proceed to employ it to analyze various well-known theories:
Maxwell electromagnetism, together with the (hard) Proca action in section \ref{sec:vectors}
and the Palatini formulation of gravity in section \ref{sec:gravity}.
Their study is cornerstone to understand the
Maxwell-Proca theory~\cite{ErrastiDiez:2019trb,ErrastiDiez:2019ttn} and paves the way to its consistent coupling to gravity.
This is discussed in section \ref{sec:MPrel}.

We conclude in section \ref{sec:final}, restating the instances when our method is most convenient
and emphasizing two crucial aspects that are sometimes overlooked.

\vspace*{0.5cm}

\hspace*{-0.6cm}{\bf Conventions.}\\
We work on a $d$-dimensional spacetime manifold $\mathcal{M}$ of the topology $\mathcal{M}\cong\mathbb{R}\times\Sigma$.
Namely, we assume $\mathcal{M}$ admits a foliation along a time-like direction.
This is true for all (pseudo-)Riemannian manifolds.
For simplicity, we consider $\Sigma$ has no boundary.
The dimension $d$ is taken to be arbitrary but finite, with the lower bound $d\geq 2$.
Spacetime indices are denoted by the Greek letters $(\mu,\nu,\ldots)$
and raised/lowered with the metric $g_{\mu\nu}$ and its inverse $g^{\mu\nu}$.
We employ the standard short-hand notation $\partial_\mu:=\frac{\partial}{\partial x^\mu}$,
where $x^\mu:=(x^0,x^1,\ldots,x^{d-1})\equiv (x^0,x^i)$, with $i=1,2,\ldots,d-1$, are spacetime local coordinates,
naturally adapted to the foliation $\mathbb{R}\times\Sigma$.
The dot stands for derivation with respect to time, so that for local functions $f:\mathcal{M}\rightarrow \mathbb{R}$,
we write $\dot{f}:=\partial_0f$ and $\ddot{f}:=\partial_0^2f$.
Brackets indicating symmetrization and antisymmetrization of indices are defined as
$T_{(\mu\nu)}:=(T_{\mu\nu}+T_{\nu\mu})/2$ and $T_{[\mu\nu]}:=(T_{\mu\nu}-T_{\nu\mu})/2$, respectively.
As is customary, summation over repeated indices should be understood at all times.

%%%%%%%%%%%%%%%%%%%%%%%%%%%%%%%%%%%%%%%%%%%%%%%%%%%%%%%%%%%%%%%%%%%%%%%%%%%%%%%%%%%%%%%%%

%%%%%%%%%%%%%%%%%%%%%%%%%%%%%%%%%%%%%%%%%%%%%%%%%%%%%%%%%%%%%%%%%%%%%%%%%%%%%%%%%%%%%%%%%
\section{Exposition of the method}
\label{sec:method}
%%%%%%%%%%%%%%%%%%%%%%%%%%%%%%%%%%%%%%%%%%%%%%%%%%%%%%%%%%%%%%%%%%%%%%%%%%%%%%%%%%%%%%%%%

We begin by putting forward a coordinate-dependent, i.e.~non-geometrical, Lagrangian approach to obtain all the constraints present
in a manifestly first order classical field theory.
Needless to say, there exists a vast literature on the topic: some standard references are~\cite{Lagcons};
but for its elegance and concision, we particularly recommend~\cite{Lee:1990nz}.
This section serves us to fix the notation used throughout the paper and provide a self-contained derivation of all our results.
We stress that, although the method is not new per se, we are not aware of any reference where this material
is comprehensively presented in a ready to be used manner and keeping the technicalities at a bare minimum, as we do here.

Our only assumptions shall be the principle of stationary action and finite reducibility.
The first assumption is rather obviously a very mild one,
but it is worth noting that this is not an essential requirement; for instance, see~\cite{Bekenstein:2014uwa}.
We will explain the second assumption shortly.
For the time being, it suffices to note that, to our knowledge, the only known example of a classical field theory (of the kind here considered) not satisfying it
is bosonic string field theory, both in its open~\cite{Beng} and closed~\cite{BaGo} variants.

Given a Lagrangian density $\mathcal{L}$ within the above postulates, our analysis yields the \emph{constraint structure characterizing triplet} 
\begin{align}
\label{eq:tripletNdef}
t^{(N)}:=(l,g,e).
\end{align}
We stress that this is a purely Lagrangian statement,
since it collects the outcome of our subsequently proposed purely Lagrangian method.
Here, $N$ is the number of \emph{a priori independent field variables} in terms of which $\mathcal{L}$ is written.
As such, $N$ is equal to the dimension of the theory's configuration space, which we shall shortly introduce.
The other numbers $l$, $g$ and $e$ are defined below.

On shell, we obtain $l$: the total number of \emph{functionally independent Lagrangian constraints}.
Our analysis elaborates on the iterative algorithm presented in~\cite{Kamimura:1981fe}
and employed in appendix A of~\cite{ErrastiDiez:2019trb}.
It is the suitable generalization to field theory of the coordinate-dependent method used in~\cite{Diaz:2014yua} for particle systems,
which is in turn based on~\cite{Shir}.
The non-trivial geometric extension to field theory of~\cite{Diaz:2014yua} was carried out in~\cite{Diaz:2017tmy},
where the discussion was extended to the treatment of off shell constraints as well.
Thus, our discussion is complementary to all these references~\cite{Kamimura:1981fe,Diaz:2014yua,Diaz:2017tmy}.

Off shell, we shall obtain $g$ and $e$:
the number of \emph{gauge identities} and \emph{effective gauge parameters}, respectively.
Gauge identities are to be understood in the usual sense, as (differential) relations between certain functional variations of the action that identically vanish.
By effective gauge parameters we mean the number of independent gauge parameters plus their successive time derivatives
that explicitly appear in the gauge transformations.
We determine $g$ and $e$ for theories where the gauge transformations are known a priori
and provide suitable references that deal with the treatment of theories where the gauge transformations are unknown beforehand.
Notice that knowledge of the gauge transformations for the field theory is not a necessary assumption, unlike the principle of stationary action and finite reducibility.
However, this information considerably shortens the analysis and, being a feature of all the theories
we shall explicitly consider, we have opted for only developing in detail such case.

Given the triplet $t^{(N)}$, the \emph{physical degrees of freedom} $n_{\textrm{dof}}$ in the theory under study can be counted,
employing the result derived in~\cite{Diaz:2014yua}:
\begin{align}
\label{eq:dofformula}
n_{\textrm{dof}}=N-\frac{1}{2}(l+g+e).
\end{align}
We will refer to (\ref{eq:dofformula}) as the {\it master formula}, the way the authors of~\cite{Diaz:2014yua} themselves do.
The remarkable feature about the previous counting is that it is purely Lagrangian, as opposed to the usually employed Hamiltonian formula
\begin{align}
\label{eq:Dirdof}
n_{\textrm{dof}}=N-N_1-\frac{1}{2}N_2,
\end{align}
attributed to Dirac.
Here, $(N_1,N_2)$ denote the number of first and second class constraints, respectively.
As a reminder, first (second) class constraints are those which do (not) have a weakly vanishing Poisson bracket with all of the constraints present in a given theory.

Needless to say, the proven equivalence between the Lagrangian and Hamiltonian formulations
of classical theories~\cite{Kamimura:1981fe,Pons} is a most celebrated body of work.
The two given prescriptions for the degree of freedom count in (\ref{eq:dofformula}) and (\ref{eq:Dirdof})
are a particular materialization of this equivalence, which was further exploited in~\cite{Diaz:2014yua}
to develop a one-to-one mapping between the Lagrangian parameters $(l,g,e)$ and their Hamiltonian counterparts:
\begin{align}
\label{eq:HamLageq}
l=N_1+N_2-N_1^{(\textrm{P})}, \qquad g=N_1^{(\textrm{P})}, \qquad e=N_1,
\end{align}
where $N_1^{(\textrm{P})}$ stands for the number of so-called primary first class constraints,
those first class constraints that hold true off shell.
Using this information, the triplet $t^{(N)}$ defined in (\ref{eq:tripletNdef}) can be readily seen to admit the following equivalent Hamiltonian parametrization:
\begin{align}
\label{eq:LagHamtN}
t^{(N)}= (N_1^{(\textrm{P})},N_1,N_2).
\end{align}

An important comment is in order here.
Our subsequently proposed Lagrangian approach to determine $t^{(N)}$ does not guarantee $n_{\textrm{dof}}\in\mathbb{N}\cup \{0\}$.
This means that, even though all $l$, $g$ and $e$ in (\ref{eq:dofformula}) are integers by definition, their sum need not be an even number.
The reason is simple: we put forward an analytical tool, not a mechanism to detect (or even correct) ill posed theories.
If, for some Lagrangian density $\mathcal{L}$, a half-integer number of physical degrees of freedom is found upon correctly employing our prescription for $t^{(N)}$
together with (\ref{eq:dofformula}), then it must be concluded that the theory is unphysical.
The (possibly non-trivial) modifications required on $\mathcal{L}$ for it to propagate an integer number of physical modes
is a question beyond the scope of this manuscript\footnote{This
should not alarm the reader.
The same is true on the standard Hamiltonian formalism.
In (\ref{eq:Dirdof}), $N_2$ is not necessarily an even number, unless demands are made on the Hamiltonian.}.

For the renowned examples in sections \ref{sec:vectors} and \ref{sec:gravity}, we shall minutely determine the triplet $t^{(N)}$ defined in (\ref{eq:tripletNdef})
and then use (\ref{eq:dofformula}) to explicitly count physical modes.
As such, we shall perform various countings solely in Lagrangian terms.
Afterwards, we shall (partially) verify our results by comparing them to a representative subset of the Hamiltonian-based literature via (\ref{eq:Dirdof}) and (\ref{eq:HamLageq}).
Additionally, the examples of section \ref{sec:vectors} shall be worked out in two different (but dynamically equivalent) Lagrangian formulations,
based on distinct values $N$ and $\pmb{N}\neq N$ of the dimension of the configuration space.
We will then see that, even though the constraint structure characterizing triplets don't coincide,
the number of propagating modes $n_{\textrm{dof}}$ does match for both descriptions:
\begin{align}
\label{eq:NandNprime}
t^{(N)}:=(l,g,e)\neq t^{(\pmb{N})}:=(\pmb{l}, \pmb{g},\pmb{e}), \qquad N-\frac{1}{2}(l+g+e)=n_{\textrm{dof}}=\pmb{N}-\frac{1}{2}(\pmb{l}+\pmb{g}+\pmb{e}).
\end{align}
This is because $n_{\textrm{dof}}$ is a physical observable, while $(N,l,g,e)$ are not.
Obviously, the same situation arises in the Hamiltonian picture as well, which we briefly illustrate at the end of section \ref{sec:gravity}.

In the following, we explain how to obtain the constraint structure characterizing triplet $t^{(N)}$ in (\ref{eq:tripletNdef}).

%%%%%%%%%%%%%%%%%%%%%%%%%%%%%%%%%%%%%%%%%%%%%%%%%%%%%%%%%%%%%%%%%%%%%%%%%%%%%%%%%%%%%%%%%

%%%%%%%%%%%%%%%%%%%%%%%%%%%%%%%%%%%%%%%%%%%%%%%%%%%%%%%%%%%%%%%%%%%%%%%%%%%%%%%%%%%%%%%%%
\subsection{On shell Lagrangian constraints}
\label{sec:onshell}
%%%%%%%%%%%%%%%%%%%%%%%%%%%%%%%%%%%%%%%%%%%%%%%%%%%%%%%%%%%%%%%%%%%%%%%%%%%%%%%%%%%%%%%%%

Let $\mathcal{C}$ be the configuration space of a classical field theory.
As usual, we take $\mathcal{C}$ to be a differentiable Banach manifold whose points are labeled by $N$ real field variables $Q^A$:
\begin{align}
\label{eq:ConQAN}
\mathcal{C}=\textrm{span}\{Q^A\}, \qquad A=1,2,\ldots,N.
\end{align}
We stress that $A$ comprises all possible discrete indices that the real field variables have.
For instance, if one considers Yang-Mills theory, $A$ consists of both spacetime indices and color indices.
If one wishes to entertain complex Yang-Mills,
then the real and imaginary parts of each and every Yang-Mills field component must be counted separately in $A$.
So, for $SU(2)$ complex Yang-Mills theory in four spacetime dimensions, we would have that $N=2(4\cdot 3)=24$.
Notice that $Q^A$ are real smooth functions of spacetime $Q^A=Q^A(x^\mu)$, but we will suppress this dependence all along, so as to alleviate notation.
Thus, our notation matches that in~\cite{Diaz:2017tmy} and leaves out the spacetime argument compared to the condensed notation introduced by DeWitt in~\cite{dW}
and extensively used in the literature, e.g.~\cite{ParkerToms}.
Then, $T\mathcal{C}$ is the tangent bundle of $\mathcal{C}$, which is spanned by $\{Q^A,\dot{Q}^A\}$.
We refer to $(Q^A,\dot{Q}^A,\ddot{Q}^A)$ as the generalized coordinates, velocities and accelerations of the theory, respectively.

As already stated and common to most field theories, we assume that the dynamics are derivable from a principle of stationary action.
In other words, the \emph{Euler-Lagrange equations} $E_A\overset{!}{=}0$ for the field theory follow from
the requirement that the action functional
\begin{align}
\label{eq:action}
S=S[Q^A]=\int_{\mathcal{M}}d^dx\,\mathcal{L}=\int_{t_1}^{t_2}dx^0\int_{\Sigma} d^{d-1}x \,\mathcal{L},
\end{align}
remains stationary under arbitrary functional variations $\delta Q^A=\delta Q^A(x^0,x^i)$ that vanish at times $t_1$ and $t_2$ on
the spatial slice $\Sigma$:
\begin{align}
\label{eq:EAvar}
\delta S = \frac{\delta S}{\delta Q^A}\delta Q^A \equiv \int_{\mathcal{M}}d^dx\, E_A \delta Q^A\overset{!}{=}0,
\end{align}
with $\delta Q^A(t_1,x^i)=0=\delta Q^A(t_2,x^i)$.
The above variational derivative is defined as
\begin{align}
\label{eq:ELeqs}
E_A:=\partial_\mu\left(\frac{\partial \mathcal{L}}{\partial(\partial_\mu Q^A)}\right)
-\frac{\partial\mathcal{L}}{\partial Q^A}\overset{!}{=}0,
\end{align}
where the latter equality is the \emph{on shell} demand.
This on shell requirement commences the iterative algorithm we shall employ to determine the Lagrangian constraints present in the theory.
Here, $\mathcal{L}=\mathcal{L}[Q^A]$ is the Lagrangian density.
Observe that we have already restricted attention to \emph{manifestly first order} field theories,
i.e.~we consider $\mathcal{L}$ depends only on $Q^A$ and its first derivatives $\partial_\mu Q^A$.
The study of higher order field theories\footnote{One may be tempted to evade the higher order character of a theory via the Ostrogradski prescription,
i.e~introducing additional generalized coordinates in a manner that results in a manifestly first order Lagrangian density.
Such alteration of $T\mathcal{C}$ must be compensated through the inclusion of Lagrange multipliers that preserve the equivalence to the original setup.
To do so consistently, one needs to either verify the so-called 
Ostrogradsky non-singularity condition or exploit alternative methods, as detailed in~\cite{Andrzejewski:2010kz}.
In view of these non-trivial subtleties, we restrict ourselves to the study of manifestly first order theories.} 
---where $\mathcal{L}$ explicitly depends on $\partial_\mu^n Q^A$, with $n\geq2$---
lies beyond the scope of our present investigations.
We omit the possible dependence of $\mathcal{L}$ on non-dynamical field variables,
such as the spacetime metric in any special relativistic theory.
The said dependence can be easily incorporated to our analysis, but it does not arise in the theories we discuss in this work.

An important remark on notation follows.
As introduced in (\ref{eq:ConQAN}), $Q^A$ is an ordered set of a priori independent field variables;
it is neither a row nor a column vector.
The same is true for $E_A$ in (\ref{eq:ELeqs}):
this is the ordered set of Euler-Lagrange equations for the $Q^A$ field variables; not a vector.
We have opted for a notation where the set indices are always assigned the same position when ascribed to a certain ordered set
(for instance, upper position for the field variables $Q^A$ and lower position for the Euler-Lagrange equations $E_A$).
The assignation is such that the Einstein summation convention employed throughout the paper is apparent.
The only quantities that will show up in this section which have a definite character within matrix calculus are the following.
The various Hessians, their Moore-Penrose pseudo-inverses and the Jacobians are all matrices.
The null vectors of the Hessians are row vectors.
Their transposed column vectors also show up.
The row or column character of the ordered sets is then straightforwardly fixed according to dimensional analysis in all formulae.

As a practical starting point for our iterative method, it is convenient to recast the Euler-Lagrange equations (\ref{eq:ELeqs}) in the form
\begin{align}
\label{eq:ELeqs2}
E_B=\ddot{Q}^AW_{AB}+\alpha_B\overset{!}{=}0,
\end{align}
where we have defined the so-called \emph{primary Hessian} $W_{AB}:=\partial_{\dot{A}}\partial_{\dot{B}}\mathcal{L}$, as well as
\begin{align}
\label{eq:alphaB}
\alpha_B:=(\partial_{\dot{B}}\partial^i_A\mathcal{L}+\partial^i_B\partial_{\dot{A}}\mathcal{L})\partial_i\dot{Q}^A
+(\partial^i_B\partial^j_A\mathcal{L})\partial_i\partial_jQ^A +(\partial^i_B\partial_A\mathcal{L})\partial_iQ^A
+(\partial_{\dot{B}}\partial_A\mathcal{L})\dot{Q}^A-\partial_B\mathcal{L}.
\end{align}
To alleviate notation, we have introduced the following short-hands:
\begin{align}
\partial_{\dot{A}}:=\frac{\partial}{\partial \dot{Q}^A}, \qquad \partial^i_A:=\frac{\partial}{\partial (\partial_i Q^A)}, \qquad
\partial_A:=\frac{\partial}{\partial Q^A},
\end{align}
which we shall extensively employ henceforth.

We focus on singular (or constrained) field theories next\footnote{We leave out non-singular field theories because the subsequent analysis is redundant for them:
in this case $\textrm{det}(W_{AB})\neq0$, which implies $l=0$ and one can directly move on to section \ref{sec:offshell}.
Within our framework, scalar field theories in flat spacetime constitute a prominent example of non-singularity.}.
That is, we look at field theories described by a Lagrangian density whose primary Hessian has a vanishing determinant $\textrm{det}(W_{AB})=0$.
This means that the rank of $W_{AB}$ (the number of linearly independent rows or columns) is not equal to its dimension $N$;
instead, it is reduced.

By definition it follows that, for singular Lagrangians, the $N$ number of Euler-Lagrange equations in (\ref{eq:ELeqs2})
can be split into two types.
First, \emph{primary equations of motion}: these are the $\mathcal{R}_1:=\textrm{rank}(W_{AB})$ number of
on shell second order differential equations that explicitly involve the generalized accelerations $\ddot{Q}^A$.
Second, \emph{primary Lagrangian constraints}: these are the $M_1:=\textrm{dim}(W_{AB})-\textrm{rank}(W_{AB})=N-\mathcal{R}_1$ number of on shell relations
between the generalized coordinates $Q^A$ and their generalized velocities $\dot{Q}^A$.
We stress an explicit dependence on $\dot{Q}^A$ ($Q^A$) is not necessary for the primary Lagrangian constraints,
they can be relations between the $Q^A$'s ($\dot{Q}^A$'s) only.
Consistency requires that these constraints are preserved under time evolution.

In the following, we obtain the said constraints and ensure the consistency of the field theory by means of an iterative algorithm.
We refer to each iteration in the algorithm as a \emph{stage}.
In every stage, the above specified notions of equations of motion and Lagrangian constraints will arise.
The algorithm closes when the preservation under time evolution of all Lagrangian constraints is guaranteed.
Equivalently, when all $n$-th stage Lagrangian constraints are stable, for some finite integer $n\geq 2$.
An $n$-th stage Lagrangian constraint is said to be \emph{stable} if its time derivative does not lead to a new (i.e.~functionally independent)
Lagrangian constraint in the subsequent $(n+1)$-th stage.
Below, we explain in detail the different manners in which the necessary stability of the functionally independent Lagrangian constraints may manifest itself.

\vspace*{0.5cm}

\hspace*{-0.6cm}{\bf Primary stage.}\\
In order to determine the subset of $M_1$ number of primary Lagrangian constraints out of the set of all $N$ number of Euler-Lagrange equations in (\ref{eq:ELeqs2}),
we first introduce a set of $M_1$ number of linearly independent null vectors $\gamma_I$ associated to the primary Hessian $W_{AB}$:
\begin{align}
(\gamma_I)^AW_{AB}=0, \qquad I=1,2,\ldots,M_1.
\label{eq:nullvecs}
\end{align}
We require that these form an orthonormal basis of the kernel of $W_{AB}$,
which amounts to imposing the normalization condition
\begin{align}
(\gamma_{I})^A(\gamma^J)_A=\delta_I{}^J, \qquad \textrm{where } \gamma^I:=(\gamma_I)^T, \label{eq:ortho}
\end{align}
with $T$ denoting the transpose operation.
We stress that, even though in all the examples considered in sections \ref{sec:vectors} and \ref{sec:gravity}
we have chosen null vectors that are constant, this is not a required feature for our formalism.
Rather, this is just a possible choice in all the given examples that has been opted for due to its computational convenience.
Only the normalization (\ref{eq:ortho}) is an essential requirement for the null vectors.
In full generality, the null vectors of all stages can have an explicit dependence on the field variables $Q^A$ and their first derivatives $\partial_\mu Q^A$.

Then, the $M_1$ primary Lagrangian constraints are obtained by contracting the Euler-Lagrange equations $E_A$ in (\ref{eq:ELeqs2})
with the above null vectors\footnote{The complementary subset of $\mathcal{R}_1=N - M_1$ primary equations of motion
can be obtained by contracting $E_A$ with the basis vectors of the image of $W_{AB}$.
Here, we concentrate only on the Lagrangian constraints.}.
Namely, by performing the contraction with $\gamma_I$:
\begin{align}
\label{eq:onprim}
\varphi_I\equiv (\gamma_I)^{A}E_{A}=(\gamma_{I})^A\alpha_{A}\overset{!}{=} 0.
\end{align}
Notice that the last equality is a direct consequence of the on shell demand in (\ref{eq:ELeqs}) or equivalently in (\ref{eq:ELeqs2}).
Hence, the primary Lagrangian constraints are on shell constraints by definition.
One can also see this through equivalence to the more familiar Hamiltonian analysis.
It is common knowledge, e.g.~\cite{Cari}, that primary Lagrangian constraints relate to
secondary constraints in the Hamiltonian framework,
which are on shell constraints by definition.

The primary Lagrangian constraints in (\ref{eq:onprim}) need not be functionally independent from each other\footnote{This is in contrast to the primary equations of motion,
which are guaranteed by construction to be functionally independent among themselves.}.
When they are, the field theory is said to be \emph{irreducible} at the primary stage.
Otherwise, the theory is reducible at the primary stage.
Before we carry on, we must restrict attention to the functionally independent primary Lagrangian constraints
$\varphi_{I^\prime} \overset{!}{=} 0$, where $I^\prime=1,2,\ldots M_1^\prime\leq M_1$.
Their number is given by $M_1^\prime=\textrm{rank}(J_{I\Lambda})$, where the \emph{Jacobian matrix} $J_{I\Lambda}$ is defined as
\begin{align}
\label{eq:primJac}
J_{I\Lambda}:=\frac{\partial\varphi_I}{\partial X^\Lambda}, \qquad X^\Lambda=\{Q^A,\dot{Q}^A\}.
\end{align}
This test can be easily related to the standard Hamiltonian framework:
it is the pullback of the phase space regularity conditions in~\cite{Miskovic:2003tn}.
For the theories we are concerned with in this work, we verify $M_1^\prime=M_1$.
Hence, all of the primary Lagrangian constraints in (\ref{eq:onprim}) must be considered in the following\footnote{If $M_1^\prime<M_1$
and the functionally independent constraints are not straightforwardly identifiable, more work is required.
Indeed, there exists an iterative algorithm to extract the functionally independent subset of Lagrangian constraints from (\ref{eq:onprim}).
This is explained in section IID of~\cite{Diaz:2017tmy} and subsequently exemplified.
When the said algorithm requires a(n) finite (infinite) number of iterations, we face a(n) finitely (infinitely) reducible theory.
As already pointed out, the procedure here described requires, at the very least, the closure of the reducibility algorithm to proceed.
Thus, infinitely reducible theories cannot be studied with the present formalism.
We restate bosonic string field theory~\cite{Beng,BaGo} is the only physically relevant example of an infinitely reducible theory
we are aware of. \label{fn:red}}.

The vanishing of all the functionally independent primary Lagrangian constraints defines
the so-called \emph{primary constraint surface} $T\mathcal{C}_1$, which is a subspace of the \emph{moduli space} $T\mathcal{C}_0$ of the field theory:
\begin{align}
\label{eq:TC_1,0}
T\mathcal{C}_1 := \{ (Q^A, \dot{Q}^A) \in T\mathcal{C}_0 \hspace{1 mm} \rvert  \hspace{1mm} \varphi_{I}{=} 0 \} \subseteq T \mathcal{C}_0, \qquad 
T\mathcal{C}_0 := \{ (Q^A, \dot{Q}^A) \in T\mathcal{C} \hspace{1 mm} \rvert  \hspace{1mm} E_A{=} 0 \} \subset T \mathcal{C}.
\end{align}
For brevity, we write
\begin{align}
\label{eq:firstcs}
\varphi_{I}\overset{!}{\underset{1}{:\approx}}0.
\end{align}
Equalities that hold true in $T\mathcal{C}_1$ (and not in the entire of the moduli space) shall be denoted $\underset{1}{\approx}$
and referred to as \emph{primary weak equalities}.

As previously noted, consistency requires us to not only enforce the primary Lagrangian constraints (\ref{eq:firstcs}),
but also to ensure that these are preserved under time evolution.
Explicitly, $\widetilde{E}_J:=\dot{\varphi}_J\overset{!}{\underset{1}\approx}0$.
This requirement starts the second iteration in the algorithm.

\vspace*{0.5cm}

\hspace*{-0.6cm}{\bf Secondary stage.}\\
The freshly introduced demands $\widetilde{E}_J\overset{!}{\underset{1}\approx}0$\footnote{For clarity,
we will use a notation where tilde quantities belong to the secondary stage and hat quantities pertain to the tertiary stage.
This will be particularly helpful in section \ref{sec:Palatini}. \label{fn:not}}
are known as the secondary Euler-Lagrange equations.
In order to split them into secondary equations of motion and secondary Lagrangian constraints, it is convenient to write them as
\begin{align}
\label{eq:secELeqs}
\widetilde{E}_J=\ddot{Q}^A(\gamma^I)_A\widetilde{W}_{IJ}+\widetilde{\alpha}_J\overset{!}{\underset{1}\approx}0,
\end{align}
where we have defined
\begin{align}
\label{eq:secWandA}
\widetilde{W}_{IJ}:=(\gamma_I)^A\partial_{\dot{A}}\varphi_J, \qquad
\widetilde{\alpha}_J:=\left( -\alpha_A M^{AB}\partial_{\dot{B}} +\dot{Q}^A\partial_A + (\partial_{i}\dot{Q}^{A})\partial^i_A \right) \varphi_J.
\end{align}
We point out that, in obtaining this expressions, we have employed the on shell statement (\ref{eq:ELeqs2}),
so as to eliminate from (\ref{eq:secELeqs}) as much dependence on the generalized accelerations $\ddot{Q}^A$ as possible\footnote{
In the equivalent and more familiar Hamiltonian approach, this corresponds to solving as many generalized velocities as possible
in terms of generalized coordinates and conjugate momenta: $\dot{Q}^A=\dot{Q}^A(Q^A,\Pi_A)$.}.
Here, $\widetilde{W}_{IJ}$ is the so-called secondary Hessian and the auxiliary matrix $M^{AB}$ is the \emph{Moore-Penrose pseudo-inverse}
(as detailed in~\cite{Bou}) of the primary Hessian.
The latter is ensured to always exist and be unique.
Its defining relations are\footnote{In~\cite{Kamimura:1981fe}, the first relation is referred to as completeness relation.
There, both equations in (\ref{eq:Mdef}) are further used to obtain the explicit form of the functionally independent secondary equations of motion.
Unlike at the primary stage, functional independence is not guaranteed by construction.
As in the first iteration  earlier on, our interest lies in the form of the secondary Lagrangian constraints exclusively.}
\begin{align}
\label{eq:Mdef}
M^{AB}W_{BC}-\delta^A{}_C+(\gamma^I)_C(\gamma_I)^A=0, \qquad M^{AB}(\gamma^I)_B=0.
\end{align}
To gain some more intuition into $M^{AB}$, we note that it constitutes a generalization of the standard matrix inverse.
It is introduced so that $W_{AB}M^{BC}$ and $M^{AB}W_{BC}$ are orthogonal projections onto the image of $W_{AB}$ and $M^{AB}$, respectively.
For regular square matrices, the Moore-Penrose pseudo-inverse is equivalent to the standard matrix inverse: $M=W^{-1}$ iff $\textrm{det}(W)\neq0$.

If $\textrm{rank}(\widetilde{W}_{IJ})=\textrm{dim}(\widetilde{W}_{IJ})=M_1$, no secondary Lagrangian constraints arise and thus the primary Lagrangian constraints are stable.
In this case, we say that the consistency of the primary Lagrangian constraints (\ref{eq:firstcs}) under time evolution is \emph{dynamically ensured},
by a set of $M_1$ (necessarily functionally independent) secondary equations of motion
$\widetilde{E}_J=\widetilde{E}_J(\ddot{Q}^A)$.
As a result, the total number of functionally independent Lagrangian constraints present in such field theories is $l=M_1^\prime$.
However, this is not what happens in the theories of our interest.

Generically, the rank of the secondary Hessian is smaller than its dimension.
Consequently, $M_2:=\textrm{dim}(\widetilde{W}_{IJ})-\textrm{rank}(\widetilde{W}_{IJ})$ of the equations in (\ref{eq:secELeqs})
are secondary Lagrangian constraints, whose consistency under time evolution must be ensured.
This is done exactly as in the primary stage before.
In other words, the analysis from equation (\ref{eq:nullvecs}) onwards is to be repeated.

In details, the $M_2$ number of linearly independent null vectors $\widetilde{\gamma}_R$ of the secondary Hessian must be obtained:
\begin{align}
(\widetilde{\gamma}_R)^I\widetilde{W}_{IJ}=0, \qquad R=1,2,\ldots,M_2,
\end{align}
and chosen so that the normalization condition
\begin{align}
(\widetilde{\gamma}_R)^I(\widetilde{\gamma}^S)_I=\delta_R{}^S, \qquad \textrm{with }\,\,\widetilde{\gamma}^S:=(\widetilde{\gamma}_S)^T,
\end{align}
is satisfied.
Then, these must be contracted with the secondary Euler-Lagrange equations in (\ref{eq:secELeqs})
to yield the secondary Lagrangian constraints in the theory,
\begin{align}
\label{eq:seccons}
\widetilde{\varphi}_R\equiv (\widetilde{\gamma}_R)^I \widetilde{E}_I = (\widetilde{\gamma}_R)^I \widetilde{\alpha}_I \overset{!}{\underset{1}\approx}0.
\end{align}
If the secondary Lagrangian constraints vanish when evaluated on the first constraint surface $\widetilde{\varphi}_R\underset{1}{\approx}0$,
then the total number of functionally independent Lagrangian constraints is $l=M_1^\prime$.
Again, this is not what happens in (all of) the theories of our interest.

As a consequence, we must proceed with the algorithm.
First, we need to obtain the (subset of) $\widetilde{\varphi}_R$'s which are
functionally independent among themselves when evaluated on the first constraint surface.
Their number $M_2^\prime \leq M_2$ is given by
\begin{align}
M_2^\prime =\textrm{rank}(\widetilde{J}_{R\Lambda}), \qquad\qquad \textrm{where }\quad
\widetilde{J}_{R\Lambda}:=\frac{\partial}{\partial X^\Lambda}\Big(\widetilde{\varphi}_R\Big|_{T\mathcal{C}_1}\Big)
\end{align}
and $X^\Lambda$ was introduced in (\ref{eq:primJac}).
When $M_2^\prime\neq0$, we verify $M_2^\prime=M_2$ for the theories we shall consider ---so that they are irreducible theories at the secondary stage.
Thus, all secondary Lagrangian constraints in (\ref{eq:seccons}) must be considered subsequently\footnote{When $0<M_2^\prime<M_2$,
the iterative algorithm referenced in footnote \ref{fn:red} must be employed to extract the functionally independent
secondary Lagrangian constraints from (\ref{eq:seccons}).}.

The vanishing of the functionally independent secondary Lagrangian constraints defines the secondary constraint surface
$T\mathcal{C}_2\subset T\mathcal{C}_1$; which we write as $\widetilde{\varphi}_R\overset{!}{\underset{2}{:\approx}}0$.
Equalities holding true in $T\mathcal{C}_2$ shall be denoted $\underset{2}{\approx}$ and referred to as secondary weak equalities.
It should be obvious that the secondary Lagrangian constraints are on shell constraints by definition.

\vspace*{0.5cm}

\hspace*{-0.6cm}{\bf Tertiary stage.}\\
Let $\widehat{W}_{RS}:=(\widetilde{\gamma}_R)^I(\gamma_I)^A\partial_{\dot{A}}\widetilde{\varphi}_S$ be the tertiary Hessian.
When the tertiary Hessian's rank does not match its dimension,
the consistency under time evolution of $M_3:=\textrm{dim}(\widehat{W}_{RS})-\textrm{rank}(\widehat{W}_{RS})$ number of the functionally independent secondary Lagrangian constraints
is not (dynamically) guaranteed.
Instead, it must be enforced through a third iteration of the just described procedure.
We stress that it is essential to close the iterative algorithm in order to find the correct number $l$ of functionally independent Lagrangian constraints.

For completeness, we provide the explicit expressions for all relevant quantities at some arbitrary stage of the algorithm in appendix \ref{app:genex}.
These have not appeared in the literature, as far as we know.

\vspace*{0.5cm}

\hspace*{-0.6cm}{\bf Closure of the algorithm.}\\
In full generality and as already anticipated, our algorithm stops when all functionally independent Lagrangian constraints have been stabilized.
This can happen in either of the following different manners:
\begin{enumerate}[label=\roman*]
\item Dynamical closure.\\
Firstly, it may happen when $M_n:=\textrm{dim} (W^{(n)})-\textrm{rank}(W^{(n)})=0$ for some $n$-th stage Hessian $W^{(n)}$, with $n\geq 2$.
This implies that no Lagrangian constraints arise at the $n$-th stage, since in this case $W^{(n)}$ has full rank and hence admits no null vector.
Here, the consistency under time evolution of the previous stage's functionally independent Lagrangian constraints ${\varphi}^{(n-1)}$ is dynamically ensured,
i.e.~through the (necessarily functionally independent) $n$-th stage equations of motion.
In other words, the functionally independent ${\varphi}^{(n-1)}$'s are stable.
This closure of the algorithm is exemplified in section \ref{sec:Procavec}.
\label{it:I}
\item Non-dynamical closure.\\
Secondly, it may happen when $M_n>0$, but $M_n^\prime=0$, again with $n\geq 2$.
This implies that the $n$-th stage functionally independent Lagrangian constraints $\varphi^{(n)}$'s do not define a new constraint surface,
so that $T\mathcal{C}_{n}\equiv T\mathcal{C}_{n-1}$.
We differentiate two algebraically distinct scenarios:
\label{it:IIb}
\begin{enumerate}[label*=\alph*]
\item The $\varphi^{(n)}$'s vanish identically in the $(n-1)$-th constraint surface: $\varphi^{(n)}\underset{n-1}{\approxident} 0$.
Such $\varphi^{(n)}$'s are known as \emph{Lagrangian identities}.
Clearly, Lagrangian identities are trivially stable.
The example of section \ref{sec:Maxwellvec} illustrates this closure of the algorithm.
\label{it:II}
\item The $\varphi^{(n)}$'s functionally depend on the $(n-1)$-th stage functionally independent Lagrangian constraints.
Schematically, $\varphi^{(n)}\underset{n-2}{\approx}(f_1+f_2\partial_i)\varphi^{(n-1)}$, where $(f_1,f_2)$
are arbitrary real smooth functions of the generalized coordinates and velocities $(Q^A,\dot{Q}^A)$, such that $(f_1,f_2)$ are naturally defined in $T\mathcal{C}_{n-2}$.
Then, it readily follows that $\varphi^{(n)}\underset{n-1}{\approx}0$ and it is obvious that such Lagrangian constrains are stable.
This closure happens in both of the examples in section \ref{sec:gravity}.
\label{it:III}
\end{enumerate}
\end{enumerate}

In all the detailed cases, the total number of functionally independent Lagrangian constraints is given by
\begin{align}
\label{eq:lfinal}
\displaystyle l=\sum_{a=1}^{n-1}M^\prime_a,
\end{align}
where $M^\prime_a$ counts the number of functionally independent $a$-th stage Lagrangian constraints and $n\geq 2$.
We are not aware of any physically relevant example of a field theory where $n$ is infinite.

\vspace*{0.5cm}

\hspace*{-0.6cm}{\bf Noteworthy considerations.}\\
We restate that it is of utmost importance to close the iterative procedure in order to determine $l$.
If the algorithm is not closed (only some or none of the constraints are stabilized), one can only give a lower bound on $l$.
While this may be enough to ensure the absence of Ostrogradsky instabilities~\cite{Ostr} in the field theory,
it is insufficient to guarantee the propagation of a definite number of degrees of freedom.
In such case, one can only infer an upper bound on $n_{\textrm{dof}}$.
This observation is further discussed and exemplified in section \ref{sec:final}.

We also point out that, in general, the different stabilizations of the functionally independent Lagrangian constraints
that we listed  are all present in a given field theory.
Namely, some functionally independent Lagrangian constraints in the theory are stabilized dynamically,
while others are stabilized non-dynamically.
This is indeed what happens in our examples of sections \ref{sec:Pleb} and \ref{sec:gravity}.

Besides, we warn the readers against deceiving themselves regarding the ease of the  exposed iterative algorithm.
Even though our methodology is sound and rigorous and its logic is easy to follow,
there can be no misapprehension as to the algebraic complexity of its implementation in concrete theories, most significantly those involving gravity.
From this point of view, the examples in section \ref{sec:vectors} are uninvolved, while that in section \ref{sec:Palatini} is quite challenging.
The example in section \ref{sec:Palatinid2} constitutes an intermediate difficulty case.
We comment further on this important (from a practical point of view) topic in section \ref{sec:final}.

At last, we remark that the algorithm just exposed does not break covariance.
Namely, if a field theory within our postulates is covariant, its study under the outlined iterative methodology will preserve this feature.
Nonetheless, a suitable space and time decomposition of the a priori independent field variables
and an evaluation of the Lagrangian constraints in the various constraint surfaces will generically break manifest covariance.
This should not be confused with the loss of covariance.

%%%%%%%%%%%%%%%%%%%%%%%%%%%%%%%%%%%%%%%%%%%%%%%%%%%%%%%%%%%%%%%%%%%%%%%%%%%%%%%%%%%%%%%%%

%%%%%%%%%%%%%%%%%%%%%%%%%%%%%%%%%%%%%%%%%%%%%%%%%%%%%%%%%%%%%%%%%%%%%%%%%%%%%%%%%%%%%%%%%
\subsection{Off shell gauge identities}
\label{sec:offshell}
%%%%%%%%%%%%%%%%%%%%%%%%%%%%%%%%%%%%%%%%%%%%%%%%%%%%%%%%%%%%%%%%%%%%%%%%%%%%%%%%%%%%%%%%%

We now obtain $g$ and $e$, the two remaining numbers in the triplet $t^{(N)}$ defined in (\ref{eq:tripletNdef}) of our interest.
To begin with, we notice that in the principle of stationary action (\ref{eq:EAvar}), we have so far only considered that $\delta S=0$ follows from the $E_A$ piece.
However, $\delta S=0$ may also follow from the $\delta Q^A$ piece.
Subsequently, we briefly review the latter scenario: how the vanishing of $\delta S$ may be a consequence of off shell identities
stemming from a strict symmetry of the action.
This kind of symmetry --- gauge invariance--- is only  manifest through specific field variations $\delta_\theta Q^A$,
in contrast to our previous consideration in section \ref{sec:onshell} of arbitrary $\delta Q^A$'s.
Correspondingly, we will differentiate between $\delta_\theta S$ and $\delta S$ as well.

There are different methods to obtain the said off shell identities, but it is not our goal to provide an overview of them here.
Our subsequent discussion summarizes and employs the approach put forward in~\cite{Samanta:2007fk}
and later on adapted to exhibit manifest covariance in~\cite{McKeon:2010nf}.
This adaptation makes it straightforward to apply~\cite{Samanta:2007fk} to any manifestly first order classical field theory,
which is our framework.

Consider the field transformations $ Q^A\rightarrow Q^A+\delta_\theta Q^A$. Let the changes $\delta_\theta Q^A$ be of the form
\begin{align}
\label{eq:fieldvarform}
\delta_\theta Q^A=\sum_{s=0}^n(-1)^s\left(\partial_{\mu_1}\partial_{\mu_2} \dots \partial_{\mu_s} \theta^\beta\right)
(\Omega_{\beta}{}^{A})^{\mu_1\mu_2\ldots \mu_s},
\end{align}
where $n\in \mathbb{N} \cup\{0\}$, $\beta$ is an (possibly collective) index that is to be summed over and the $\theta^\beta$'s and $\Omega_{\beta}{}^{A}$'s are
known as the \emph{gauge parameters} and \emph{gauge generators} of the transformation, respectively.
The the $\theta^\beta$'s are real smooth functions of the spacetime coordinates $x^\mu$, while the $\Omega_{\beta}{}^{A}$'s are defined in $T\mathcal{C}$
and as such are real smooth functions of $(Q^A,\dot{Q}^A)$.
The former are unspecified, while the latter are to be determined.
Introducing the above in (\ref{eq:EAvar}) and operating, one finds that
\begin{align}
\label{eq:defvarrho}
\delta_\theta S=\int_{\mathcal{M}}d^dx\, \theta^\beta\varrho_\beta, \qquad
\varrho_\beta:=\sum_{s=0}^n\partial_{\mu_1} \dots \partial_{\mu_s} \big[E_A(\Omega_{\beta}{}^{A})^{\mu_1\mu_2\ldots \mu_s}\big].
\end{align}
If, under the field variations (\ref{eq:fieldvarform}) for some $\Omega_{\beta}{}^{A}$'s, the action remains invariant $\delta_\theta S\equiv0$, then we have that
\begin{align}
\label{eq:gaugeid}
\varrho_\beta\equiv0
\end{align}
holds true off shell (i.e.~without making use of $E_A\overset{!}{=}0$).
In such a case, (\ref{eq:fieldvarform}) and (\ref{eq:gaugeid}) are known as the \emph{gauge transformations} and \emph{gauge identities}
in the theory, respectively.

Given (\ref{eq:fieldvarform}), $g$ is equal to the number of different $\theta$ parameters there present.
Equivalently, $g$ is the number of linearly independent gauge identities \eqref{eq:gaugeid}.
On the other hand, $e$ is equal to the total number of distinct parameters plus their successive time derivatives
$(\theta,\dot{\theta},\ddot{\theta},\ldots)$ that appear in (\ref{eq:fieldvarform}).
Obviously, $e\geq g$.

The recursive construction of the gauge generators $\Omega_{\beta}{}^{A}$ has been a subject of vivid interest for decades.
The approach in~\cite{Banerjee:1999sz} is perhaps the most befitting to our own exposition,
requiring only a suitable adaptation from particle systems to manifestly first order field theories that is devoid of conceptual subtleties.
We shall not present the corresponding discussion here because, for the theories at hand,
the explicit form of the gauge transformations is already known.
This a priori knowledge allows us to effortlessly infer the generators $\Omega_{\beta}{}^{A}$ in all the subsequent examples.

We stress that the determination of $g$ and $e$ is possible and has been made systematic in theories for which the gauge transformations
are unknown from the onset.
The calculations in such theories are more involved, but there is no theoretical obstacle that has to be overcome.
To illustrate this point, the reader can consult~\cite{Samanta:2007fk}
for the explicit derivation of the gauge generators in Yang-Mills theory
and both the metric and Palatini formulations of General Relativity, by means of the formalism put forward in~\cite{Banerjee:1999sz}.

\begin{figure}[ht]
\centering
\includegraphics[width=1\linewidth]{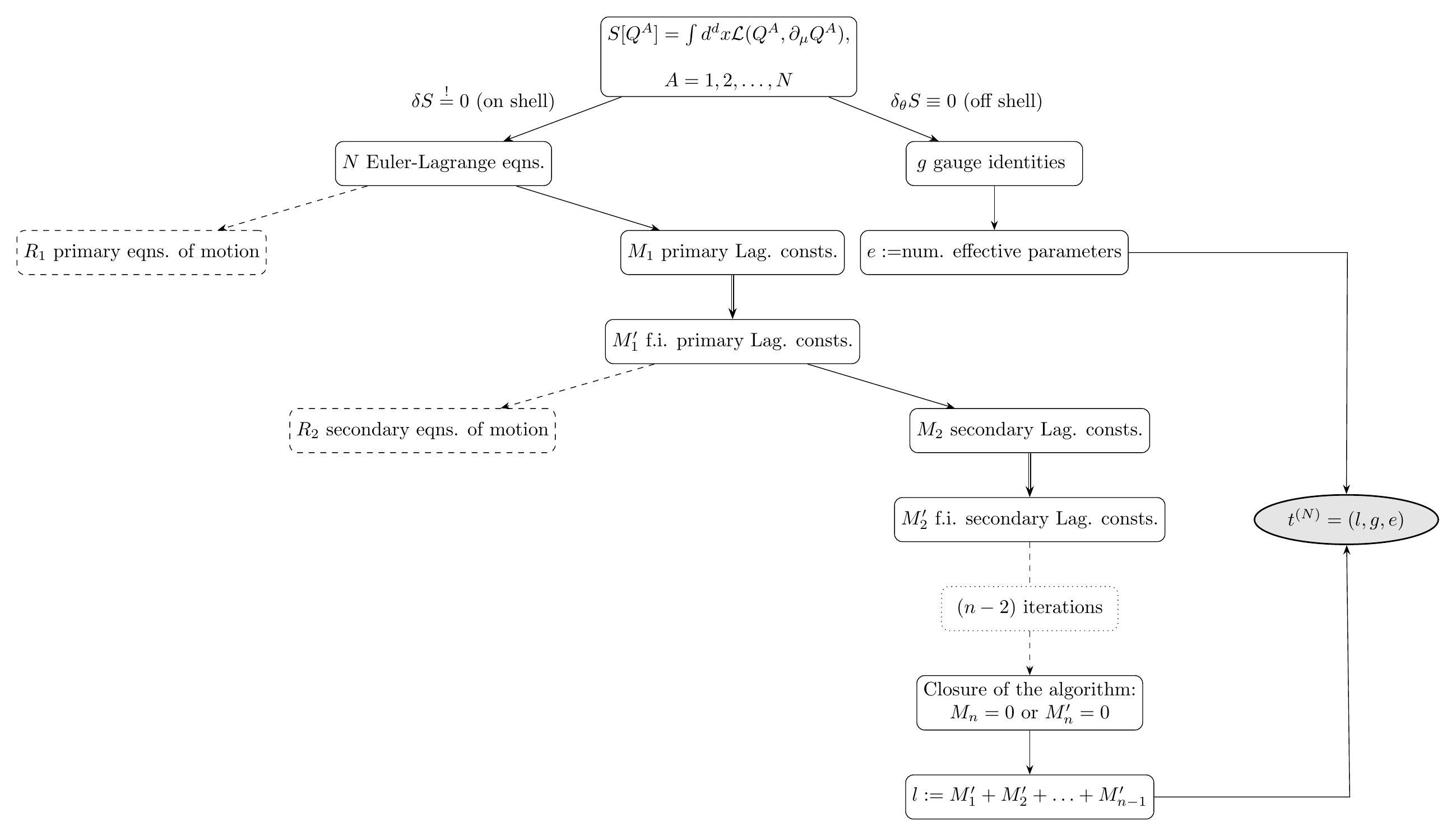}
\caption{\label{fig:scheme}Schematics of section \ref{sec:method}.
Here, (eqns., Lag. consts., f.i., num.) stand for equations, Lagrangian constraints, functionally independent and number, respectively.
The computational challenge of the steps relating Lagrangian constraints to functionally independent Lagrangian constraints (represented with a double arrow),
as well as the relevance of closing the iterative algorithm are further discussed in section \ref{sec:final}.}
\end{figure}

For the ease of the reader, we have schematically depicted the main line of reasoning behind this section \ref{sec:method} in figure \ref{fig:scheme}.

%%%%%%%%%%%%%%%%%%%%%%%%%%%%%%%%%%%%%%%%%%%%%%%%%%%%%%%%%%%%%%%%%%%%%%%%%%%%%%%%%%%%%%%%%

%%%%%%%%%%%%%%%%%%%%%%%%%%%%%%%%%%%%%%%%%%%%%%%%%%%%%%%%%%%%%%%%%%%%%%%%%%%%%%%%%%%%%%%%%
\section{Simple examples: vector field theories}
\label{sec:vectors}
%%%%%%%%%%%%%%%%%%%%%%%%%%%%%%%%%%%%%%%%%%%%%%%%%%%%%%%%%%%%%%%%%%%%%%%%%%%%%%%%%%%%%%%%%

This section is devoted to the study of some of the constraint structure characterizing triplets $t^{(N)}$ that are possible for the theories
(within the framework of section \ref{sec:method}) describing the dynamics of a single vector field.
Recall there are only two distinct types of vector fields that one can entertain classically: massless and massive.
For simplicity, we will restrict to real Abelian vector fields and focus on their most elementary actions: Maxwell electromagnetism and the (hard) Proca theory, respectively.
We shall consider two equivalent formulations of each of these theories, based on different numbers $N$ and $\pmb{N}\neq N$ of a priori independent field variables.
Our forthcoming detailed analyses are based on the purely Lagrangian method described in the previous section \ref{sec:method} and thus serve to illustrate it.

Besides and as we shall explain in section \ref{sec:MPrel}, our forthcoming elementary calculations turn out to be enough to understand
the complete set of manifestly first order (self-)interactions among an arbitrary number of both Maxwell
and (generalized) Proca~\cite{Tasinato:2014eka} fields
in four-dimensional flat spacetime~\cite{ErrastiDiez:2019trb,ErrastiDiez:2019ttn}.
This hints to the convenience of the proposed method, compared to other possible approaches;
a point that shall be reinforced in the more elaborate examples of the next section \ref{sec:gravity}
and discussed in the concluding section~\ref{sec:final}.

In the remaining of this section, we shall work on $d$-dimensional Minkowski spacetime, still for finite $d\geq2$.
We will choose Cartesian coordinates with the mostly positive signature, so that $g_{\mu\nu}=\eta_{\mu\nu}=\textrm{diag}(-1,1,1,\ldots,1)$.
Subsequently, all spacetime indices shall be raised/lowered by $\eta_{\mu\nu}$ and its inverse $\eta^{\mu\nu}$.

%%%%%%%%%%%%%%%%%%%%%%%%%%%%%%%%%%%%%%%%%%%%%%%%%%%%%%%%%%%%%%%%%%%%%%%%%%%%%%%%%%%%%%%%%

%%%%%%%%%%%%%%%%%%%%%%%%%%%%%%%%%%%%%%%%%%%%%%%%%%%%%%%%%%%%%%%%%%%%%%%%%%%%%%%%%%%%%%%%%
\subsection{Maxwell electromagnetism}
\label{sec:Maxwellvec}
%%%%%%%%%%%%%%%%%%%%%%%%%%%%%%%%%%%%%%%%%%%%%%%%%%%%%%%%%%%%%%%%%%%%%%%%%%%%%%%%%%%%%%%%%

This renowned manifestly first order singular field theory describes an Abelian massless vector field and its linear interactions with sources
in terms of $N=d$ number of a priori independent field variables.
As already stated, we take the Maxwell vector field (which we denote $A_\mu$) to be real
and consider the particularly simple case when there are no sources.

\vspace*{0.5cm}

\hspace*{-0.6cm}{\bf Lagrangian constraints.}\\
The canonically normalized Lagrangian density of sourceless classical electromagnetism is
\begin{align}
\label{eq:covMax}
\mathcal{L}_{\textrm{M}}=-\frac{1}{4}A_{\mu\nu}A^{\mu\nu}, \qquad \textrm{with } A_{\mu\nu}:=\partial_\mu A_\nu-\partial_\nu A_\mu=-A_{\nu\mu}.
\end{align}
The components of the Maxwell field constitute the generalized coordinates for this theory: $Q^A=\{A_\mu\}$,
so that $A=1,2,\ldots,d=\textrm{dim}(\mathcal{C})\equiv N$, as already announced.
As is well-known and can be easily calculated by means of (\ref{eq:ELeqs}), the Euler-Lagrange equations following from (\ref{eq:covMax}) are
\begin{align}
\label{eq:Maxeom}
E_A\equiv -\partial_\mu A^{\mu\nu}\overset{!}{=}0.
\end{align}

If we decompose the Maxwell field into its space and time components $A_\mu:=(A_0,A_i)\equiv (A,A_i)$ with $i=1,2,\dots,d-1$;
the Lagrangian \eqref{eq:covMax} can be conveniently rewritten as
\begin{align}
\label{eq:MaxLag}
\mathcal{L}_{\textrm{M}}=\frac{1}{2}\big[\dot{A}_i^2+(\partial_iA)^2-2\dot{A}_i\partial_iA\big]-\frac{1}{4}A_{ij}^2,
\end{align}
where sum over repeated indices is to be understood and we have been careful to lower all indices with the flat metric $\eta_{\mu\nu}$.
It is then easy to see that the primary Hessian following from (\ref{eq:MaxLag}) is $W_{AB}=\delta_{AB}-\delta_A{}^1\delta_B{}^1$.
and therefore manifestly possesses the symmetry dictated by its very definition: $W_{AB}=W_{BA}$.
Further, its Moore-Penrose pseudo-inverse is given by $M^{AB}=\delta^{AB}-\delta_1{}^A\delta_1{}^B$.
Since the primary Hessian takes such an uncomplicated form, it readily follows that $\mathcal{R}_1 =3$ and thus $M_1=1(=M_1^\prime)$ in this case.
A convenient choice for the null vector of $W_{AB}$ amounts to $(\gamma_1)^A=\delta_1{}^A$.
Then, the one and only primary Lagrangian constraint for the theory can be effortlessly calculated to take the explicit form
\begin{align}
\label{eq:Maxcons1}
\varphi_1=\partial_i A_{i0}\overset{!}{\underset{1}{:\approx}}0.
\end{align}
This is the familiar Gauss law, telling us that, in the absence of sources, the electric field is divergenceless.
Note that this is an on shell statement by construction.

The Gauss law constraint straightforwardly yields a vanishing secondary Hessian $\widetilde{W}_{11}\equiv0$,
so that $M_2=M_1=1$ and we choose $(\widetilde{\gamma}_1)^1=1$.
With all this information, it is a matter of easy algebra to find the only secondary Lagrangian constraint:
\begin{align}
\label{eq:MaxLagid}
\widetilde{\varphi}_1\underset{1}{\approxident}0.
\end{align}
Therefore, $M_2^\prime=0$ and the end of the iterative algorithm   is signalled
according to the non-dynamical prescription in case \ref{it:II}.
We have thus found that the total number of Lagrangian constraints for Maxwell electromagnetism is just $l=M_1^\prime+M_2^\prime=1$.

\vspace*{0.5cm}

\hspace*{-0.6cm}{\bf Gauge identities.}\\
Maxwell's theory enjoys an apparent $U(1)$ gauge symmetry.
Indeed, under the transformation $A_\mu\rightarrow A_\mu+\partial_\mu\theta$, the Lagrangian (\ref{eq:covMax}) remains invariant.
Here, $\theta$ is the only gauge parameter, while $(\theta,\dot{\theta})$ are the sole two effective gauge parameters
present in the fields' transformation.
Consequently, we have that $g=1$ and $e=2$.

For completeness, we point out that the said transformation, when compared to (\ref{eq:fieldvarform}) immediately allows us
to read off the gauge generator of the symmetry.
This is $(\Omega^A)^\nu=-\delta^{A\nu}$.
When combined with the primary Euler-Lagrange equations (\ref{eq:Maxeom}) as indicated in (\ref{eq:defvarrho}),
we can right away verify the off shell gauge identity we counted: $\varrho=\partial_\mu\partial_\nu A^{\mu\nu}\equiv0$.

\vspace*{0.5cm}

\hspace*{-0.6cm}{\bf Physical degrees of freedom.}\\
According to our prior analysis, which shows that the constraint structure of classical electromagnetism in its standard formulation with $N=d$ is
\begin{align}
\label{eq:tripletM}
t_{\textrm{M}}^{(N)}=(l=1,g=1,e=2),
\end{align}
and making use of the master formula (\ref{eq:dofformula}),
we count $n_{\textrm{dof}}=d-2$ propagating modes.
In $d=4$, these correspond to the two polarizations of the photon.
Exploiting the equalities in (\ref{eq:HamLageq}), we check that our counting corresponds to two first class constraints,
one primary and one secondary.
Therefore, our purely Lagrangian investigation is in perfect agreement with the standard literature, e.g.~\cite{Sunder}.
It also matches the Hamiltonian definition of the Maxwell field given in~\cite{ErrastiDiez:2019trb}:
``a real Abelian vector field $[\ldots]$ associated with two first class constraints''.
This latter correspondence will play a role in section \ref{sec:MPrel}.

%%%%%%%%%%%%%%%%%%%%%%%%%%%%%%%%%%%%%%%%%%%%%%%%%%%%%%%%%%%%%%%%%%%%%%%%%%%%%%%%%%%%%%%%%

%%%%%%%%%%%%%%%%%%%%%%%%%%%%%%%%%%%%%%%%%%%%%%%%%%%%%%%%%%%%%%%%%%%%%%%%%%%%%%%%%%%%%%%%%
\subsection{The (hard) Proca theory}
\label{sec:Procavec}
%%%%%%%%%%%%%%%%%%%%%%%%%%%%%%%%%%%%%%%%%%%%%%%%%%%%%%%%%%%%%%%%%%%%%%%%%%%%%%%%%%%%%%%%%

We turn our attention to the Proca theory next, in the modern formulation of the original proposal in~\cite{Proca}.
Namely, we focus on the (manifestly first order) field theory of a real Abelian vector field of mass $m$ in the absence of any source
described by $N=d$ a priori independent field variables.
The remark (hard) is to avoid ambiguity with respect to the Generalized Proca theory, discussed in section \ref{sec:MPrel}.
We refer to the Proca field as $B_\mu$.

\vspace*{0.5cm}

\hspace*{-0.6cm}{\bf Lagrangian constraints.}\\
The Lagrangian density of the said Proca theory is
\begin{align}
\label{eq:covPro}
\mathcal{L}_{\textrm{P}}=-\frac{1}{4}B_{\mu\nu}B^{\mu\nu}-\frac{1}{2}m^2B_\mu B^\mu, \qquad \textrm{with } B_{\mu\nu}:=\partial_\mu B_\nu-\partial_\nu B_\mu=-B_{\nu\mu}.
\end{align}
As in the Maxwell case earlier on, the components of the Proca field are the generalized coordinates: $Q^A=\{B_\mu\}$.
We thus see that $A=1,2,\ldots,d=\textrm{dim}(\mathcal{C})\equiv N$ here as well.
The Euler-Lagrange equations following from (\ref{eq:covPro}) can be easily obtained as indicated in (\ref{eq:ELeqs}).
The result is
\begin{align}
\label{eq:Proeom}
E_A\equiv -\partial_\mu B^{\mu\nu}+m^2B^\nu\overset{!}{=}0.
\end{align}

At this point, it is straightforward to see that the primary Hessian ---and hence also its Moore-Penrose pseudo-inverse---
is the same as for the Maxwell theory earlier on.
This implies $M_1=1(=M_1^\prime)$ and the associated null vector can again be chosen as $(\gamma_1)^A=\delta_1{}^A$.
The primary Lagrangian constraint differs, though:
\begin{align}
\label{eq:Procprim}
\varphi_1=\partial_i B_{i0}-m^2B\overset{!}{\underset{1}{:\approx}}0,
\end{align}
where we have introduced $B_\mu:=(B_0,B_i)\equiv (B,B_i)$.

The above once more leads to a vanishing secondary Hessian, so that $M_2=M_1=1$ and $(\widetilde{\gamma}_1)^1=1$.
The secondary Lagrangian constraint in this case takes the form
\begin{align}
\label{eq:Procsec}
\widetilde{\varphi}_1=-m^2\dot{B}\overset{!}{\underset{2}{:\approx}}0.
\end{align}
Contrary to the Maxwell theory, \eqref{eq:Procsec} is obviously not a Lagrangian identity, so the algorithm is not closing here according to the prescription in case \ref{it:II}.
Notice as well that $\varphi_1$ and $\widetilde{\varphi}_1$ are functionally independent from each other,
so that we are not in case \ref{it:III} of the general method either.
Instead, we have $M_2^\prime=M_2=1$ and we must move on to the tertiary stage.

It is easy to check that the tertiary Hessian following from (\ref{eq:Procsec}) is $\widehat{W}_{11}=-m^2$.
As such, its dimension and rank match ($M_3=0=M_3^\prime$) and the algorithm closes according to the dynamical prescription in case \ref{it:I}.
Namely, the consistency of (\ref{eq:Procsec}) under time evolution is ensured via a tertiary equation of motion and there are no tertiary constraints.
As a result, we have obtained $l=M_1^\prime+M_2^\prime+M_3^\prime=2$ functionally independent Lagrangian constraints in the (hard) Proca theory.

\vspace*{0.5cm}

\hspace*{-0.6cm}{\bf Gauge identities.}\\
The mass term for the Proca field explicitly breaks the $U(1)$ gauge invariance of Maxwell electromagnetism.
In our conventions, this means that there is no field transformation of the form (\ref{eq:fieldvarform})
that leaves the action invariant.
Therefore, there are no off shell identities associated to (\ref{eq:covPro}) and we have $g=0=e$.

\vspace*{0.5cm}

\hspace*{-0.6cm}{\bf Physical degrees of freedom.}\\
Using the (hard) Proca constraint structure for $N=d$
\begin{align}
\label{eq:tripletP}
t_{\textrm{P}}^{(N)}=(l=2,g=0,e=0)
\end{align}
  obtained before in the master formula (\ref{eq:dofformula}),
we count $n_{\textrm{dof}}=d-1$ degrees of freedom in the theory.
By means of (\ref{eq:HamLageq}), it is immediate to certify that this corresponds to two second class constraints;
as explicitly shown, for instance, in~\cite{Darabi:2011st}.
As with the Maxwell field before, we thus find agreement with the Proca field's definition given in~\cite{ErrastiDiez:2019trb}:
``a real Abelian vector field $[\ldots]$ associated with two second class constraints''.
We will further comment on this connection in section \ref{sec:MPrel} later on.

%%%%%%%%%%%%%%%%%%%%%%%%%%%%%%%%%%%%%%%%%%%%%%%%%%%%%%%%%%%%%%%%%%%%%%%%%%%%%%%%%%%%%%%%%

%%%%%%%%%%%%%%%%%%%%%%%%%%%%%%%%%%%%%%%%%%%%%%%%%%%%%%%%%%%%%%%%%%%%%%%%%%%%%%%%%%%%%%%%%
\subsection{The Schwinger-Plebanski reformulation of Maxwell and Proca}
\label{sec:Pleb}
%%%%%%%%%%%%%%%%%%%%%%%%%%%%%%%%%%%%%%%%%%%%%%%%%%%%%%%%%%%%%%%%%%%%%%%%%%%%%%%%%%%%%%%%%

In this section, we reanalyze the constraint structures of the above massless and massive vector field theories in a formulation with $\pmb{N}\neq N=d$ a priori degrees of freedom. 
Specifically, we entertain the reformulation of sourceless classical electromagnetism originally proposed by Schwinger~\cite{Schwinger:1953}
and later on popularized by Plebanski~\cite{Pleb}
and employ it for the (hard) Proca theory simultaneously.
In this setup, the real Abelian (covariant) vector field $C_\mu$ ---be it massless or massive--- and its antisymmetric (contravariant) field strength $F^{\mu\nu}$
are regarded as independent at the onset: 
\begin{align}
\label{eq:QAVec}
Q^A=\{C\equiv C_0,F^{ij}=-F^{ji},F^i\equiv F^{0i}=-F^{i0},C_i\}, \qquad 
A=1,2,\ldots, \pmb{N}= d(d+1)/2.
\end{align}

The aim of this section \ref{sec:Pleb} is to determine the constraint structure characterizing triplets $t_{\textrm{M}}^{(\pmb{N})}$ and $t_{\textrm{P}}^{(\pmb{N})}$,
so as to illustrate in a simple double-example the general claim in (\ref{eq:NandNprime}).
Namely, these triplets differ from the previously determined ones $t_{\textrm{M}}^{(N)}$ and $t_{\textrm{P}}^{(N)}$, but yield the same
number of propagating degrees of freedom.

A clarifying remark follows.
Classical electromagnetism as written in~\cite{Schwinger:1953} is commonly called the manifestly first order formulation of electrodynamics.
This refers to the order of its primary Euler-Lagrange equations, contrarily to our convention here, where the order refers to the Lagrangian density.
For us, all examples in sections \ref{sec:vectors} and \ref{sec:gravity} are manifestly first order
and as such can be investigated by means of the methodology in section \ref{sec:method}.
In view of this dissonance, we can already anticipate that there will be no primary equations of motion in our subsequent examples.
The primary Euler-Lagrange equations, being first order, will not involve the generalized accelerations $\ddot{Q}^A$ and so they will all be primary Lagrangian constraints.
Further, this is possible iff the primary Hessian of the theories identically vanishes, as we shall see it does.

\vspace*{0.5cm}

\hspace*{-0.6cm}{\bf Lagrangian constraints.}\\
Inspired by~\cite{Schwinger:1953}, we take the Lagrangian density
\begin{align}
\label{eq:Lag.Max-1st-ord}
\mathcal{L}_{\textrm{V}}=\mathcal{L}_{\textrm{V}}[C_\mu, F^{\mu\nu}]= -\frac{1}{2} (\partial_\mu C_\nu - \partial_\nu C_\mu) F^{\mu \nu} +
\frac{1}{4} F_{\mu \nu} F^{\mu \nu}-\frac{1}{2}m^{2}C_{\mu}C^{\mu} \qquad \textrm{with }\quad m\geq0,
\end{align}
as our starting point.
When $m=0$,  \eqref{eq:Lag.Max-1st-ord} describes classical electromagnetism.
For $m\neq 0$, the (hard) Proca theory is portrayed.
The Euler-Lagrange equations following from (\ref{eq:Lag.Max-1st-ord}) are
\begin{align}
\label{eq:ELPleb}
E_{(C_\nu)}:=-\partial_\mu F^{\mu \nu} +m^2 C^\nu\overset{!}{=} 0, \qquad E_{(F^{\mu\nu})}:=F_{\mu \nu} -\partial_\mu C_\nu + \partial_\nu C_\mu\overset{!}{=}0.
\end{align}
Solving the latter for $F_{\mu\nu}$ and substituting the result into the former, we recover Maxwell's (\ref{eq:Maxeom}) or Proca's (\ref{eq:Proeom}) equations of motion,
depending on the value of $m$.
Then, we say both formulations, in (\ref{eq:Lag.Max-1st-ord}) and in (\ref{eq:covMax}) or (\ref{eq:covPro}) as pertinent, are dynamically equivalent, as foretold.

We proceed to explicitly confirm our predictions.
The primary Hessian following from \eqref{eq:Lag.Max-1st-ord} vanishes identically $W_{AB} \equiv 0$,
so $\mathcal{R}_1=0$ and $M_1 = \pmb{N}$.
We can choose its appropriate null vectors as $(\gamma_I)^A=\delta_I{}^A$.
As a result, the primary Lagrangian constraints coincide with the primary Euler-Lagrange equations.
These can be readily seen to be functionally independent among themselves.
Consequently, the first constraint surface $T\mathcal{C}_1$ coincides with the moduli space in this case.
This set of circumstances can be summarized as
\begin{align}\label{eq:E-Lalphaphi}
0\overset{!}{=} E_A=\alpha_A=(\gamma_I)^A\alpha_A=\varphi_I \overset{!}{\underset{1}{:\approx}}0
\end{align}
or simply as $M_1^\prime=M_1=\pmb{N}$.
Notice that $W_{AB}\equiv 0$ immediately makes its Moore-Penrose pseudo-inverse vanish as well: $M^{AB}=0$.
We encounter this same situation of a zero primary Hessian in both of the theories analyzed in section \ref{sec:gravity}.

We briefly depart from the application of the iterative algorithm in order to introduce an extremely useful notation  that will be recurrent from now on.
We wish to be able to refer to each kind of field variables in \eqref{eq:QAVec} individually.
To this aim, we shall henceforth understand that the index $A$ therein decomposes into two distinct sets of indices $A\equiv \mathcal{A}_1\mathcal{A}_2$,
the first referring to the type of field variable and the second to the spacetime structure of each type of field variable.
In this way, $\mathcal{A}_1=1,2,\ldots,4$ and we have
\begin{align}
\label{eq:[]not}
[Q^{1}]\equiv C, \quad \qquad [Q^2]^{ij}\equiv F^{ij}, \quad \qquad [Q^{3}]^{i} \equiv F^{i}, \quad \qquad  [Q^4]_{i}\equiv C_{i}.
\end{align}
Observe that we have employed the symbol $[\cdot]$ to visually split the $\mathcal{A}_1$ index from the $\mathcal{A}_2$ one.

Back to the algorithm and putting into practice the above notation, we write the primary Lagrangian constraints as
\begin{align}
\displaystyle
\label{eq:PrimCts1stMax}
\begin{array}{llllllllll}
&[\varphi_1]&\hspace*{-0.3cm}=&\hspace*{-0.3cm}\partial_i F^{i}-m^{2}C,
&\qquad [\varphi_2]_{ij}&\hspace*{-0.3cm}=&\hspace*{-0.3cm}F_{ij} -2\partial_{[i} C_{j]}, \vspace*{0.2cm}\\
&[\varphi_3]_i&\hspace*{-0.3cm}=&\hspace*{-0.3cm}F_{i}+\dot{C}_i-\partial_i C, 
&\qquad [\varphi_4]^i&\hspace*{-0.3cm}=&\hspace*{-0.3cm}-\dot{F}^{i}-\partial_j F^{ij}+m^2C^{i}.
\end{array}
\end{align}
Notice that $[\varphi_2]_{ij}=-[\varphi_2]_{ji}$, as required by definition.

We go on to the secondary stage next.
The secondary Hessian $\widetilde{W}_{IJ}=\partial_{\dot{I}}\varphi_{J}$, can be portrayed in our recently introduced notation as follows:
\begin{align}
\label{eq:secHess-MW}
\widetilde{W}_{IJ}=
\left(
\begin{array}{cccccc}
\vspace{1mm}
[\widetilde{W}_{11}] & [\widetilde{W}_{12}]_{ij} & [\widetilde{W}_{13}]_i & [\widetilde{W}_{14}]^{i} \\
\vspace{1mm}
{}_{ij}[\widetilde{W}_{21}] & {}_{ij}[\widetilde{W}_{22}]_{kl} & {}_{ij}[\widetilde{W}_{23}]_k & {}_{ij}[\widetilde{W}_{24}]^{k} \\
\vspace{1mm}
{}_i[\widetilde{W}_{31}] & {}_i[\widetilde{W}_{32}]_{jk} & {}_i[\widetilde{W}_{33}]_j & {}_i[\widetilde{W}_{34}]^{j} \\
\vspace{1mm}
{}^{i}[\widetilde{W}_{41}] & {}^{i}[\widetilde{W}_{42}]_{jk} & {}^{i}[\widetilde{W}_{43}]_j & {}^{i}[\widetilde{W}_{44}]^{j} \\
\end{array}
\right),
\end{align}
where, for each entry of the secondary Hessian, we have placed the space-like tensorial indices of the field variables
(primary Lagrangian constraints) labeled by $I$ ($J$) to the left (right).
A few explicit examples that should clarify our notation are
\begin{align}
\displaystyle
[\widetilde{W}_{11}]:=\frac{ \partial [\varphi_{ 1}]  }{\partial [Q^{1}] }\equiv \frac{ \partial [\varphi_{ 1}]  }{ \partial C}, \qquad
{}_{ij}[\widetilde{W}_{21}]:= \frac{ \partial [\varphi_{ 1}] }{\partial [Q^{2}]^{ij} }\equiv \frac{ \partial [\varphi_{ 1} ]  }{ \partial F^{ij}}, \qquad
{}^i[\widetilde{W}_{42}]_{jk}:=\frac{ \partial [\varphi_{ 2}]_{jk} }{\partial [Q^{4}]_{i} }\equiv \frac{ \partial [\varphi_{ 2} ]_{jk}  }{ \partial C_{i}}.
\end{align}
The only non-zero components in (\ref{eq:secHess-MW}) are
\begin{align}
\quad {}^j[\widetilde{W}_{43}]_i=\delta_i^j=-\, {}_i[\widetilde{W}_{34}]^j,
\end{align}
which lead to a simple secondary Moore-Penrose pseudo-inverse $\widetilde{M}_{AB}$ with non-zero elements
$ {}_{j} [ \widetilde{M}_{43} ]^{i}=-\delta^{i}_{j} =  {}^{i}[\widetilde{M}_{34} ]_{j}$.
This corresponds to the transpose of \eqref{eq:secHess-MW}.
It is easy to see that $\mathcal{R}_2 :=\text{rank}(\widetilde{W}_{IJ})=2 (d-1)$, which in turn implies that $M_2=(d^2 - 3d + 4)/2$.
We choose the suitably normalized linearly independent null vectors for (\ref{eq:secHess-MW}) as $(\widetilde{\gamma}_R)^I=\delta_R{}^I$. 

The above results can be employed to determine the functionally independent secondary Lagrangian constraints $\widetilde{\varphi}_R\overset{!}{\underset{2}{:\approx}}0$.
First, we calculate $\widetilde{\varphi}_R=(\widetilde{\gamma}_R)^I\dot{\varphi}_I$ and obtain
\begin{align}
\label{eq:primstrong}
[\widetilde{\varphi}_1]=\partial_i \dot{F}^i-m^2\dot{C},
\qquad [\widetilde{\varphi}_2]_{ij}=\dot{F}_{ij} -2\partial_{[i} \dot{C}_{j]},
\end{align}
where again the antisymmetry property $[\widetilde{\varphi}_2]_{ij}=-[\widetilde{\varphi}_2]_{ji}$ required by definition is apparent.
Evaluation on the first constraint surface then gives 
\begin{align}
\label{eq:1stMaxSecC}
[\widetilde{\varphi}_1] \underset{1}{\approx}  m^2( \partial_{i}C^{i}-\dot{C} ), \qquad
[\widetilde{\varphi}_2]_{ij} \underset{1}{\approx} \dot{F}_{ij} +2\partial_{[i}F_{j]},
\end{align}
which respects the noted symmetry, as it must.
In more detail, the evaluation has been carried out as follows.
By setting to zero all $\varphi$'s in (\ref{eq:PrimCts1stMax}), solving for $(\dot{F}_i,\dot{C}_i)$ and plugging the resulting expressions into (\ref{eq:primstrong}).
Next, we need to select only the functionally independent secondary constraints.
It is obvious that the mass $m$ plays a crucial role here,
as could easily be anticipated in view of our results in the previous sections \ref{sec:Maxwellvec} and \ref{sec:Procavec}.
Indeed, if $m=0$, then one constraint identically vanishes in the first constraint surface $[\widetilde{\varphi}_1] \underset{1}{\approxident}0$.
It is thus a Lagrangian identity,
meaning that $[{\varphi}_1]$ is non-dynamically (trivially) stabilized at the secondary stage in this case.
We therefore see that 
\begin{align}
\label{eq:M2max}
{M^\prime_2=}
\begin{cases}
M_2-1 &\quad \textrm{if } m=0,  \\
M_{2}      &\quad \textrm{if }  m>0.
\end{cases}
\end{align}

We turn to the time evolution of the functionally independent secondary constraints, i.e.~we commence the tertiary stage.
The tertiary Hessian can be succinctly expressed as 
\begin{align}
\widehat{W}_{RS}=\left(
\begin{array}{ccccc}
\vspace{1mm}
[\widehat{W}_{11}] & [\widehat{W}_{12}]_{ij} \\
\vspace{1mm}
{}_{ij}[\widehat{W}_{21}] & {}_{ij}[\widehat{W}_{22}]_{kl}\\
\end{array}
\right),
\end{align}
where we have made use of the same notation as in (\ref{eq:secHess-MW}) earlier on, so that
\begin{align}
[\widehat{W}_{11}]\equiv \frac{\partial[\widetilde{\varphi}_1]}{\partial \dot{C}}, \qquad
[\widehat{W}_{12}]_{ij}\equiv \frac{\partial[\widetilde{\varphi}_2]_{ij}}{\partial \dot{C}}, \qquad
{}_{ij}[\widehat{W}_{21}]\equiv \frac{\partial[\widetilde{\varphi}_1]}{\partial \dot{F}^{ij}}, \qquad
{}_{ij}[\widehat{W}_{22}]_{kl}\equiv \frac{\partial[\widetilde{\varphi}_2]_{kl}}{\partial \dot{F}^{ij}}.
\end{align}
Notice that, for $m=0$, the first row $[\widehat{W}_{1R}]$ should not be present, as $[\widetilde{\varphi}_1]$ weakly vanishes in this case.
However, we keep it along here, so that both Maxwell and (hard) Proca theories can be reanalyzed simultaneously.
The non-zero components are explicitly given by
\begin{align}
[\widehat{W}_{11}]=-m^2, \qquad {}_{ij}[\widehat{W}_{22}]_{kl}=2\delta_{i[k} \delta_{l]j}.
\end{align}
Hence, the tertiary Hessian has full rank $\mathcal{R}_3=M^\prime_2$ and consequently $M_3=0$.
Observe that this is true for both the $m=0$ and the $m>0$ cases.
The functionally independent secondary constraints' consistency under time evolution is at this point dynamically ensured
and the algorithm closes according to the prescription in case \ref{it:I}.
The total number of functionally independent Lagrangian constraints is
\begin{align}
\pmb{l}=M^\prime_1+M^\prime_2=
\begin{cases}
d(d-1)+1 &\quad \textrm{if } m=0,  \\
d(d-1)+2      &\quad \textrm{if }  m>0.
\end{cases}
\end{align} 
We see that the mass $m$ gives rise to one more functionally independent Lagrangian constraint,
exactly as in the previous sections, where we found that $l=1$ for electromagnetism, while $l=2$ for the (hard) Proca theory.

Here, the remaining constraints that $\pmb{l}$ counts are associated to the field strength $F^{\mu\nu}$,
as a result of having promoted it to a set of a priori independent field variables.
Notice that there are $d(d-1)$ number of such supplementary constraints, two times the number of independent components in $F^{ij}$.
This duplicity makes it manifest that these fields are superfluous when describing the dynamics of the theory.
In other words, no initial data is needed for them: $F^{ij}$ and $\dot{F}^{ij}$ need not be specified at some initial time $t_1$
when solving their associated equations of motion.
Yet another way to understand this is to map them to the Hamiltonian picture, where they correspond to second class constraints, as we shall shortly see.
 
\vspace*{0.5cm}

\hspace*{-0.6cm}{\bf Gauge identities.}\\
Consider the following transformations of the field variables: $C_\mu\rightarrow C_\mu+\delta_\theta C_\mu$ and $F^{\mu\nu}\rightarrow F^{\mu\nu}+\delta_\theta F^{\mu\nu}$, with
\begin{align}
\label{eq:deltaCF}
\delta_\theta C_\mu = \partial_\mu \theta, \qquad \delta_\theta F^{\mu \nu}=0.
\end{align}
Here, $\theta$ is an arbitrary parameter.
It can be easily checked that, under the said transformations, the Lagrangian (\ref{eq:Lag.Max-1st-ord}) remains invariant iff $m=0$.
Therefore, these are the very same gauge transformations of the massless theory that we noted in section \ref{sec:Maxwellvec},
while the massive theory does not exhibit any kind of gauge symmetry.
Straightforwardly, we count
\begin{align}
\pmb{g}=
\begin{cases}
1 &\quad \textrm{if } m=0,  \\
0 &\quad \textrm{if }  m>0,
\end{cases}
\qquad \qquad \pmb{e}=
\begin{cases}
2 &\quad \textrm{if } m=0,  \\
0 &\quad \textrm{if }  m>0.
\end{cases}
\end{align}

For completeness, we  provide the gauge identity and generators for $m=0$ next.
Comparing (\ref{eq:fieldvarform}) and \eqref{eq:deltaCF}, we can immediately read off the non-zero generators:
\begin{align}
\label{eq:omegaspleb}
\delta_\theta C_\mu=-(\partial_{\mu_1}\theta)\big(\Omega_\mu\big)^{\mu_1}, \qquad \big(\Omega_\mu\big)^{\mu_1}=-\delta^{\mu_1}_\mu.
\end{align}
Notice that here we have dropped the, in this case, single-valued $\beta$ index from (\ref{eq:fieldvarform}).
Putting together (\ref{eq:ELPleb}) and (\ref{eq:omegaspleb}) as indicated in (\ref{eq:defvarrho}), we readily confirm the gauge identity:
\begin{align}
\label{eq:GIvec}
\varrho=\partial_{\mu_1}\Big[E_{(C_\mu)}\big(\Omega_\mu\big)^{\mu_1}\Big]\equiv0.
\end{align}

\hspace*{-0.6cm}{\bf Physical degrees of freedom.}\\
We have now achieved our goal.
Namely, we have shown that the constraint structure characterizing triplet for (\ref{eq:Lag.Max-1st-ord}) is
\begin{align}\label{eq:tVec}
t_{\textrm{V}}^{(\pmb{N})}=
\begin{cases}
t_{\textrm{M}}^{(\pmb{N})}=\Big(\pmb{l}=d(d-1)+1,\pmb{g}=1,\pmb{e}=2\Big) &\quad \textrm{if } m=0, \vspace*{0.3cm}\\
t_{\textrm{P}}^{(\pmb{N})}=\Big(\pmb{l}=d(d-1)+2,\pmb{g}=0,\pmb{e}=0\Big) &\quad \textrm{if }  m>0.
\end{cases}
\end{align}
Substituting  the quantities \eqref{eq:tVec} into the master formula (\ref{eq:dofformula}), we count 
\begin{align}
n_{\textrm{dof}}=
\begin{cases}
d-2 &\quad \textrm{if } m=0, \\
d-1 &\quad \textrm{if }  m>0,
\end{cases}
\end{align}
propagating degrees of freedom.
This counting coincides with the ones performed in sections \ref{sec:Maxwellvec} and \ref{sec:Procavec}, where appropriate.
We have thus verified (\ref{eq:NandNprime}) in two simple examples.
Exploiting the equalities in (\ref{eq:HamLageq}), we see the following relation to the Hamiltonian side.
The massless theory exhibits two first class constraints, one of which is a primary first class constraint, and $d(d-1)$ second class constraints.
On the other hand, the massive theory has only second class constraints, $d(d-1)+2$ of them.
Our purely Lagrangian investigation is thus in perfect agreement with the standard Hamiltonian literature, e.g.~\cite{Sundermeyer:1982gv}.

%%%%%%%%%%%%%%%%%%%%%%%%%%%%%%%%%%%%%%%%%%%%%%%%%%%%%%%%%%%%%%%%%%%%%%%%%%%%%%%%%%%%%%%%%

%%%%%%%%%%%%%%%%%%%%%%%%%%%%%%%%%%%%%%%%%%%%%%%%%%%%%%%%%%%%%%%%%%%%%%%%%%%%%%%%%%%%%%%%%
\section{A comprehensive constraint analysis of Palatini theories}
\label{sec:gravity}
%%%%%%%%%%%%%%%%%%%%%%%%%%%%%%%%%%%%%%%%%%%%%%%%%%%%%%%%%%%%%%%%%%%%%%%%%%%%%%%%%%%%%%%%%

In the following, we apply the general framework presented in section \ref{sec:method}
to the Palatini action.
We split our calculations into the $d>2$ and the $d=2$ cases, as these are physically distinct theories.
As we shall see, the former case is much more algebraically involved than the latter.
However, compared to their equivalent Hamiltonian investigations, our Lagrangian approach shall prove much simpler in both instances.

For concreteness, we specify our framework to be that of the metric-affine Palatini formulation of General Relativity,
ordinarily ascribed to Palatini but firstly suggested by Einstein himself~\cite{Palatini1919,Ferraris:1982}.
As such, we shall study a manifestly first order formulation of gravity based on $N=d(d+1)^2/2$ number of a priori independent degrees of freedom.
Even though alternative manifestly first order formulations do exist, such as the tetradic-Palatini action
(for example, see~\cite{Castellani:1981ue} and its recent canonical study~\cite{Montesinos:2019bkc}),
inconvenient subtleties to our aims arise in those frameworks due to their geometric construction.
For instance, unlike the metric, vielbeine are not required to be invertible.
In such scenario, the strict equivalence between Palatini and Einstein's gravity is lost due to a singular vielbein and, in general,
ends up in a dynamical manifestation of torsion~\cite{Kaul:2016zbn}. 
Similar situations might arise in other manifestly first order formulations, like the Barbero-Holst action~\cite{Holst:1995pc}
or BF-like models~\cite{Lewandowski:1999se}, which happen to enclose the most celebrated Plebanski action.
For a complete review on these topics, we refer the interested reader to~\cite{Celada:2016jdt}.
 
%%%%%%%%%%%%%%%%%%%%%%%%%%%%%%%%%%%%%%%%%%%%%%%%%%%%%%%%%%%%%%%%%%%%%%%%%%%%%%%%%%%%%%%%%

%%%%%%%%%%%%%%%%%%%%%%%%%%%%%%%%%%%%%%%%%%%%%%%%%%%%%%%%%%%%%%%%%%%%%%%%%%%%%%%%%%%%%%%%%
\subsection{Palatini in $d>2$}
\label{sec:Palatini}
%%%%%%%%%%%%%%%%%%%%%%%%%%%%%%%%%%%%%%%%%%%%%%%%%%%%%%%%%%%%%%%%%%%%%%%%%%%%%%%%%%%%%%%%%

The Palatini action in $d>2$ is a well-known (re)formulation of the Einstein-Hilbert action, which is dynamically equivalent to it.
This is explained shortly.
Most significantly for us, Palatini is a manifestly first order formulation of General Relativity,
which treats the spacetime metric $g_{\mu\nu}=g_{\nu\mu}$ and the affine connection $\Gamma^\rho_{\mu\nu}=\Gamma^\rho_{\nu\mu}$
as a priori independent variables.
As such, and unlike Einstein-Hilbert, it readily allows for the application of the methodology introduced in section \ref{sec:method}.

\vspace*{0.5cm}

\hspace*{-0.6cm}{\bf Lagrangian constraints.}\\
The Palatini action is of the general form given in (\ref{eq:action})
and its Lagrangian density can be written as~\cite{Horava:1990ba}
\begin{align}
\label{eq:PalatiniL}
\mathcal{L}_{\textrm{Pa}}=-(\partial_\rho h^{\mu\nu})G^\rho_{\mu\nu}+h^{\mu\nu}(c\,G^\rho_{\rho\mu}G^{\sigma}_{\sigma\nu}
-G^{\rho}_{\sigma\mu}G^{\sigma}_{\rho\nu}), \qquad \textrm{with } c:=\frac{1}{d-1}.
\end{align}
Here, the independent variables $h^{\mu\nu}$ and $G^\rho_{\mu\nu}$
are defined exclusively in terms of the spacetime metric and affine connection, respectively:
\begin{align}
\label{eq:defhG}
h^{\mu\nu}:=\sqrt{-\textrm{det}(g_{\mu\nu})}g^{\mu\nu}, \qquad
G^\rho_{\mu\nu}:=\Gamma^\rho_{\mu\nu}-\delta^\rho_{(\mu}\Gamma^\sigma_{\nu)\sigma}
\end{align}
and thus inherit their symmetry properties.

The primary Euler-Lagrange equations for $h^{\mu\nu}$ and $G^{\lambda}_{\mu\nu}$ following from (\ref{eq:PalatiniL}) are
\begin{align}
\label{eq:FEqsPalatini}
E_{(h^{\mu\nu})}:=\partial_{\rho}G^{\rho}_{\mu\nu}+c\,G^{\rho}_{\rho\mu}G^{\sigma}_{\sigma\nu}
-G^{\rho}_{ \sigma\mu}G^{\sigma}_{\rho\nu}\overset{!}{=}0, \qquad
E_{(G^\rho_{\mu\nu})}:=-\partial_{\rho}h^{\mu\nu}+2c\,G^{\lambda}_{\lambda\sigma}h^{\sigma(\mu}\delta^{\nu)}_{\rho}
-2G^{(\mu}_{\rho\sigma}h^{\nu)\sigma}\overset{!}{=}0.
\end{align}
Notice that these vanishings are on shell statements.
Multiplying the second set of field equations by $h_{\mu\nu}$ and employing the identity $h_{\mu\nu}h^{\nu\rho}=\delta_\mu{}^\rho$,
one finds that
\begin{align}
\label{eq:trGeq}
2 ( c-1) {G}^{\mu}_{\mu\rho} -  h_{\mu\nu} \partial_{\rho}h^{\mu\nu}\overset{!}{=}0.
\end{align}
Solving \eqref{eq:trGeq} implies that $G^{\rho}_{\mu\nu}$ is fixed (on shell) to be a function of $h^{\mu\nu}$ and its first derivatives.
The substitution of the resulting expression into \eqref{eq:PalatiniL} yields the second order formulation of General Relativity
and we say $d>2$ Palatini is dynamically equivalent to it.

It is natural and convenient to decompose the variables in (\ref{eq:defhG}) as follows:
\begin{align}
\displaystyle
\label{eq:vari}
\begin{array}{llllll}
&h\equiv h^{00}, &\qquad h^i\equiv h^{0i}, &\qquad G\equiv G^{0}_{00}, &\qquad G_i\equiv G^{0}_{0i},\\
&G_{ij}\equiv G^{0}_{ij}, &\qquad \mathcal{G}^i\equiv G^{i}_{00}, &\qquad \mathcal{G}^i_j\equiv G^i_{0j}, &\qquad \mathcal{G}^i_{jk}\equiv G^i_{jk}.
\end{array}
\end{align}
The explicit form of the Lagrangian (\ref{eq:PalatiniL}) in terms of the above variables is
\begin{align}
\displaystyle
\label{eq:PalL}
\hspace*{-0.5cm}
\begin{array}{llll}
&\mathcal{L}_{\textrm{Pa}}=&\hspace*{-0.3cm}
-\dot{h}G-2\dot{h}^iG_i-\dot{h}^{ij}G_{ij}-(\partial_ih)\mathcal{G}^i
-2(\partial_ih^j)\mathcal{G}_j^i-(\partial_ih^{jk})\mathcal{G}^i_{jk}
+h\big[(c-1)G^2+2\big(cG\mathcal{G}_i^i-G_i\mathcal{G}^i\big)\\
&&\hspace*{-0.3cm}
+c\mathcal{G}_i^i\mathcal{G}_j^j-\mathcal{G}_j^i\mathcal{G}^j_i\big]
+2h^i\big[(c-1)GG_i+cG\mathcal{G}_{ij}^j+cG_i\mathcal{G}^j_j-G_j\mathcal{G}_i^j-G_{ij}\mathcal{G}^j+c\mathcal{G}^j_j\mathcal{G}_{ik}^k
-\mathcal{G}_j^k\mathcal{G}_{ik}^j\big]\\
&&\hspace*{-0.3cm}
+h^{ij}\big[(c-1)G_iG_j+2\big(cG_{(i}\mathcal{G}_{j)k}^k-G_{k(i}\mathcal{G}_{j)}^k\big)+c\mathcal{G}_{ik}^k\mathcal{G}_{jl}^l
-\mathcal{G}_{il}^k\mathcal{G}_{jk}^l\big].
\end{array}
\end{align}

We express the generalized coordinates of the Palatini Lagrangian in (\ref{eq:PalL}) as
\begin{align}
\label{eq:QAPal}
Q^A=\{h,h^i,h^{ij},G,G_i,G_{ij},\mathcal{G}^i,\mathcal{G}^i_j,\mathcal{G}^i_{jk}\}.
\end{align}
Notice that $A$ comprises all possible indices of our chosen field variables, so that
\begin{align}
\label{eq:NinPal}
A=1,2,\ldots,d(d+1)^2/2=\textrm{dim}(\mathcal{C})\equiv N.
\end{align}
Henceforth, we shall employ the notation $[\cdot]$ introduced in section \ref{sec:Pleb} for the collective index $A$ above.
In particular, see (\ref{eq:[]not}) and explanations around.
This notation shall prove of utmost convenience.
For instance, in this way, it is obvious that
\begin{align}
G\equiv [Q^4]\neq Q^4=
\begin{cases}
h^{11} \qquad &\textrm{if } d=3,\\
h^3 \qquad &\textrm{otherwise}.
\end{cases}
\end{align}

The primary Hessian following from (\ref{eq:PalL}) vanishes identically: $W_{AB}\equiv0$, as a result of having promoted the affine connection
to a set of a priori independent field variables.
This parallels the reformulations of classical electromagnetism and the (hard) Proca theory in section \ref{sec:Pleb}.
In passing, we note that the primary Hessian is symmetric $W_{AB}=W_{BA}$, as it should by definition.
Obviously, $\textrm{rank}(W_{AB})=0$ and we have $M_1=N=d(d+1)^2/2$.
This trivialization of the primary Hessian has a number of direct implications.
First, it allows us to straightforwardly pick its suitably normalized null vectors to be
$(\gamma_I)^A=\delta_I{}^A$.
Second, it immediately makes its Moore-Penrose pseudo-inverse vanish as well: $M^{AB}=0$.
Third, it becomes apparent that the primary Euler-Lagrange equations coincide with the primary Lagrangian constraints.
All of these constraints turn out to manifestly be functionally independent from each other in this specific theory.
In other words, the moduli space is the primary constraint surface in this case and we have $M_1^\prime=M_1$.
Thus,
\begin{align}
\label{eq:primELeqs}
0\overset{!}{=} E_A=\alpha_A=(\gamma_I)^A\alpha_A=\varphi_I \overset{!}{\underset{1}{:\approx}}0,
\end{align}
exactly as in our examples of section \ref{sec:Pleb} before, see (\ref{eq:E-Lalphaphi}).
By means of the notation employed in (\ref{eq:PrimCts1stMax}), the explicit form of the $\varphi_I$'s is
\begin{align}
\displaystyle
\label{eq:primalph}
\hspace*{-0.5cm}
\begin{array}{llll}
&[\varphi_1]&=&-\big[\dot{G}+\partial_{i}\mathcal{G}^{i}+\bm{G}G+2\big(cG\mathcal{G}^{i}_{i}-G_i\mathcal{G}^{i}\big)
+\mathcal{G}^i_j\bm{\mathcal{G}}^j_i\big],\\
&\frac{[\varphi_2]_{i}}{2}&=&-\big[\dot{G}_{i}+\partial_{j}\mathcal{G}^{j}_{i}+\bm{G}G_{i}+cG\mathcal{G}^{j}_{ij}+G_j\bm{\mathcal{G}}^j_i-G_{ij}\mathcal{G}^{j}
+\bm{\mathcal{G}}^j_k\mathcal{G}^k_{ij}\big],\\
&[\varphi_3]_{ij}&=&-\big[\dot{G}_{ij}+\partial_{k}\mathcal{G}^{k}_{ij}+\bm{G}_{i}G_{j}+2\big(cG_{(i}\mathcal{G}^{k}_{j)k}
-G_{k(i}\mathcal{G}^{k}_{j)}\big)+\mathcal{G}^l_{ik}\bm{\mathcal{G}}^k_{lj}\big],\\
&[\varphi_4]&=& \dot{h}-2\big[h\bm{G}+ch\mathcal{G}^{i}_{i}+h^i\bm{G}_{i}+ch^i\mathcal{G}^{j}_{ij}\big], \\
&\frac{[\varphi_5]^{i}}{2}&=&\dot{h}^{i}+h\mathcal{G}^{i}-\big[h^i\bm{G}+h^j\bm{\mathcal{G}}^{i}_{j}
+h^{ij}\bm{G}_{j}+ch^{ij}\mathcal{G}^{k}_{jk}\big],\\
&[\varphi_6]^{ij}&=&\dot{h}^{ij}+2\big(h^{(i}\mathcal{G}^{j)}+h^{k(i}\mathcal{G}^{j)}_{k}\big),\\
&[\varphi_7]_{i}&=& \partial_{i}h+2\big(hG_{i}+h^jG_{ij}\big),\\
&\frac{[\varphi_8]_{i}^{j}}{2}&=& \partial_{i}h^{j}-\big[c\big(hG+h^kG_k\big)\delta_{i}^{j}+h\bm{\mathcal{G}}^j_i
+h^k\bm{\mathcal{G}}^{j}_{ik}\big]+h^jG_{i}+h^{jk}G_{ik},\\
&[\varphi_9]^{jk}_{i}&=& \partial_{i}h^{jk}-2\big[h^{(j}\big(cG\delta^{k)}_i+\bm{\mathcal{G}}^{k)}_i\big)
+h^{l(j}\big(cG_l\delta^{k)}_i+\bm{\mathcal{G}}^{k)}_{il}\big)\big],
\end{array}
\end{align}
where we have defined
\begin{align}
\label{eq:bmvar}
\bm{G}:=(c-1)G,\qquad \bm{G}_i:=(c-1)G_i,\qquad
\bm{\mathcal{G}}^i_j:=c\mathcal{G}^k_k\delta^i_j-\mathcal{G}^i_j, \qquad
\bm{\mathcal{G}}_{jk}^i:=c\mathcal{G}^{l}_{kl}\delta^i_j-\mathcal{G}^{i}_{jk}.
\end{align}
Note that the $\varphi_I$'s in \eqref{eq:primalph} are manifestly symmetric where appropriate, i.e.~$[\varphi_3]_{ij}=[\varphi_3]_{ji}$,
$[\varphi_6]^{ij}=[\varphi_6]^{ji}$ and $[\varphi_9]^{jk}_{i}=[\varphi_9]^{kj}_{i}$.

We now turn to the secondary stage, where we inspect the consistency under time evolution
of the functionally independent primary Lagrangian constraints.
The secondary Hessian is given by $\widetilde{W}_{IJ}=\partial_{\dot{I}}\varphi_J$.
With the conventions introduced below (\ref{eq:secHess-MW}), we may succinctly write it as
\begin{align}
\label{eq:secHess}
\widetilde{W}_{IJ}=
\left(
\begin{array}{cccccc}
[\widetilde{W}_{11}] & \ldots & [\widetilde{W}_{19}]_i^{jk} \\
\vdots & \ddots & \vdots  \\
{}_i^{jk}[\widetilde{W}_{91}] & \ldots & {}_i^{jk}[\widetilde{W}_{99}]_l^{mn} \\
\end{array}
\right).
\end{align}
For clarity, we provide a few examples of what is meant by our notation:
\begin{align}
[\widetilde{W}_{11}]\equiv \frac{\partial[\varphi_1]}{\partial \dot{h}}, \qquad
{}_i[\widetilde{W}_{23}]_{jk}\equiv \frac{\partial [\varphi_3]_{jk} }{\partial \dot{h^i}}, \qquad
[\widetilde{W}_{45}]^{i}\equiv \frac{\partial [\varphi_5]^{i} }{\partial \dot{G}}, \qquad
{}_i^{jk}[\widetilde{W}_{99}]_l^{mn}\equiv \frac{\partial [\varphi_9]^{mn}_{l}}{\partial \dot{\mathcal{G}}^i_{jk}}.
\end{align}
In (\ref{eq:secHess}), the only non-zero components are
\begin{align}
[\widetilde{W}_{14}]=-1=-[\widetilde{W}_{41}], \qquad
\,{}_i[\widetilde{W}_{25}]^j=-2\delta^j_i=-\,{}^j[\widetilde{W}_{52}]_i, \qquad
{}_{ij}[\widetilde{W}_{36}]^{kl}=-\delta^k_{(i}\delta^l_{j)}=-\, {}^{kl}[\widetilde{W}_{63}]_{ij}.
\end{align}
Notice that the secondary Hessian is antisymmetric $\widetilde{W}_{IJ}=-\widetilde{W}_{JI}$, as it should by definition.
It is easy to see that $\mathcal{R}_2:=\textrm{rank}(\widetilde{W}_{IJ})=d(d+1)$, thus yielding $M_2=d(d^2-1)/2$.
This means that $\mathcal{R}_2$ number of the functionally independent primary Lagrangian constraints
are being dynamically stabilized at the secondary stage,
while the remaining $M_2$ primary Lagrangian constraints are not stable: they lead to secondary Lagrangian constraints, which we proceed to determine.

To this aim, we first choose the suitably normalized linearly independent null vectors associated to the secondary Hessian as
\begin{align}
\label{eq:tilde-gamma-Palatini}
(\widetilde{\gamma}_R)^I=(0,0,\ldots,0,1,0,0,\ldots,0),
\end{align}
where the non-vanishing vector component is at $I=\mathcal{R}_2+R$.
Notice that all $M_2$ null vectors have length $M_1$ and their first $\mathcal{R}_2$ components are zero.

All our results so far can be used to obtain the secondary Lagrangian constraints
$\widetilde{\varphi}_R\overset{!}{\underset{1}{\approx}}0$.
These can readily be seen to be functionally independent from each other, as well as with respect to the primary constraints, so that $M_2^\prime=M_2$.
This means their vanishing defines the secondary constraint surface: $\widetilde{\varphi}_R\overset{!}{\underset{2}{:\approx}}0$.
In our $[\cdot]$ notation, we have
\begin{align}
\displaystyle
\label{eq:secin1}
\hspace*{-0.15cm}
\begin{array}{llll}
\frac{[\widetilde{\varphi}_1]_i}{2} \underset{1}{\approx}
h(\widetilde{\tau}_1)_i+h^j(\widetilde{\tau}_2)_{ij}, \quad
\frac{[\widetilde{\varphi}_2]_i^j}{2} \underset{1}{\approx}
h^j(\widetilde{\tau}_1)_i+h^{jk}(\widetilde{\tau}_2)_{ik}+h(\widetilde{\tau}_3)^j_i+h^k(\widetilde{\tau}_4)^j_{ik}, \quad\frac{[\widetilde{\varphi}_3]_i^{jk}}{2} \underset{1}{\approx}
h^{(j}(\widetilde{\tau}_3)^{k)}_i+h^{l(j}(\widetilde{\tau}_4)^{k)}_{il},
\end{array}
\end{align}
where we have defined
\begin{align}
\displaystyle
\label{eq:tildetaus}
\hspace*{-0.55cm}
\begin{array}{lllll}
&(\widetilde{\tau}_1)_i&:=&
\bm{\partial}_iG+\bm{\partial}_{k\cdot i}^l\mathcal{G}^k_l
+\bm{G}\mathbb{G}_i-\bm{G}G_i
+(c-2)G_k\bm{\mathcal{G}}^k_i+\mathcal{G}^k_l\bm{\widetilde{\mathcal{G}}}^l_{ik},
\vspace*{0.1cm}\\
&(\widetilde{\tau}_2)_{ij}&:=&
\bm{\partial}_iG_j+\bm{\partial}_{k\cdot i}^l\mathcal{G}^k_{lj}
+\bm{G}_j\mathbb{G}_i-\bm{G}G_{ij}
+\bm{G}_k\bm{\mathcal{G}}^k_{ij}
-G_{jk}\bm{\mathcal{G}}^k_i+\mathcal{G}^k_{jl}\bm{\widetilde{\mathcal{G}}}^l_{ik},
\vspace*{0.1cm}\\
&(\widetilde{\tau}_3)^j_i&:=&
-\bm{\delta}_{i\cdot m}^{j\cdot l}\dot{\mathcal{G}}^m_l
+\bm{\partial}_{i\cdot l}^j\mathcal{G}^l
+G\bm{\mathcal{G}}^j_i
-G_k\bm{\mathcal{G}}^{k\cdot j}_{i}
+\mathcal{G}^l\bm{\mathcal{G}}_{il}^j
-\mathcal{G}^m_l\bm{\mathcal{G}}^{l\cdot j}_{m\cdot i},
\vspace*{0.1cm}\\
&(\widetilde{\tau}_4)^j_{ik}&:=&
-\bm{\delta}_{i\cdot m}^{j\cdot l}\dot{\mathcal{G}}^m_{lk}
+\bm{\partial}_{i\cdot l}^j\mathcal{G}^l_k
-G_{kl}\bm{\mathcal{G}}^{l\cdot j}_{i}
+\mathcal{G}^l_k\bm{\mathcal{G}}_{il}^j
-\mathcal{G}^m_{kl}\bm{\mathcal{G}}^{l\cdot j}_{m\cdot i}
\end{array}
\end{align}
in terms of (\ref{eq:bmvar}) as well as the following quantities:
\begin{align}
\displaystyle
\label{eq:bmops}
&\hspace*{-0.4cm}
\begin{array}{llllll}
&\bm{\partial}_i:=(c-1)\partial_i,
&\quad\bm{\partial}^k_{i\cdot j}:=c\delta^k_i\partial_j-\delta^k_j\partial_i,
&\quad\bm{\delta}_{i\cdot j}^{k\cdot l}:=c\delta^k_i\delta^l_j-\delta^k_j\delta^l_i,
&\,\mathbb{G}_i:=c\big(G_i+\mathcal{G}^k_{ik}\big),\\
&\bm{\mathcal{G}}^{i\cdot j}_{k\cdot l}:=c\mathcal{G}^i_{k}\delta^j_l-\mathcal{G}^j_{k}\delta^i_l,
&\quad\bm{\mathcal{G}}^{i\cdot j}_{k}:=c\mathcal{G}^i\delta^j_k-\mathcal{G}^j\delta^i_k,
&\quad\bm{\widetilde{\mathcal{G}}}^k_{ij}:=c(c-1)\mathcal{G}^l_{il}\delta^k_j-\bm{\mathcal{G}}^k_{ij}.
&{}
\end{array}
\end{align}
Observe that the appropriate symmetry $[\widetilde{\varphi}_3]_i^{jk}=[\widetilde{\varphi}_3]_i^{kj}$ is manifest.
To obtain the above, we have first computed $\widetilde{\varphi}_R=(\widetilde{\gamma}_R)^I\dot{\varphi}_I$.
Then, we have evaluated the result in the first constraint surface.
In practice, this means that we have substituted $(\dot{G},\dot{G}_i,\ldots,\partial_ih^{jk})$
for their suitable weak expressions in terms of the generalized coordinates $Q^A$, which follow from setting to zero (\ref{eq:primalph}).

To conclude the secondary stage, we calculate the Moore-Penrose pseudo-inverse of $\widetilde{W}_{IJ}$.
It can be easily checked that this is $\widetilde{M}^{IJ}=-(\widetilde{W}_{IJ})^T$.

Next, the consistency under time evolution of the above functionally independent secondary Lagrangian constraints is to be inspected
at the tertiary stage.
The first step is to calculate the tertiary Hessian $\widehat{W}_{RS}=(\widetilde{\gamma}_R)^I\partial_{\dot{I}}\widetilde{\varphi}_S$.
Employing the same conventions as in (\ref{eq:secHess}) before, we write
\begin{align}
\label{eq:terHess}
\widehat{W}_{RS}=
\left(
\begin{array}{cccccc}
\vspace{1mm}
{}_i[\widehat{W}_{11}]_j &{}_i[\widehat{W}_{12}]^k_j & {}_i[\widehat{W}_{13}]_j^{kl} \\
\vspace{1mm}
{}_i^{j}[\widehat{W}_{21}]_l & {}^{j}_i[\widehat{W}_{22}]_l^m  & {{}_i^{j}[\widehat{W}_{23}]_l^{mn}} \\
{}_i^{jk}[\widehat{W}_{31}]_l & {{}^{jk}_i[\widehat{W}_{32}]_l^m}  & {}_i^{jk}[\widehat{W}_{33}]_l^{mn} \\
\end{array}
\right)
\end{align}
where the non-zero components are
\begin{align}
\displaystyle
&{}_i^j[\widehat{W}_{22}]_k^l\equiv \frac{\partial [\widetilde{\varphi}_2]_k^l}{\partial \dot{\mathcal{G}}_j^i}
=-2h\big(c\delta^{j}_i\delta^l_k-\delta^l_i\delta^j_k\big),\qquad\quad
{}_i^{jk}[\widehat{W}_{33}]_l^{mn}\equiv\frac{\partial[\widetilde{\varphi}_3]_l^{mn}}{\partial\dot{\mathcal{G}}^i_{jk}}
=2\big(\delta^{(m}_ih^{n)(j}\delta^{k)}_l-c\delta^{(m}_lh^{n)(j}\delta^{k)}_i\big), \nonumber\\
&{}_i^j[\widehat{W}_{23}]_k^{lm}\equiv\frac{\partial[\widetilde{\varphi}_3]_k^{lm}}{\partial\dot{\mathcal{G}}^i_j}
=-2h^{(l}\big(c\delta^j_i\delta^{m)}_k-\delta^{m)}_i\delta^j_k\big)
=\frac{\partial[\widetilde{\varphi}_2]_i^j}{\partial\dot{\mathcal{G}}^k_{lm}}\equiv{}_k^{lm}[\widehat{W}_{32}]_i^j.
\end{align}
Therefore, the tertiary Hessian obviously satisfies component-wise the symmetry properties that ensure
$\widehat{W}_{RS}=\widehat{W}_{SR}$, as required by definition.
It is not hard to see that $\mathcal{R}_3:=\textrm{rank}(\widehat{W}_{RS})=d(d^2-3)/2$ and hence $M_3=d$.
In more detail, the rank is equal to all non-zero rows (equivalently, columns) of the Hessian, minus one.
There is the obvious set of $d-1$ number of zero rows given by ${}_i[\widehat{W}_{1R}=0$.
But the rank of the Hessian is further reduced by one because the linear combination of rows ${}^i_i[\widehat{W}_{2R}$ is zero.
This vanishing is a direct consequence of the velocity independence of $[\widetilde{\varphi}_2]^i_i$ in $T\mathcal{C}_1$,
as can be readily verified from (\ref{eq:secin1}) and (\ref{eq:tildetaus}).
It follows that $\mathcal{R}_3$ number of the functionally independent secondary Lagrangian constraints
are being dynamically stabilized at the tertiary stage.
The remaining $M_3$ secondary Lagrangian constraints are not stable: they lead to the tertiary Lagrangian constraints that we shall find next.

The suitably normalized linearly independent null vectors of the tertiary Hessian can be chosen as follows.
Associated to ${}_i[\widehat{W}_{1R}=0$, we pick
\begin{align}
\label{eq:hat-gamma-Palatini}
(\widehat{\gamma}_U)^R=(0,0,\ldots,0,1,0,0,\ldots,0), \qquad \textrm{with } U=1,2,\dots,M_3-1,
\end{align}
where the non-vanishing vector component is at $R=U$.
These null vectors have length $M_2$ and their last $\mathcal{R}_3$ components are all zero.
Corresponding to ${}^i_i[\widehat{W}_{2R}=0$, we select the null vector
\begin{align}
\label{eq:hat-gamma-Palatini-U=M3}
(\widehat{\gamma}_{U = M_3})^R &= \frac{1}{\sqrt{d-1}} (0, \dots, 0, 1, 0, \dots, 0, 1, 0, \dots, 0),
\end{align}
where the $(d-1)$ number of non-vanishing vector components are at $R = M_3, 2M_3, \dots, (d-1)M_3$.\footnote{To determine the null vectors of
$\widehat{W}_{RS}$ in \eqref{eq:terHess},
we considered the ansatz $(\widehat{\gamma}_U)^R = (a^i, a^i_j, a^i_{jk})$, with $a^i_{jk} = a^i_{kj}$.
Then, the equation $(\widehat{\gamma}_U)^R \widehat{W}_{RS} =0$ results in  $a^j_i = c a^l_l \delta^j_i$ and $a^j_{il} = c a^k_{kl} \delta^j_i$, 
but does not impose any condition on $a^i$.
The first equation implies $a^i_j =0$ for $i \neq j$.
Setting $j=l$ in the second equation yields $a^j_{ij} =0$, which in turn implies $a^i_{jk}=0$ for all $i,j,k$.}

The tertiary Lagrangian constraints are given by the requirement $\widehat{\varphi}_U=(\widehat{\gamma}_U)^R\dot{\widetilde{\varphi}}_R\overset{!}{\underset{2}{\approx}}0$.
In this case, the derivation with respect to time is particularly simple and coincides with the naively expected one, so that
\begin{align}
\widehat{\varphi}_U=(\widehat{\gamma}_U)^R\big[\partial_i\dot{Q}^A\partial^i_A+\dot{Q}^A\partial_A\big]\widetilde{\varphi}_R
\overset{!}{\underset{2}{\approx}}0.
\end{align}
In our short-hand notation, we find it convenient to express these constraints as follows:
\begin{align}
\label{eq:tertcons}
\frac{[\widehat{\varphi}_1]_i}{2}=\dot{h}(\widetilde{\tau}_1)_i+\dot{h}^j(\widetilde{\tau}_2)_{ij}+\mathcal{O}\big[h(\widetilde{\tau}_1)_i+h^j(\widetilde{\tau}_2)_{ij}\big],
\qquad \frac{[\widehat{\varphi}_{2}]}{2}=\dot{h}^{i}(\widetilde{\tau}_{1})_{i}+\dot{h}^{ij}(\widetilde{\tau}_2)_{ij}
+\mathcal{O}\big[h^{i}(\widetilde{\tau}_1)_i+h^{ij}(\widetilde{\tau}_2)_{ij}\big],
\end{align}
where the operator $\mathcal{O}$ is defined as
\begin{align}
\mathcal{O}:=&(\partial_k \dot{G})\frac{\partial}{\partial(\partial_kG)}
+(\partial_k \dot{G}_l)\frac{\partial}{\partial(\partial_kG_l)}
+(\partial_k \dot{\mathcal{G}}^l_m)\frac{\partial}{\partial(\partial_k\mathcal{G}^l_m)}
+(\partial_k \dot{\mathcal{G}}^l_{mn})\frac{\partial}{\partial(\partial_k\mathcal{G}^l_{mn})}\nonumber \\
&+\dot{G}\frac{\partial}{\partial G}+\dot{G_k}\frac{\partial}{\partial G_k}+\dot{G}_{kl}\frac{\partial}{\partial G_{kl}}
+\dot{\mathcal{G}}^k_l\frac{\partial}{\partial \mathcal{G}^k_l}+\dot{\mathcal{G}}^k_{lm}\frac{\partial}{\partial \mathcal{G}^k_{lm}}.
\end{align}
Recall that $(\widetilde{\tau}_1)_i$ and $(\widetilde{\tau}_2)_{ij}$ are as introduced in (\ref{eq:tildetaus}).

Following the procedure described under (\ref{eq:bmops}), the tertiary Lagrangian constraints in (\ref{eq:tertcons}) can be evaluated
on the first constraint surface $T\mathcal{C}_1$.
After tedious algebraic manipulations, the weak tertiary Lagrangian constraints can be written exclusively in terms of the functionally independent
secondary constraints as
\begin{align}
\displaystyle
\label{eq:tertsin1dif}
\begin{array}{lllll}
&[\widehat{\varphi}_1]_i &\hspace*{-0.2cm}\underset{1}{\approx}&\hspace*{-0.2cm}
-(\delta^j_iG-2\mathcal{G}^j_i)[\widetilde{\varphi}_{1}]_j
-\big[\delta^j_i(G_k+\partial_k)-\tfrac{1}{2}\delta^j_k\partial_i-\mathcal{G}_{ik}^j\big][\widetilde{\varphi}_{2}]_j^k
-G_{jk}[\widetilde{\varphi}_{3}]_i^{jk},\vspace*{0.2cm}\\
&\frac{[\widehat{\varphi}_2]}{2}&\hspace*{-0.2cm}\underset{1}{\approx}&\hspace*{-0.2cm}
2\mathcal{G}^i[\widetilde{\varphi}_1]_i
+\mathcal{G}_i^j[\widetilde{\varphi}_2]_j^i
+\partial_j[\widetilde{\varphi}_{3}]^{ij}_i.
\end{array}
\end{align}
The above is a non-trivial result.
Indeed, it becomes increasingly computationally challenging to evaluate Lagrangian constraints on constraint surfaces as one goes to higher stages.
We elaborate on this topic and advice on how to handle the evaluations in section \ref{sec:final}.

Our results in (\ref{eq:tertsin1dif}) must be further evaluated on the second constraint surface $T\mathcal{C}_2$.
Namely, in the subspace of $T\mathcal{C}_1$ defined by the vanishing of (\ref{eq:secin1}).
We thus see that
\begin{align}
\label{eq:tertzero}
[\widehat{\varphi}_1]_i \underset{2}{\approx}0, \qquad [\widehat{\varphi}_2]\underset{2}{\approx}0,
\end{align}
which implies $T\mathcal{C}_3\equiv T\mathcal{C}_2$ and there are no functionally independent tertiary constraints $M_3^\prime=0$.
Consequently, the algorithm closes non-dynamically, according to case \ref{it:III}.

We are finally able to obtain the result of interest from the analysis here presented.
The number of functionally independent Lagrangian constraints for the Palatini theory in $d>2$, when described in terms of
$N=d(d+1)^2/2$ number of a priori independent field variables, is equal to
\begin{align}
\label{eq:linPal}
l=M_1^\prime+M_2^\prime+M_3^\prime=M_1+M_2+0=\frac{d}{2}(d+1)^2+\frac{d}{2}(d^2-1)+0=d^2(d+1).
\end{align}

\pagebreak

\hspace*{-0.6cm}{\bf Gauge identities.}\\
It is well-known (for instance, see~\cite{McKeon:2010nf}) that the Palatini action corresponding to
the Lagrangian density (\ref{eq:PalatiniL}) remains invariant under the following transformations of its independent variables:
$h^{\mu\nu}\rightarrow h^{\mu\nu}+\delta_\theta h^{\mu\nu}$ and $G^\rho_{\mu\nu}\rightarrow G^\rho_{\mu\nu}+\delta_\theta G^\rho_{\mu\nu}$, with
\begin{align}
\label{eq:trans1}
\delta_\theta h^{\mu\nu}=2h^{\rho(\mu}\partial_\rho\theta^{\nu)}-\partial_\rho\big(h^{\mu\nu}\theta^\rho\big), \qquad
\delta_\theta G^\rho_{\mu\nu}=-\partial_\mu\partial_\nu\theta^\rho+\delta^\rho_{(\mu}\partial_{\nu)}\partial_\sigma\theta^\sigma-\theta^\sigma\partial_\sigma G^\rho_{\mu\nu}
+G^\sigma_{\mu\nu}\partial_\sigma\theta^\rho-2G^\rho_{\sigma(\mu}\partial_{\nu)}\theta^\sigma,
\end{align}
where $\theta^\mu$ are the (unspecified) gauge parameters.
Notice that the pertinent symmetries $\delta_\theta h^{\mu\nu}=\delta_\theta h^{\nu\mu}$ and
$\delta_\theta G_{\mu\nu}^\rho=\delta_\theta G_{\nu\mu}^\rho$ are apparent in the precedent expressions.
Of course, (\ref{eq:trans1}) is just the Palatini (re)formulation of the renowned diffeomorphism invariance of the Einstein-Hilbert action.
This holds true off shell.

It is easy to see in (\ref{eq:trans1}) that the gauge parameters $\theta^\mu$ appear explicitly in all the gauge transformations $\forall\mu$.
Similarly, we note that the effective gauge parameters $(\theta^\mu,\dot{\theta}^\mu, \ddot{\theta}^\mu)$ are manifestly present
in the gauge transformations $\forall\mu$ as well.
By definition, it follows that
\begin{align}
\label{eq:geinPal}
g=d, \qquad e=3d,
\end{align}
which are the off shell parameters we aimed to obtain in this short analysis.

For completeness, we provide the gauge generators and confirm the gauge identities of $d>2$ Palatini next.
A direct comparison between (\ref{eq:fieldvarform}) and (\ref{eq:trans1}), allows us to rewrite the latter as
\begin{align}
\displaystyle
\begin{array}{lllllll}
&\delta_\theta h^{\mu\nu}&\hspace*{-0.2cm}=&\hspace*{-0.2cm}
\theta^\beta[(\Omega_\beta)^{\mu\nu}]-(\partial_{\mu_1}\theta^\beta)[(\Omega_\beta)^{\mu\nu}]^{\mu_1}, \vspace*{0.2cm}\\
&\delta_\theta G_{\mu\nu}^\rho&\hspace*{-0.2cm}=&\hspace*{-0.2cm}
\theta^\beta[(\Omega_\beta)^\rho_{\mu\nu}]-(\partial_{\mu_1}\theta^\beta)[(\Omega_\beta)^\rho_{\mu\nu}]^{\mu_1}
+(\partial_{\mu_1}\partial_{\mu_2}\theta^\beta)[(\Omega_\beta)^\rho_{\mu\nu}]^{\mu_1\mu_2},
\end{array}
\end{align}
where we have introduced a bracket $(\cdot)$ to visually split the (in general collective) indices $\beta$ and $A$ for latter convenience.
In view of these transformations, the gauge generators can easily be identified to be
\begin{align}
\displaystyle
\hspace*{-0.4cm}
\begin{array}{llllllllllll}
&[(\Omega_\beta)^{\mu\nu}]&\hspace*{-0.1cm}=&\hspace*{-0.2cm}
-\partial_\beta h^{\mu\nu},
&\,\, [(\Omega_\beta)^{\mu\nu}]^{\mu_1}&\hspace*{-0.1cm}=&\hspace*{-0.2cm}
-2h^{\mu_1(\mu}\delta^{\nu)}_\beta+h^{\mu\nu}\delta^{\mu_1}_\beta,
&\hfill&\hfill&\hfill\\
&[(\Omega_\beta)^\rho_{\mu\nu}]&\hspace*{-0.1cm}=&\hspace*{-0.2cm}
-\partial_\beta G^\rho_{\mu\nu},
&\,\,[(\Omega_\beta)^\rho_{\mu\nu}]^{\mu_1}&\hspace*{-0.1cm}=&\hspace*{-0.2cm}
2G^\rho_{\beta(\mu}\delta^{\mu_1}_{\nu)}-G^{\mu_1}_{\mu\nu}\delta^\rho_\beta,
&\,\, [(\Omega_\beta)^\rho_{\mu\nu}]^{\mu_1\mu_2}&\hspace*{-0.1cm}=&\hspace*{-0.2cm}
\delta^\rho_{(\mu}\delta^{(\mu_1}_{\nu)}\delta^{\mu_2)}_\beta-\delta^{\mu_1}_{(\mu}\delta^{\mu_2}_{\nu)}\delta^\rho_\beta.
\end{array}
\end{align}
Combining (\ref{eq:FEqsPalatini}) with the above as prescribed in (\ref{eq:defvarrho}) and working through, the gauge identities are obtained:
\begin{align}
\begin{array}{lllllll}
\rho_\beta &\hspace*{-0.2cm}=&\hspace*{-0.2cm} E_{(h^{\mu\nu})}[(\Omega_\beta)^{\mu\nu}]+\partial_{\mu_1}\Big(E_{(h^{\mu\nu})}[(\Omega_\beta)^{\mu\nu}]^{\mu_1}\Big)\\
&&\hspace*{-0.3cm}+E_{(G^\rho_{\mu\nu})}[(\Omega_\beta)^\rho_{\mu\nu}]+\partial_{\mu_1}\Big(E_{(G^\rho_{\mu\nu})}[(\Omega_\beta)^\rho_{\mu\nu}]^{\mu_1}\Big)
+\partial_{\mu_1}\partial_{\mu_2}\Big(E_{(G^\rho_{\mu\nu})}[(\Omega_\beta)^\rho_{\mu\nu}]^{\mu_1\mu_2}\Big)\equiv0.
\end{array}
\end{align}

\vspace*{0.5cm}

\hspace*{-0.6cm}{\bf Physical degrees of freedom.}\\
Putting everything together, we can finally count the number of propagating modes present in the theory.
Namely, employing (\ref{eq:NinPal}), (\ref{eq:linPal}) and (\ref{eq:geinPal}) in the master formula (\ref{eq:dofformula}), we get
\begin{align}
\label{eq:dofinPal}
n_{\textrm{dof}}=\frac{d}{2}(d-3).
\end{align}
When $d=4$, we have that $n_{\textrm{dof}}=2$, corresponding to the two massless tensor's polarizations of the graviton.
For $d=3$, the widely known triviality is recovered, with no physical degrees of freedom being propagated.

Our result is in perfect agreement with the counting performed in~\cite{Ghalati:2007sv,McKeon:2010nf}, where a purely Hamiltonian analysis was done.
We have thus carried out another (non-trivial) explicit verification of the already noted equivalence
between (\ref{eq:dofformula}) and (\ref{eq:Dirdof}).
This equivalence can be further verified as follows.
It is explicitly shown in~\cite{Ghalati:2007sv,McKeon:2010nf} that $N_1=3d$, $N_2=d(d-1)(d+2)$ and $N_1^{(\textrm{P})}=d$
for the $d>2$ Palatini theory when (\ref{eq:NinPal}) holds true.
Substitution of these results in (\ref{eq:HamLageq}) readily confirms our own counting in (\ref{eq:linPal}) and (\ref{eq:geinPal}).
Besides, a direct comparison between the calculations in~\cite{Ghalati:2007sv,McKeon:2010nf} and those presented in this
section \ref{sec:Palatini} unequivocally shows that our purely Lagrangian computation is an algebraically much simpler way to derive
(\ref{eq:tripletPa}), from which the number of physical modes follows readily.

To sum up, we have derived the constraint structure characterizing triplet $t_{\textrm{Pa}}^{(N)}$, with $N=d(d+1)^2/2$, of the Palatini theory in $d>2$ dimensions
in a purely Lagrangian approach and ratified its equivalence with a representative Hamiltonian analysis performed in the past.
Mathematically,
\begin{align}
\label{eq:tripletPa}
t_{\textrm{Pa}}^{(N)}=\big(l=d^2(d+1),g=d,e=3d\big) 
\end{align}
in the Lagrangian picture, while
\begin{align}
t_{\textrm{Pa}}^{(N)}=\big(N_1^{(\textrm{P})}=d,N_1=3d,N_2=d(d-1)(d+2)\big)
\end{align}
in the Hamiltonian side ---recall (\ref{eq:LagHamtN})---, both of which imply (\ref{eq:dofinPal}).

%%%%%%%%%%%%%%%%%%%%%%%%%%%%%%%%%%%%%%%%%%%%%%%%%%%%%%%%%%%%%%%%%%%%%%%%%%%%%%%%%%%%%%%%%

%%%%%%%%%%%%%%%%%%%%%%%%%%%%%%%%%%%%%%%%%%%%%%%%%%%%%%%%%%%%%%%%%%%%%%%%%%%%%%%%%%%%%%%%%
\subsection{A special case: Palatini in $d=2$}
\label{sec:Palatinid2}
%%%%%%%%%%%%%%%%%%%%%%%%%%%%%%%%%%%%%%%%%%%%%%%%%%%%%%%%%%%%%%%%%%%%%%%%%%%%%%%%%%%%%%%%%

General Relativity, in its standard second order formulation, behaves drastically different in two dimensions.
Specifically, it can be shown that
\begin{align}
\label{eq:d2RG}
S_{\textrm{EH}}=\int_{\mathcal{M}_{2}}{ d^2{x} \sqrt{-g}R}\,\propto \,\chi (\mathcal{M}_{2}).
\end{align} 
Namely, the Einstein-Hilbert action is proportional to the Euler characteristic $\chi$
of the spacetime manifold $\mathcal{M}_{2}$, see e.g.~\cite{Grumiller:2002nm}.
The above implies that General Relativity is a topological theory in $d=2$ and, accordingly, propagates no degrees of freedom;
a fact that we shall explicitly verify in the following.

Turning to the Palatini Lagrangian in \eqref{eq:PalatiniL} for $d=2$,
we restate that this is not dynamically equivalent to two-dimensional Einstein's gravity (\ref{eq:d2RG}).
To see this, consider its corresponding Euler-Lagrange equations in (\ref{eq:FEqsPalatini}).
These are valid for $d\geq 2$.
However, recall that $c:=(d-1)^{-1}$, so that $c=1$ in two dimensions.
In this particular case, it is obvious that (\ref{eq:trGeq}) cannot be solved as we said, i.e.~$G^\rho_{\mu\nu}\overset{!}{=}G^\rho_{\mu\nu}(h^{\mu\nu},\partial_\rho h^{\mu\nu})$.
As a result, the dynamical equivalence to Einstein's gravity is lost.
A more general yet detailed argumentation can be found in~\cite{Deser:1995cs}. 

Correspondingly, the dynamics of the two-dimensional Palatini action does not constitute a smooth limit of its higher dimensional counterpart.
Namely, the Lagrangian \eqref{eq:PalatiniL} in $d=2$ does not describe the evolution of the same family of fields as that very same Lagrangian in $d>2$:
these are two physically different theories.
The easiest way to ratify this second inequivalence is to note that the counting of degrees of freedom in \eqref{eq:dofinPal},
when we set $d=2$, yields a negative number of propagating modes, which is an unphysical result.
Thus, a different constraint structure characterizing triplet
\begin{align}
t_{\textrm{2Pa}}^{(N=9)}\neq \lim_{d\rightarrow 2}t_{\textrm{Pa}}^{(N)}, \qquad \textrm{where }\quad N=d(d+1)^2/2,
\end{align}
is then to be expected.
We proceed to determine this $t_{\textrm{2Pa}}^{(N=9)}$ next.

\vspace*{0.5cm}

\hspace*{-0.6cm}{\bf Lagrangian constraints.}\\
As a starting point, we express the generalized coordinates of the Palatini theory in $d=2$ as
\begin{align}
Q^A=\{h,h^1,h^{11},G,G_1,G_{11},\mathcal{G}^1,\mathcal{G}^1_1,\mathcal{G}^{1}_{11}\}, \qquad A=1,2,\ldots,9=\textrm{dim}(\mathcal{C})\equiv N,
\end{align}
in direct analogy to (\ref{eq:QAPal}) earlier on.
Next, we compute the first stage quantities associated to the $d=2$ version of the Palatini Lagrangian \eqref{eq:PalL}.
One can verify that the set of primary Lagrangian constraints thus obtained
matches the consistent $d=2$ evaluation of those for a generic dimension in (\ref{eq:primELeqs}) and (\ref{eq:primalph}). 
Comparatively, these two-dimensional constraints have a much simpler form, given by the vanishing of
\begin{align}
\displaystyle
\label{eq:primalph2D}
\hspace*{-0.5cm}
\begin{array}{llllllllll}
&[\varphi_{1}]&\hspace*{-0.2cm}=&\hspace*{-0.2cm}
-\big[\dot{G}+\partial_{1}\mathcal{G}^{1}+2(G \mathcal{G}^{1}_{1}-G_{1}\mathcal{G}^{1})\big], &\qquad \quad
&\frac{ [\varphi_{2}]_{1}}{2}&\hspace*{-0.2cm}=&\hspace*{-0.2cm}
-\big[\dot{G}_{1}+\partial_{1}\mathcal{G}^{1}_{1}+G\mathcal{G}^{1}_{11}-G_{11}\mathcal{G}^{1}\big],\\
&[\varphi_{3}]_{11}&\hspace*{-0.2cm}=&\hspace*{-0.2cm}
-\big[\dot{G}_{11}+\partial_{1}\mathcal{G}^{1}_{11}+2(G_{1}\mathcal{G}^{1}_{11}-G_{11}\mathcal{G}^{1}_{1})\big], &\qquad \quad
&[\varphi_{4}]&\hspace*{-0.2cm}=&\hspace*{-0.2cm}
\dot{h}-2(h\mathcal{G}^{1}_{1}+h^1\mathcal{G}^{1}_{11}),\\
&\frac{[\varphi_{5}]^{1}}{2}&\hspace*{-0.2cm}=&\hspace*{-0.2cm}
\dot{h}^{1}+h\mathcal{G}^{1}-h^{11}\mathcal{G}^{1}_{11}, &\qquad \quad
&[\varphi_{6}]^{11}&\hspace*{-0.2cm}=&\hspace*{-0.2cm}
\dot{h}^{11}+2(h^1\mathcal{G}^{1}+h^{11}\mathcal{G}^{1}_{1}),\\
&[\varphi_{7}]_{1}&\hspace*{-0.2cm}=&\hspace*{-0.2cm}
\partial_{1}h+2(hG_{1}+h^1G_{11}), &\qquad \quad
&\frac{[\varphi_{8}]^{1}_{1}}{2}&\hspace*{-0.2cm}=&\hspace*{-0.2cm}
\partial_{1}h^{1}-hG+h^{11}G_{11},\\
&[\varphi_{9}]_{1}^{11}&\hspace*{-0.2cm}=&\hspace*{-0.2cm}
\partial_{1}h^{11}-2(h^1G+h^{11}G_{1}).
\end{array}
\end{align}
The demand that the above be zero constitutes a set of nine scalar primary Lagrangian constraints (${M}_1=9$), whose functional independence is rather obvious ($M'_1=9$)
---and can be ratified through the Jacobian test in (\ref{eq:primJac}).
Therefore, such vanishing defines the primary constraint surface $T\mathcal{C}_1$ of the theory, which coincides with the moduli space due to the primary Hessian being zero,
as in the $d>2$ case before.
In other words, (\ref{eq:primELeqs}) holds true here as well.

The progress to the subsequent stage parallels that of the $d>2$ case.
The secondary Hessian is given by the skew-symmetric constant matrix 
\begin{align}
\label{eq:secHess2D}
\widetilde{W}_{IJ}=
\left(
\begin{array}{ccccccccc}
0 & -\omega & 0\\
\omega & 0 & 0\\
0 & 0 & 0
\end{array}
\right),\qquad \text{with} \quad\omega:=\text{diag}(1,2,1).
\end{align}
The Hessian \eqref{eq:secHess2D} has rank $\mathcal{R}_{2}=6$ and so $M_{2}=3$.
This means that six of the primary Lagrangian constraints are dynamically stabilized by the (functionally independent) secondary equations of motion.
For the remaining three primary Lagrangian constraints, the algorithm must be pursued.

We choose the suitably normalized linearly independent null vectors of (\ref{eq:secHess2D}) as
\begin{align}
\label{eq:secnullvecs2D}
(\widetilde{\gamma}_{R})^{I}=\delta_{R+6}{}^{I}, \qquad \text{with }\quad R=1,2,3.
\end{align}
Using (\ref{eq:primalph2D}) and (\ref{eq:secnullvecs2D}),
we obtain the three secondary Lagrangian constraints as the vanishing of 
\begin{align}
\displaystyle
\label{eq:rawsecs2D}
\begin{array}{llllll}
&[\widetilde{\varphi}_{1}]_{1}&\hspace*{-0.2cm}=&\hspace*{-0.2cm}
\partial_1 \dot{h}+2(\dot{h}G_{1}+\dot{h}^1G_{11}+h\dot{G}_{1}+h^{1}\dot{G}_{11}), \vspace*{0.1cm}\\
&[\widetilde{\varphi}_{2}]^{1}_{1}&\hspace*{-0.2cm}=&\hspace*{-0.2cm}
2(\partial_1\dot{h}^{1}-\dot{h}G+\dot{h}^{11}G_{11}-h\dot{G}+h^{11}\dot{G}_{11}), \vspace*{0.1cm}\\
&[\widetilde{\varphi}_{3}]^{11}_{1}&\hspace*{-0.2cm}=&\hspace*{-0.2cm}
\partial_1\dot{h}^{11}-2(\dot{h}^{1}G+\dot{h}^{11}G_1+h^{1}\dot{G}+h^{11}\dot{G}_{1}).
\end{array}
\end{align}
Notice that the above are the total time derivatives of $[\varphi_{7}]_{1}$, $[\varphi_{8}]^{1}_{1}$ and $[\varphi_{9}]_{1}^{11}$ in (\ref{eq:primalph2D}), respectively.
It is easy to check that the secondary Lagrangian constraints are functionally dependent on the primary Lagrangian constraints.
Specifically,
\begin{align}
\displaystyle
\label{eq:secondaries2D}
\begin{array}{llllll}
&[\widetilde{\varphi}_{1}]_{1}&\hspace*{-0.2cm}=&\hspace*{-0.2cm}
\partial_{1}[\varphi_4]-h[\varphi_{2}]_{1}-2h^1[\varphi_{3}]_{11}+2G_1[\varphi_{4}]+G_{11}[\varphi_{5}]^{1}
+2\mathcal{G}^{1}_{1}[\varphi_{7}]_{1}+\mathcal{G}^{1}_{11}[\varphi_{8}]^{1}_{1}, \vspace*{0.1cm}\\
&[\widetilde{\varphi}_{2}]^{1}_{1}&\hspace*{-0.2cm}=&\hspace*{-0.2cm}
\partial_{1}[\varphi_5]^{1}+2(h[\varphi_{1}]-h^{11}[\varphi_{3}]_{11} -G[\varphi_{4}]+G_{11}[\varphi_{6}]^{11}
-\mathcal{G}^{1}[\varphi_{7}]_{1}+\mathcal{G}^{1}_{11}[\varphi_{9}]^{11}_{1}), \vspace*{0.1cm}\\
&[\widetilde{\varphi}_{3}]^{11}_{1}&\hspace*{-0.2cm}=&\hspace*{-0.2cm}
\partial_{1}[\varphi_6]^{11}+2h^1[\varphi_{1}]+h^{11}[\varphi_{2}]_{1}-(G[\varphi_{5}]^{1}
+2G_{1}[\varphi_{6}]^{11}+\mathcal{G}^{1}[\varphi_{8}]^{1}_{1}+2\mathcal{G}^{1}_{1}[\varphi_{9}]^{11}_{1}).
\end{array}
\end{align}
Therefore, the secondary constraints vanish on $T\mathcal{C}_{1}$,
\begin{align}
\label{eq:Pal2DsecctsTC1}
[\widetilde{\varphi}_{1}]_{1}\underset{1}{\approx} 0, \qquad [\widetilde{\varphi}_{2}]^{1}_{1}\underset{1}{\approx}0, \qquad
[\widetilde{\varphi}_{3}]^{11}_{1}\underset{1}{\approx}0,
\end{align}
and so $T\mathcal{C}_2\equiv T\mathcal{C}_1$.
This in turn implies that there are no functionally independent secondary constraints: $M'_2=0$.
Here, the algorithm closes non-dynamically, as described in case \ref{it:III}.
It follows that the total number of functionally independent Lagrangian constraints is $l=M'_1+M'_2=9$.
We note that this result does not correspond to setting $d=2$ in (\ref{eq:linPal}).

\vspace*{0.5cm}

\hspace*{-0.6cm}{\bf Gauge identities.}\\
Given the already pointed out inequivalence between the $d>2$ and $d=2$ Palatini theories,
it is not too surprising that the gauge transformations \eqref{eq:trans1} preserving the action (\ref{eq:PalL}) meet a non-smooth limit for $d=2$.
The argument is more subtle than that of the purely on shell inequivalence; for example, see~\cite{Kiriushcheva:2005sk}.
We shall touch upon it shortly.

It has been proven, e.g.~\cite{Kiriushcheva:2005sk,McKeon:2016vuq}, that the two-dimensional Palatini action is invariant under the field transformations
$h^{\mu\nu}\rightarrow h^{\mu\nu}+\delta_\theta h^{\mu\nu}$ and $G_{\mu\nu}^\rho\rightarrow G_{\mu\nu}^\rho+ \delta_\theta G_{\mu\nu}^\rho$, with
\begin{align}
\label{eq:GaugePa2D}
\delta_\theta h^{\mu \nu}=2\epsilon^{\rho(\mu}h^{\nu)\sigma}\theta_{\rho \sigma}, \qquad
\delta_\theta G^{\rho}_{\mu \nu}=\epsilon^{\rho\sigma}\partial_{\sigma}\theta_{\mu \nu}+2\epsilon^{\sigma\lambda}G^{\rho}_{\sigma( \mu}\theta_{\nu)\lambda}.
\end{align}
Here, $\epsilon^{\mu\nu}$ stands for the two-dimensional Levi-Civita symbol (we work with the convention $\epsilon^{01}=1$)
and $\theta_{\mu\nu}=\theta_{\nu\mu}$, so there are three arbitrary gauge parameters that characterize the transformation.
It is obvious that all the gauge parameters explicitly appear in the gauge transformations.
Hence, $g=3$.
Their first time derivatives also show up, adding to a total number of effective gauge parameters $e=6$.
We point out that $g$ does not match the value predicted in (\ref{eq:geinPal}) for $d=2$.
There is a match for $e$, but this is purely coincidental.

These numbers $(g=3,e=6)$, in contrast to the naively expected ones $(g=2,e=6)$ from the diffeomorphism transformations (\ref{eq:geinPal}) in $d>2$ Palatini,
reflect the fact that $d=2$ Palatini is associated to a comparatively larger symmetry group.
Its connection to the $d>2$ gauge group is not obvious,
but finds its origin in the underlying two-dimensional geometry.
Briefly recall the conformal flatness of two-dimensional spacetimes, i.e.
\begin{align}
g_{\mu\nu}=\Omega^{2} \eta_{\mu\nu} ,\qquad g^{\mu\nu}=\Omega^{-2} \eta^{\mu\nu}, \qquad \textrm{with } \mu,\nu=0,1,
\end{align}
for some conformal factor $\Omega=\Omega(x^\mu)$.
Given this property, the variable $h^{\mu\nu}$ introduced in (\ref{eq:defhG}) simplifies to
\begin{align}
h^{\mu\nu}:=\sqrt{\textrm{det}(-g_{\mu\nu})}g^{\mu\nu}=\sqrt{\textrm{det}(-\Omega^2\,\eta_{\mu\nu})}\Omega^{-2}\eta^{\mu\nu}=\eta^{\mu\nu}.
\end{align}
Consequently, in the conformal frame, $h^{\mu\nu}$ is flat and $\textrm{det}(h^{\mu\nu})=-1$, independent of $\textrm{det}(g^{\mu\nu})$.
This latter equality can be expressed as an algebraic constraint: 
\begin{align}
\label{eq:det2dcf}
hh^{11}-(h^{1})^{2}+1=0,
\end{align}
referred to as the metricity condition.
We will soon get back to such condition.
For a richer discussion on this topic, though, we refer the reader to~\cite{Kiriushcheva:2006gp}.

In analogy to the higher dimensional case before, we provide the gauge generators and identities of $d=2$ Palatini next.
Comparing (\ref{eq:fieldvarform}) to (\ref{eq:GaugePa2D}), we can conveniently rewrite the latter as
\begin{align}
\delta_\theta h^{\mu\nu}=\theta_{\alpha\beta}[(\Omega^{\alpha\beta})^{\mu\nu}], \qquad
\delta_\theta G^\rho_{\mu\nu}=\theta_{\alpha\beta}[(\Omega^{\alpha\beta})^\rho_{\mu\nu}]-(\partial_{\mu_1}\theta_{\alpha\beta})[(\Omega^{\alpha\beta})^\rho_{\mu\nu}]^{\mu_1},
\end{align}
with the gauge generators readily recognized as
\begin{align}
\label{eq:gens2p}
[(\Omega^{\alpha\beta})^{\mu\nu}]=2\big(\epsilon^{\alpha(\mu}h^{\nu)\beta}+\epsilon^{\beta(\mu}h^{\nu)\alpha}\big), \qquad 
[(\Omega^{\alpha\beta})^\rho_{\mu\nu}]=4\epsilon^{\sigma(\alpha|}G^\rho_{\sigma(\mu}\delta^{|\beta)}_{\nu)}, \qquad 
[(\Omega^{\alpha\beta})^\rho_{\mu\nu}]^{\mu_1}=2\epsilon^{\rho\mu_1}\delta^\alpha_{(\mu}\delta^\beta_{\nu)},
\end{align}
where the bar $|$ notation delimits the symmetrized indices. Besides the the generators, the other element needed to determine the gauge identities are the primary Euler-Lagrange equations.
These are given by the straightforward evaluation of (\ref{eq:FEqsPalatini}) for $d=2$.
Let us refer to them as $E_{(h^{\mu\nu})}^{(2)}$ and $E_{(G^\rho_{\mu\nu})}^{(2)}$, respectively.
Then, their merging together with (\ref{eq:gens2p}) as indicated in (\ref{eq:defvarrho}) yields the gauge identities for $d=2$ Palatini we were seeking,
after some tedious yet elementary algebra:
\begin{align}
\varrho^{\alpha\beta}=E_{(h^{\mu\nu})}^{(2)}[(\Omega^{\alpha\beta})^{\mu\nu}]+E_{(G^\rho_{\mu\nu})}^{(2)}[(\Omega^{\alpha\beta})^\rho_{\mu\nu}]
+\partial_{\mu_1}\Big(E_{(G^\rho_{\mu\nu})}^{(2)}[(\Omega^{\alpha\beta})^\rho_{\mu\nu}]^{\mu_1}\Big)\equiv 0.
\end{align}
Observe that the manifest symmetry under the exchange $\alpha\leftrightarrow\beta$ makes the number of independent gauge identities coincide with our earlier counting: $g=3$.

\vspace*{0.5cm}

\hspace*{-0.6cm}{\bf Physical degrees of freedom.}\\
Altogether, we have now obtained the constraint structure characterizing triplet of Palatini in $d=2$ in terms of Lagrangian quantities:
\begin{align}
\label{eq:triplet2Pa}
t_{2\textrm{Pa}}^{(N)}=(l=9,g=3,e=6),
\end{align}
where $N=9$.
Employing (\ref{eq:HamLageq}) and (\ref{eq:LagHamtN}), it is immediate to rewrite this triplet in Hamiltonian terms:
\begin{align}
t_{2\textrm{Pa}}^{(N)}=(N_1^{(\textrm{P})}=3,N_{1}=6,N_{2}=6).
\end{align}
Plugging \eqref{eq:triplet2Pa} in the master formula \eqref{eq:dofformula}, we confirm the well-known fact that there are no physical degrees of freedom propagated by the theory:
$n_{\textrm{dof}}=0$.
We restate that the above cannot be obtained by simply setting $d=2$ in (\ref{eq:dofinPal}).

To wrap up this section, we check our results are in good agreement with some of the previously carried out Hamiltonian calculations.
We begin our comparisons by looking into the approach closest to our own, the one in~\cite{Kiriushcheva:2005sk}.
There, the quantities $(h^{\mu\nu},G^{\rho}_{\mu\nu})$ were regarded as the $N=9$ a priori independent field variables for $d=2$ Palatini, exactly as we did here.
Following the Dirac-Bergmann procedure, it was shown that $N_1^{(\textrm{P})}=3$, $N_1=6$ and $N_2=6$, which readily confirms our own independent findings.
In~\cite{Kiriushcheva:2006gp}, the metricity condition (\ref{eq:det2dcf}) was taken into account from the onset.
As a result of incorporating this information in the form of additional terms preceded by two Lagrange multipliers in the Hamiltonian,
their setup had $\pmb{N}=11\neq N=9$ a priori independent field variables.
It was there shown that, in such formulation, $\pmb{N}_1^{(\textrm{P})}=5$, $\pmb{N}_1=7$ and $\pmb{N}_2=8$,
yielding no propagating degrees of freedom and thus corroborating again our own final result.
This latter comparison provides another example for our point around (\ref{eq:NandNprime}), this time in Hamiltonian terms.
Namely, 
\begin{align}
\label{eq:twot2pa}
t_{2\textrm{Pa}}^{(N)}=(N_1^{(\textrm{P})}=3,N_1=6,N_2=6)\neq t_{2\textrm{Pa}}^{(\pmb{N})}=(\pmb{N}_1^{(\textrm{P})}=5,\pmb{N}_1=7,\pmb{N}_2=8),
\end{align}
but we find that $n_{\textrm{dof}}=0$ for both sets of numbers upon employing (\ref{eq:Dirdof}). 
As a last remark, we notice that our calculations in this section \ref{sec:Palatinid2}
are comparatively simpler than those in~\cite{Kiriushcheva:2005sk,Kiriushcheva:2006gp}.
Namely, our approach is certainly to be preferred if the goal is to determine the constraint structure of the theory and thereby manifestly count its propagating modes.

%%%%%%%%%%%%%%%%%%%%%%%%%%%%%%%%%%%%%%%%%%%%%%%%%%%%%%%%%%%%%%%%%%%%%%%%%%%%%%%%%%%%%%%%%

%%%%%%%%%%%%%%%%%%%%%%%%%%%%%%%%%%%%%%%%%%%%%%%%%%%%%%%%%%%%%%%%%%%%%%%%%%%%%%%%%%%%%%%%%
\section{Contextualization and potentiality of our results}
\label{sec:lit}
%%%%%%%%%%%%%%%%%%%%%%%%%%%%%%%%%%%%%%%%%%%%%%%%%%%%%%%%%%%%%%%%%%%%%%%%%%%%%%%%%%%%%%%%%

The study of constrained systems was initiated in the thirties by Rosenfeld, in a sometimes overlooked work~\cite{Rosenfeld:1930aa},
nowadays acknowledged and revisited~\cite{Salisbury:2017oev}.
It was later greatly developed during the fifties~\cite{DiracBergmann} and has since been a very active field of theoretical research.
As such, one may have the impression that the investigation of manifestly first order singular classical field theories must be an already closed subject.
This is not true.
There are ongoing advances in this fundamental topic, particularly within the Lagrangian picture.
Besides the references already provided in section \ref{sec:method}, the recent work~\cite{Heidari:2020cil} stands as a neat example.
The methodology there put forward is equivalent to our own proposal, as we shall show in the next section \ref{sec:comir}.

To further reassure the reader of the topicality of our formalism,
in section \ref{sec:MPrel} we explain how our method lends itself to a conversion from an analytic machinery to a constructive one.
Indeed, the Lagrangian building principle originally put forward in~\cite{ErrastiDiez:2019trb,ErrastiDiez:2019ttn}
finds in the contents of section \ref{sec:method} a solid footing for attempting the construction of novel theories.
This argumentation is carried out in terms of a concrete application for clarity, but the general proposal is much broader.
In particular, we explain that the less elaborated upon procedure in~\cite{ErrastiDiez:2019trb,ErrastiDiez:2019ttn}
was cornerstone for the development of the so-called Maxwell-Proca theory.
This discussion justifies an interest in the calculations of section \ref{sec:vectors} well beyond a simple exemplification of the explicit usage of the proposed method.
When gravity is to be involved, the examples in section \ref{sec:gravity} provide a useful possible basis.

%%%%%%%%%%%%%%%%%%%%%%%%%%%%%%%%%%%%%%%%%%%%%%%%%%%%%%%%%%%%%%%%%%%%%%%%%%%%%%%%%%%%%%%%%

%%%%%%%%%%%%%%%%%%%%%%%%%%%%%%%%%%%%%%%%%%%%%%%%%%%%%%%%%%%%%%%%%%%%%%%%%%%%%%%%%%%%%%%%%
\subsection{On a recent equivalent Lagrangian approach}
\label{sec:comir}
%%%%%%%%%%%%%%%%%%%%%%%%%%%%%%%%%%%%%%%%%%%%%%%%%%%%%%%%%%%%%%%%%%%%%%%%%%%%%%%%%%%%%%%%%

During the preparation of this manuscript,
a novel Lagrangian approach to obtain the functionally independent Lagrangian constraints and count propagating modes in constrained systems
(of the kind here considered) appeared~\cite{Heidari:2020cil}.
The method therein is physically equivalent to that put forward in~\cite{Diaz:2014yua,Diaz:2017tmy},
which ---as already mentioned--- are complementary references to our own discussion in section \ref{sec:method}.
This equivalence can be easily verified, as both~\cite{Heidari:2020cil} and~\cite{Diaz:2014yua,Diaz:2017tmy} provide a mapping between their proposed Lagrangian parameters
and the usual numbers of different kinds of Hamiltonian constraints.
We have checked this leads to a consistent mapping between their different Lagrangian parameters.

In our understanding, the method in~\cite{Heidari:2020cil} distinguishes itself because it introduces the notion of
first and second class (functionally independent) Lagrangian constraints.
In our language, these are easy to identify.
They are the sum of the various functionally independent Lagrangian constraints arising at all prior stages
whose algorithm finalizes non-dynamically (as in cases \ref{it:II} and \ref{it:III}) and dynamically (as in case \ref{it:I}), respectively.
This abstract definition is clarified in the following, by classifying the functionally independent Lagrangian constraints we found in all the given examples
into first and second class Lagrangian constraints.

In the case of Maxwell electromagnetism, the primary Lagrangian constraint (\ref{eq:Maxcons1}) we found is a first class Lagrangian constraint.
This is because it leads to a secondary constraint (\ref{eq:MaxLagid}) that is identically satisfied and so non-dynamically stabilized by means of the closure \ref{it:II}.
In fact, this same example is worked out in~\cite{Heidari:2020cil} as well.

Next, consider the (hard) Proca theory.
There, both the primary (\ref{eq:Procprim}) and secondary (\ref{eq:Procsec}) Lagrangian constraints we determined
are second class Lagrangian constraints, since the algorithm closes dynamically at the next stage by means of case \ref{it:I}.
Such closure implies that the consistency under time evolution of the secondary constraint is determined through a
tertiary equation of motion.

We move to Schwinger-Plebanski formulation of both electromagnetism and  the (hard) Proca theory.
In both cases, the velocity independent primary constraints $[\varphi_3]_i\underset{1}{:\approx}0$ and $[\varphi_4]^i\underset{1}{:\approx}0$ in (\ref{eq:PrimCts1stMax})
are second class Lagrangian constraints.
This is because their stability is dynamically ensured, via the secondary equations of motion.
Thus, the algorithm closes as in case \ref{it:I} for them.
Similarly, the secondary constraints $[\widetilde{\varphi}_2]_{ij}\underset{2}{:\approx}0$ in (\ref{eq:1stMaxSecC}) are stabilized dynamically at the subsequent stage.
Therefore, these and also their ascendant primary constraints $[{\varphi}_2]_{ij}\underset{1}{:\approx}0$ in (\ref{eq:PrimCts1stMax})
are second class Lagrangian constraints $\forall m\geq0$.
The same is true for $[\varphi_1]\underset{1}{:\approx}0$ and $[\widetilde{\varphi}_1]\underset{2}{:\approx}0$ in the massive case.
When $m=0$, we see that $[\widetilde{\varphi}_1]\underset{1}{\approxident}0$.
This is a Lagrangian identity that closes the algorithm non-dynamically, following case \ref{it:II}.
As a result, both the said Lagrangian identity and its ascendant $[\varphi_1]\underset{1}{:\approx}0$ are first class Lagrangian constraints.

Turning to $d=2$ Palatini, we see it is rather simple to reclassify the nine functionally independent Lagrangian constraints we obtained into first and second class.
At the primary level, we notice that there are six velocity dependent Lagrangian constraints among the relations that follow from requiring the vanishing of \eqref{eq:primalph2D}.
These are $[\varphi_a]\,\overset{!}{{\underset{1}{\approx}}}\,0$, where $a=1,2,\dots,6$ and the tensorial indices outside the square brackets $[\cdot]$ have been omitted.
Their stability is dynamically ensured (via the secondary equations of motion) and so these are second class Lagrangian constraints.
The three remaining primary Lagrangian constraints, $[\varphi_a]\,\overset{!}{{\underset{1}{\approx}}}\,0$ with $a=7,8,9$, are manifestly velocity independent.
They show a trivial stability at the secondary stage, see (\ref{eq:Pal2DsecctsTC1}).
Accordingly, we identify these as three first class Lagrangian constraints.

At last, we reclassify the functionally independent Lagrangian constraints we found for $d>2$ Palatini into first and second class Lagrangian constraints.
Recall that we obtained $M_1=d(d+1)^2/2$ functionally independent primary Lagrangian constraints, given by the vanishing of (\ref{eq:primalph}).
Notice now that we can straightforwardly split these primary constraints into
\begin{align}
\begin{cases}
M_{1(\textrm{v})}=d(d+1) \textrm{ number of velocity dependent constraints: }
[\varphi_1]\overset{!}{\underset{1}{:\approx}}0,\,[\varphi]_i\overset{!}{\underset{1}{:\approx}}0,\,\ldots,\,[\varphi_6]^{ij}\overset{!}{\underset{1}{:\approx}}0, \\
M_{1(\textrm{nv})}=\frac{d}{2}(d^2-1) \textrm{ number of velocity independent constraints: }
[\varphi_7]_i\overset{!}{\underset{1}{:\approx}}0, \,[\varphi_8]_i^j\overset{!}{\underset{1}{:\approx}}0, \,[\varphi_9]_i^{jk}\overset{!}{\underset{1}{:\approx}}0,
\end{cases}
\end{align}
where the subscripts (n)v stand for (non-) velocity dependent constraints.
The consistency under time evolution of the velocity dependent constraints is dynamically fixed at the secondary stage and so these are second class Lagrangian constraints.
The remaining velocity independent constraints give rise to the $M_2=M_{1(\textrm{nv})}$ number of functionally independent
secondary Lagrangian constraints, which are equal to the vanishing of (\ref{eq:secin1}).
Once more, it is trivial to differentiate between
\begin{align}
\begin{cases}
M_{2(\textrm{v})}=\frac{d}{2}(d^2-3) \textrm{ number of velocity dependent constraints: }
[\widetilde{\varphi}_2]_i^j\overset{!}{\underset{1}{:\approx}}0 \textrm{ with } i\neq j,\,[\widetilde{\varphi}_3]_i^{jk}\overset{!}{\underset{1}{:\approx}}0,\\
M_{2(\textrm{nv})}=d \textrm{ number of velocity independent constraints: }
[\widetilde{\varphi}_1]_i\overset{!}{\underset{1}{:\approx}}0 ,\,[\widetilde{\varphi}_2]_i^i\overset{!}{\underset{1}{:\approx}}0 .
\end{cases}
\end{align}
The consistency under time evolution of the former is ensured by the tertiary equations of motion.
Equivalently, the algorithm closes according to the dynamical case \ref{it:I} for them. 
As a result, they are second class Lagrangian constraints.
Further, the subset of $M_{2(\textrm{v})}$ number of velocity independent primary constraints they follow from are
second class Lagrangian constraints as well.
Specifically, $[\varphi_8]_i^j\overset{!}{\underset{1}{:\approx}}0$ with $i\neq j$ and
$[\varphi_9]_i^{jk}\overset{!}{\underset{1}{:\approx}}0$ are second class.
On the other hand, the above velocity independent constraints are trivially stable, as their time evolution yields tertiary constraints that identically vanish in
$T\mathcal{C}_2$: recall (\ref{eq:tertzero}).
The algorithm closes non-dynamically as in case \ref{it:III} for them.
Consequently, they are first class Lagrangian constraints and their ascendant primary Lagrangian constraints
$[\varphi_7]_i\overset{!}{\underset{1}{:\approx}}$ and $[\varphi_8]_i^i\overset{!}{\underset{1}{:\approx}}0$ are first class too.

%%%%%%%%%%%%%%%%%%%%%%%%%%%%%%%%%%%%%%%%%%%%%%%%%%%%%%%%%%%%%%%%%%%%%%%%%%%%%%%%%%%%%%%%%

%%%%%%%%%%%%%%%%%%%%%%%%%%%%%%%%%%%%%%%%%%%%%%%%%%%%%%%%%%%%%%%%%%%%%%%%%%%%%%%%%%%%%%%%%
\subsection{Relation to the Maxwell-Proca theory and beyond}
\label{sec:MPrel}
%%%%%%%%%%%%%%%%%%%%%%%%%%%%%%%%%%%%%%%%%%%%%%%%%%%%%%%%%%%%%%%%%%%%%%%%%%%%%%%%%%%%%%%%%

As we explicitly showed in section \ref{sec:Maxwellvec}, in a purely Lagrangian formulation with as many a priori independent field variables
as the dimension of the underlying flat spacetime,
the constraint structure of the simplest theory for a single Maxwell field can be characterized by the triplet $t_{\textrm{M}}^{(N)}$ in (\ref{eq:tripletM}).
The analogous investigation in section \ref{sec:Procavec} of the most elementary theory of one (hard) Proca field yielded the constraint structure characterizing triplet
$t_{\textrm{P}}^{(N)}$ in (\ref{eq:tripletP}).
Employing the results of~\cite{Diaz:2014yua}, we also verified the corresponding Hamiltonian characterization of these two triplets.
We thus checked that the Maxwell and (hard) Proca fields are associated with
two first and second class constraints, respectively.
Although usually Maxwell and Proca fields are defined in the latter Hamiltonian manner,
in the following we take the former Lagrangian triplets as the vector fields' defining features.
We stress both definitions are equivalent.

The manifestly first order completions of the Maxwell and (hard) Proca theories analyzed in sections \ref{sec:Maxwellvec} and \ref{sec:Procavec}
are non-linear electrodynamics (NLE) and the so-called generalized Proca (GP) or vector-Galileon theory\footnote{We are aware of the recent proposal in~\cite{deRham:2020yet}.
However, the Lagrangian there put forward is not in a manifestly first order form.
The authors leave for further studies this result.
In the lack of it, their theory lies beyond our framework and we cannot address it.}, respectively.
NLE encompasses a large class of theories.
The celebrated Born-Infeld theory~\cite{Born:1934ji} is part of it,
but also the more recently proposed exponential~\cite{Hendi:2013dwa} and logarithmic~\cite{Gaete:2013dta} electrodynamics, among others.
Schematically, the Lagrangian density for NLE can be written as
\begin{align}
\mathcal{L}_{\textrm{NLE}}=\mathcal{L}_{\textrm{M}}+f(A_{\mu\nu}),
\end{align}
where $\mathcal{L}_{\textrm{M}}$ is the Maxwell Lagrangian as introduced in (\ref{eq:MaxLag}) and $f$ is a smooth real function.
Notice that the above depends on the Maxwell field $A_\mu$ exclusively through its field strength $A_{\mu\nu}$ ---up to boundary terms.
Indeed, it is well-known~\cite{Pleb} that a more involved dependence is not possible, if the $U(1)$ gauge symmetry is to be respected.
This feature remains true even when coupling the Maxwell field to General Relativity~\cite{Horndeski:1976gi}.
Only a few fine-tuned terms that contract $A_{\mu\nu}$ with the Riemann tensor are possible in such case.
It is not hard to convince oneself that the constraint structure of NLE is characterized by the triplet $t_{\textrm{M}}^{(N)}$ in (\ref{eq:tripletM}).
In other words, it has the same constraint structure as classical electromagnetism, in its standard formulation of section \ref{sec:Maxwellvec}.

The GP theory was put forward in~\cite{Tasinato:2014eka} and its complete Lagrangian was established in~\cite{Jimenez:2016isa}.
Again schematically, we may express it as
\begin{align}
\label{eq:GPsch}
\mathcal{L}_{\textrm{GP}}=\mathcal{L}_{\textrm{P}}+g(B_{\mu})
+\sum_{n=1}^{d}\mathcal{T}^{\nu_1\ldots\nu_n\rho_1\ldots\rho_n}\partial_{\nu_1}B_{\rho_1}\ldots\partial_{\rho_n}B_{\mu_n}
\end{align}
in $d$ dimensions, where $\mathcal{L}_{\textrm{P}}$ is the (hard) Proca Lagrangian in (\ref{eq:covPro}),
$g$ is a real smooth function and each $\mathcal{T}^{\nu_1\ldots\nu_n\rho_1\ldots\rho_n}$ is a certain smooth real object
constructed out of the spacetime metric $\eta_{\mu\nu}$, the $d$-dimensional Levi-Civita tensor $\epsilon_{\mu_1\ldots\mu_d}$ and
the Proca field $B_\mu$.
Although GP has only been formulated for $d=4$, its systematic construction allows for a straightforward inferring of (\ref{eq:GPsch}).
Here, the underlying key idea consists in supplementing the (hard) Proca Lagrangian with derivative self-interaction terms
of the Proca field $B_\mu$.
This implies a non-local extension of the notion of mass for the vector field.
As such, we regard GP as an effective classical field theory\footnote{The quantization of non-local field theories generically leads to
acausality, most often for phenomena beyond tree-level.
It is sometimes possible to circumvent this problem, specially in the presence of supersymmetry ---for instance, see~\cite{Addazi:2015dxa}.
The consistent quantization of GP has been investigated in~\cite{qGP}.
Although the results obtained so far justify an optimistic attitude, no complete and rigorous quantization scheme seems to have been proposed so far,
along the lines of the BRST and path integral quantization of the (hard) Proca theory in~\cite{Kim:1996gk} and~\cite{Su:1998wz}, respectively.
Therefore, we here adopt the conservative point of view that regards GP as an effective classical field theory.}.
It can be readily inferred from the calculations in~\cite{ErrastiDiez:2019trb} that the constraint structure of GP is characterized by the triplet
$t_{\textrm{P}}^{(N)}$ in (\ref{eq:tripletP}).
Namely, GP has the same constraint structure as the (hard) Proca theory, when the latter is formulated as in section \ref{sec:Procavec}.

Next, we consider a multi-field scenario, including $n_{\textrm{M}}$ number of Maxwell fields,
as well as $n_{\textrm{P}}$ number of (generalized) Proca fields.
In four-dimensional Minkowski spacetime, the Maxwell-Proca (MP) theory~\cite{ErrastiDiez:2019trb,ErrastiDiez:2019ttn}
is the complete set of manifestly first order (self-)interactions among an arbitrary number of real Abelian vector fields
that propagates the correct number of degrees of freedom.
These consistent interactions were derived by demanding that the constraint structure of each Maxwell and Proca field
is characterized by the triplets $t_{\textrm{M}}^{(N)}$  and $t_{\textrm{P}}^{(N)}$ , respectively.
Let $\mathcal{N}_1=(n_{\textrm{M}}+n_{\textrm{P}})d$.
We denote the constraint structure characterizing triplet of MP as $t_\textrm{MP}^{(\mathcal{N}_1)}$.
Then, we say that the building principle of the theory is based on the requirement:
\begin{align}
t_\textrm{MP}^{(\mathcal{N}_1)}\overset{!}{=}n_{\textrm{M}}\cdot t_{\textrm{M}}^{(N=d)} \oplus n_{\textrm{P}}\cdot t_{\textrm{P}}^{(N=d)}=
(l=n_{\textrm{M}}+2n_{\textrm{P}},g=n_{\textrm{M}},e=2n_{\textrm{M}}),
\end{align}
where in the last equality we have made use of (\ref{eq:tripletM}) and (\ref{eq:tripletP}).

At this point, it should be clear that our calculations of $t_{\textrm{M}}^{(N)}$ and $t_{\textrm{P}}^{(N)}$ in section \ref{sec:vectors},
elementary as they are, can be used as a basis for the construction of non-trivial theories.
Having a ready-to-be-used method optimized to obtain such triplets (i.e.~the method explained in section \ref{sec:method} and graphically summarized in figure \ref{fig:scheme})
is thus a powerful tool for the development
of manifestly first order classical field theories where multiple fields of different spins (self-)interact.

For instance, an interesting open question is that of the consistent coupling of the MP theory to gravity.
It is in principle possible to combine our calculations in all the previous sections to attempt this ambitious goal as follows.
Let $\mathcal{N}_2=(n_{\textrm{M}}+n_{\textrm{P}})d+d(d+1)^2/2$, with $d\geq2$ the dimension of the spacetime.
A manifestly first order Lagrangian density $\mathcal{L}_{\textrm{MP(2)Pa}}$ that describes the dynamics of $n_{\textrm{M}}$ number of Maxwell fields
and $n_{\textrm{P}}$ number of (generalized) Proca fields in the presence of Einstein's gravity in terms of $\mathcal{N}_2$
a priori independent field variables must be associated with a constraint structure characterizing triplet
$t_{\textrm{MP(2)Pa}}^{(\mathcal{N}_2)}$ satisfying
\begin{align}
\label{eq:tripletMPPa}
t_{\textrm{MP(2)Pa}}^{(\mathcal{N}_2)}\overset{!}{=}n_{\textrm{M}}\cdot t_{\textrm{M}}^{(N)} \oplus n_{\textrm{P}}\cdot t_{\textrm{P}}^{(N)}
\oplus t_{\textrm{(2)Pa}}^{(N)},
\end{align}
where all the triplets on the right-hand side have already been calculated in this work;
see (\ref{eq:tripletM}), (\ref{eq:tripletP}), (\ref{eq:tripletPa}) and (\ref{eq:triplet2Pa}).
Substituting these results, we have that 
\begin{align}
t_{\textrm{MP(2)Pa}}^{(\mathcal{N}_2)}\overset{!}{=}
\begin{cases}
(l=n_{\textrm{M}}+2n_{\textrm{P}}+d^2(d+1),g=n_{\textrm{M}}+d,e=2n_{\textrm{M}}+3d) &\qquad \textrm{if } d>2, \\
(l=n_{\textrm{M}}+2n_{\textrm{P}}+9,g=n_{\textrm{M}}+3,e=2n_{\textrm{M}}+6) &\qquad \textrm{if } d=2.
\end{cases}
\end{align}
The conversion of any of the above necessary conditions into a Lagrangian density building principle is an algebraically involved exercise beyond the scope
of our present investigations. 
We thus leave it for future works.

A last remark is due.
As we observed at the very end of section \ref{sec:onshell} and should be apparent from our calculations in section \ref{sec:Palatini},
it is in general a conceptually clear but algebraically non-trivial exercise to obtain the triplet $t^{(N)}$ of a given Lagrangian density $\mathcal{L}$ within our framework.
It is even more challenging to determine the (exhaustive) form of $\mathcal{L}$ from the necessary condition that it should be associated to a certain triplet $t^{(N)}$.
The reason is that such inversion in the logic requires solving sets of coupled non-linear partial differential equations in most cases.
Therefore, it is overwhelmingly convenient to use all freedom of choice available in order to simplify this task to the utmost.
For instance, one is advised to choose constant null vectors for the Hessians at all stages, if possible. 
For the concrete research project here proposed, it may be the case that (\ref{eq:tripletMPPa}) is not the optimal starting point.
It could happen that the equivalent demand
\begin{align}
t_{\textrm{MP(2)Pa}}^{(\mathcal{N}_3)}\overset{!}{=}n_{\textrm{M}}\cdot t_{\textrm{M}}^{(\pmb{N})} \oplus n_{\textrm{P}}\cdot t_{\textrm{P}}^{(\pmb{N})}
\oplus t_{\textrm{(2)Pa}}^{(N)} \qquad \textrm{with } \quad \mathcal{N}_3=(n_{\textrm{M}}+n_{\textrm{P}}+d+1)d(d+1)/2,
\end{align}
with the right-hand side triplets as given in (\ref{eq:tVec}), (\ref{eq:tripletPa}) and (\ref{eq:triplet2Pa}),
is a more befitting way to try to derive the set of consistent (self-)interactions of vector fields in a curved background.
For the reasons given at the beginning of section \ref{sec:gravity}, we believe that $t_{\textrm{(2)Pa}}^{(N)}$
is indeed a beneficial basis for the gravity piece above.

%%%%%%%%%%%%%%%%%%%%%%%%%%%%%%%%%%%%%%%%%%%%%%%%%%%%%%%%%%%%%%%%%%%%%%%%%%%%%%%%%%%%%%%%%

%%%%%%%%%%%%%%%%%%%%%%%%%%%%%%%%%%%%%%%%%%%%%%%%%%%%%%%%%%%%%%%%%%%%%%%%%%%%%%%%%%%%%%%%%
\section{Conclusions}
\label{sec:final}
%%%%%%%%%%%%%%%%%%%%%%%%%%%%%%%%%%%%%%%%%%%%%%%%%%%%%%%%%%%%%%%%%%%%%%%%%%%%%%%%%%%%%%%%%

In the following, we summarize the results we have put forward in this manuscript.
Then, we proceed to discuss their relevance and pertinence.
At last, we comment on the increasing (in $n$) computational difficulty of evaluating Lagrangian constraints on constraint surfaces $T\mathcal{C}_n$
and concretize the pathologies a theory may suffer from when the algorithm of section \ref{sec:onshell} is not verified to close.

\vspace*{0.5cm}

\hspace*{-0.6cm}{\bf Summary of results.}\\
In section \ref{sec:method}, we have collected and complemented results from the extensive literature on constrained systems
and presented a self-contained and ready to be used method to determine all the constraints in a theory.
By postulation, the theory is required to be described by a manifestly first order Lagrangian.
We make the mild assumptions of the principle of stationary action and finite reducibility.
When the theory is covariant, the iterative algorithm presented for the determination of the functionally independent Lagrangian constraints does not contravene this feature.
Nonetheless, manifest covariance is generically lost in our approach.
In sections \ref{sec:vectors} and \ref{sec:gravity}, we have minutely exemplified the usage of our said procedure.
In section \ref{sec:lit}, we have argued for the pertinence and contemporaneity of both the general formalism and the given examples.
Indeed, an equivalent but different methodology has been put forward lately~\cite{Heidari:2020cil}.
The examples of section \ref{sec:vectors} constitute the foundation of the also recent Maxwell-Proca theory~\cite{ErrastiDiez:2019trb,ErrastiDiez:2019ttn}
and those of section \ref{sec:gravity} can potentially form the basis for the consistent coupling of Maxwell-Proca to gravity.

\vspace*{0.5cm}

\hspace*{-0.6cm}{\bf Critical discussion of results.}\\
The procedure explained in section \ref{sec:method} presents two main appealing features.
First, it is a coordinate-dependent approach, as opposed to a geometrical one.
It thus readily allows for its application, given a Lagrangian density satisfying the initial postulates,
without having to work out any symplectic two-form.
With pragmatism in mind, section \ref{sec:method} has been written in a way that is (hopefully) accessible to a broad audience.
Even though the method stands on a rigorous footing, the discussion has been made largely devoid of mathematical technicalities.

Second, it is an intrinsically Lagrangian procedure, as opposed to a Hamiltonian or a hybrid one.
The appeal of this characteristic resides in the fact that, in many areas of high energy theoretical physics,
manifestly first order classical field theories are predominantly posed and studied in their Lagrangian formulation.
This is the case for instance in cosmology, astrophysics, black hole physics and holographic condensed matter.
In all these disciplines, GP, MP and allied theories, specially in the presence of gravity, have been convincingly argued to be of significant interest,
e.g.~\cite{Tasinato:2014eka,ErrastiDiez:2019ttn,genProca}.
As such, our proposed procedure avoids non-negligible obstacles that typically arise in the transformation from the Lagrangian to the Hamiltonian picture.
Besides, as already noted in the end of sections \ref{sec:Palatini} and \ref{sec:Palatinid2}, our Lagrangian approach is a computationally
faster and simpler way to obtain the constraint structures of these theories, compared to representative Hamiltonian analyses.
(The examples in section \ref{sec:vectors} are so effortless comparatively that they do not substantiate an analogous argumentation.)

In more detail, implementing our algorithm in section \ref{sec:onshell} is considerably easier than
carrying out a Hamiltonian counterpart algorithm based on the Dirac-Bergman~\cite{DiracBergmann} procedure.
As the attentive reader will have already noticed in our explicit examples of section \ref{sec:gravity} and we shall address shortly, 
the most demanding step in our approach consists in evaluating the $n$-th stage Lagrangian constraints in the $(n-1)$-th constraint surface, with $n\geq1$.
An analogous evaluation is necessary within the Hamiltonian picture as well, where two additional hurdles arise.
On the one hand, one must classify the Dirac constraints into first and second class.
This entails calculating the Poisson brackets of all Dirac constraints,
a generically challenging task in field theory because non-local algebras usually arise\footnote{This non-locality is
as a consequence of the distributional nature of fields.
Dirac constraints must be smeared with suitable test functions and integrated over for a correct evaluation of the Poisson brackets.}, e.g.~\cite{Ghalati:2007sv,McKeon:2010nf}.
On the other hand, in the standard Hamiltonian transition from one stage to the next,
novel constraints emerge and must be consistently included via Lagrange multipliers.
Closure of the algorithm requires the determination of as many Lagrange multipliers as possible,
which in turn implies the resolution of algebraic or even differential equations.
Even in the comparatively benign algebraic scenario, finding a solution is an increasingly (in stage) laborious and non-trivial task
that involves inverting field-dependent matrices with complicated spatial index structures.

For a suggestive utility of the examples in sections \ref{sec:vectors} and \ref{sec:gravity}, the reader is referred to section \ref{sec:MPrel}.
Recall that the proposal therein is illustrative of the general theory-construction idea outlined
in the introduction section \ref{sec:intro} and at the beginning of section~\ref{sec:lit}.

\vspace*{0.5cm}
\hspace*{-0.6cm}{\bf Two final observations.}\\
In the first of our observations, we bring to light a series of considerations that must be taken into account when applying our method.
In particular, we wish to discuss the practical complications that field theories of the kind here considered commonly exhibit when
their Lagrangian constraints are to be evaluated on the suitable constraint surface.

First, we debunk what naively may look like an ambiguity. 
Recall that any constraint surface $T\mathcal{C}_n$ for some finite $n\geq1$
is defined by the weak vanishing of the functionally independent Lagrangian constraints at all prior stages:
\begin{align}
{\varphi_{I}(Q^A,\dot{Q}^A,\partial_{i}Q^A)}\overset{!}{\underset{1}{:\approx}} 0,\qquad 
{\widetilde{\varphi}_{R}(Q^A,\dot{Q}^A,\partial_{i}Q^A)}\overset{!}{\underset{2}{:\approx}} 0, \qquad 
{\widehat{\varphi}_{U}(Q^A,\dot{Q}^A,\partial_{i}Q^A)}\overset{!}{\underset{3}{\approx}} 0, \qquad \textrm{etc.}
\end{align} 
As a direct consequence of the above, one can determine a maximal set of functionally independent relations of the form
\begin{align}
\label{eq:funcrel}
\dot{Q}^{B}\overset{!}{\underset{n}{\approx}}\dot{Q}^{B}(Q^{A},\dot{Q}^{A},\partial_{i}Q^{A}),\qquad
\partial_{i}{Q}^{B}\overset{!}{\underset{n}{\approx}}\partial_{i}{Q}^{B}(Q^{A}, \dot{Q}^{A},\partial_{i}Q^{A}), \qquad
Q^{B}\overset{!}{\underset{n}{\approx}} {Q}^{B}(Q^{A} ).
\end{align}
Though it should be clear by now, we confirm the different 
role played by the generalized velocities $\dot{Q}^A$ and the spacelike derivatives of the generalized coordinates $\partial_i Q^A$.
The former are independent coordinates on $T\mathcal{C}$, while the latter are functionally related to the generalized coordinates $Q^A$.
This clarification becomes pertinent when evaluating the secondary Lagrangian constraints in $T\mathcal{C}_1$ already.
At this point (and in subsequent stages), derivatives of the form $\partial_{i}\dot{Q}^{A}$ generically show up.
In such expressions, one must first replace the primary weak expression for the generalized velocity $\dot{Q}^A$ ---if pertinent--- and then apply the spatial derivative on it.

Having clarified this point, we notice that its consistent implementation leads to the following nested situation.
Substitution of $\dot{Q}^{A}$ according to (\ref{eq:funcrel}) in $\partial_{i} \dot{Q}^{A}$ normally leads to the presence of terms of the form $\partial_i Q^B$ in \eqref{eq:funcrel}.
These are again prone to be evaluated in $T\mathcal{C}_n$ and can in turn contribute terms depending on $Q^B$'s in \eqref{eq:funcrel}; etc.
We emphasize that one must reach an expression where this nesting ceases to occur, before proceeding with the algorithm.
Not doing so would imply a wrong evaluation of the Lagrangian constraints in $T\mathcal{C}_n$, may lead to a misidentification of the functionally independent Lagrangian constraints
and will almost invariably yield wrong results at the following $(n+1)$-th stage.
In fact, a wrong evaluation will typically land the researcher in a physically inequivalent theory from the one he/she started with.
 
Additionally and normally, when evaluating some Lagrangian constraints in $T\mathcal{C}_n$, potentially
contrived functions of the previous stages' functionally independent Lagrangian constraints also show up.
To understand the difficulty their appearance implies, consider the tertiary Lagrangian constraints (\ref{eq:tertcons}) we found for $d>2$ Palatini.
Their raw expressions, prior to any evaluation in a constraint surface, contain quantities $f=f(Q^A,\dot{Q}^A,\partial_iQ^A)$ that vanish in $T\mathcal{C}_1$.
However, recognizing such $f$'s as primary weak zeros is a challenging task.
Specifically,
\begin{align}
[\widehat{\varphi}_1]_i\supset \mathcal{G}^jf_{ij}, \qquad 
f_{ij}=
\partial_{[i}[\varphi_{7}]_{j]}-2G_{(i}[\varphi_{7}]_{j)}-G_{k(i}[\varphi_{8}]^{k}_{j)}\underset{1}{\approx}0,
\end{align}
where in the $\supset$ relation we have omitted numerical factors and the $\varphi$'s are as given in (\ref{eq:primalph}). 
In the expression for $f_{ij}$, the first equality is non-trivial, while the subsequent primary weak equality is obvious.
An analogous situation arises with other $f$'s that are based on both functionally independent primary (\ref{eq:primalph}) and secondary \eqref{eq:secin1}
Lagrangian constraints.
A brute force resolution to identify all such $f$'s consists in putting forward the most general ansatz compatible with the tensorial character of each of the Lagrangian constraints
one is trying to evaluate and comparing it to their explicit expressions.
This is indeed how we laboriously arrived at (\ref{eq:tertsin1dif}).

For the second and last observation, the reader should heed (\ref{eq:dofformula}) and (\ref{eq:lfinal}).
We already stressed the importance of closing the iterative algorithm for obtaining the functionally independent Lagrangian constraints 
towards the end of section \ref{sec:onshell}.
Now, we are equipped to better grasp the implications of not doing so, mentioned in the introductory section \ref{sec:intro}.
Most often, failure to close the algorithm will give rise to the propagation of unphysical modes.
These are Ostrogradski instabilities~\cite{Ostr}, but we shall loosely refer to them as ghosts.
Even after ensuring ghost-freedom, not closing the algorithm can lead to trouble: it may overconstrain the theory, so that
fewer than the desired number of degrees of freedom are propagated.

Let us consider the MP theory~\cite{ErrastiDiez:2019trb,ErrastiDiez:2019ttn} discussed in section \ref{sec:MPrel} as a concrete framework to clarify the above two unwanted scenarios.
For our present purposes, it will suffice to consider the case when there are no Maxwell fields $n_{\textrm{M}}=0$
and there are an arbitrary but finite number of Proca fields $n_{\textrm{P}}$.
Recall that, in the standard formulation, we already saw in section \ref{sec:Procavec} that a Proca field is associated to $l=M_1^\prime+M_2^\prime=1+1=2$
number of functionally independent Lagrangian constraints.
Bear in mind that this is also true for a generalized Proca field.

We denote the natural generalization of the GP theory in (\ref{eq:GPsch}) to a multi-field setup as $\mathcal{L}_{\textrm{PP}}$.
$\mathcal{L}_{\textrm{PP}}$ automatically leads to $M_1^\prime=n_{\textrm{P}}$ number of functionally independent primary constraints.
The consistency under time evolution of these constraints
does not generically yield the $M_2^\prime=n_{\textrm{P}}$ number of functionally independent secondary constraints one would naively expect.
Only a fine-tuned subset of terms in $\mathcal{L}_{\textrm{PP}}$ does, precisely the terms that are part of the MP theory.
For all those terms, it was shown that no tertiary constraints arise ($M_3=0$) and the algorithm closes dynamically giving rise to $l=2n_{\textrm{P}}$.
Therefore, the correct number of physical modes $n_{\textrm{dof}}=d-n_{\textrm{P}}$ are present in the theory.
(To obtain this result, notice that, since there are no gauge identities, $g=0=e$.)

Notice that, if one studies only the primary stage for $\mathcal{L}_{\textrm{PP}}$, one will be deceived into thinking that the theory is valid,
as it suitably extends the primary stage of GP.
However, $\mathcal{L}_{\textrm{PP}}$ has $M_2^\prime<n_{\textrm{P}}$ in general and therefore $l<2n_{\textrm{P}}$ and $n_{\textrm{dof}}>d-n_{\textrm{P}}$.
The additional propagating modes are precisely the ghosts of the first scenario we warn against.

If one studies both the primary and secondary stages for $\mathcal{L}_{\textrm{PP}}$, then one can fine-tune the Lagrangian density so that $M_2^\prime=n_{\textrm{P}}$ as desired.
But these functionally independent secondary constraints in the fine-tuned theory are at this point not necessarily stable.
Their consistency under time evolution could in principle lead to further functionally independent tertiary constraints, so that
$l>2n_{\textrm{P}}$ and $n_{\textrm{dof}}<d-n_{\textrm{P}}$.
This would place us in the second unwelcome scenario.
For the given example, it so happens that the fine-tuned $\mathcal{L}_{\textrm{PP}}$ is associated to a full rank tertiary Hessian.
Consequently, the functionally independent secondary constraints are dynamically stabilized without further fine-tunings of the theory.
However, this cannot be assumed, it has to be checked, so as to ensure the theory is not overconstrained.

It is interesting to point out that in~\cite{Crisostomi:2017aim} our very same admonition against the overconstrained scenario is made, albeit in a different context.
The authors look into second order field theories with no gauge symmetry and derive the necessary conditions for such Lagrangians to not propagate ghosts.
They show that, in the presence of Lorentz symmetry, the existence of any number $M_1^\prime>0$ of functionally independent Lagrangian constraints automatically leads to
the same number $M_2^\prime=M_1^\prime$ of functionally independent Lagrangian constraints.
They unequivocally recognize our second scenario: those $M_2^\prime$ are not necessarily stable, so one could be facing an overconstrained theory.

%%%%%%%%%%%%%%%%%%%%%%%%%%%%%%%%%%%%%%%%%%%%%%%%%%%%%%%%%%%%%%%%%%%%%%%%%%%%%%%%%%%%%%%%%

\vspace*{1cm}

\noindent
{\bf Acknowledgements:}
The authors are indebted to Dieter L\"{u}st for his careful review of the manuscript.
We also very much thank Angnis Schmidt-May for enlightening discussions in the early stages of the project.
In particular, for her insights regarding the relation between first and second order formulations of (modified) gravity theories.
We recognize Brage Gording for the sharpening of our statements with respect to the Maxwell-Proca theory
and his lucid pondering over our results.
VED is grateful to Marina Krstic Marinkovic for her incisive dialogues and bringing up~\cite{Su:1998wz}.
This work is supported by a grant from the Max Planck Society.
MM would like to express his great appreciation for the Max Planck Institute for Physics for employing him during his masters thesis, of which this paper is a natural follow-up.
JAMZ was partially funded by the grant ``Convocatoria para estancias postdoctorales Max-Planck-CONACyT 2018''.
MTT would like to thank the hospitality of the Max Planck Institute for Physics during his visit in October 2019. 

\noindent
VED lovingly dedicates this work to the memory of her late mother.

%%%%%%%%%%%%%%%%%%%%%%%%%%%%%%%%%%%%%%%%%%%%%%%%%%%%%%%%%%%%%%%%%%%%%%%%%%%%%%%%%%%%%%%%%

\pagebreak

%%%%%%%%%%%%%%%%%%%%%%%%%%%%%%%%%%%%%%%%%%%%%%%%%%%%%%%%%%%%%%%%%%%%%%%%%%%%%%%%%%%%%%%%%
\appendix
%%%%%%%%%%%%%%%%%%%%%%%%%%%%%%%%%%%%%%%%%%%%%%%%%%%%%%%%%%%%%%%%%%%%%%%%%%%%%%%%%%%%%%%%%

%%%%%%%%%%%%%%%%%%%%%%%%%%%%%%%%%%%%%%%%%%%%%%%%%%%%%%%%%%%%%%%%%%%%%%%%%%%%%%%%%%%%%%%%%
\section{Formulae at an arbitrary stage of the algorithm}
\label{app:genex}
%%%%%%%%%%%%%%%%%%%%%%%%%%%%%%%%%%%%%%%%%%%%%%%%%%%%%%%%%%%%%%%%%%%%%%%%%%%%%%%%%%%%%%%%%

In this appendix, we show the explicit expressions of all quantities involved in an arbitrary $a$-th stage
of the iterative algorithm for irreducible theories presented in section \ref{sec:onshell}.
Needless to say, in the appropriate limit, the general expressions here given yield the primary and secondary stages' formulae there shown.

Let $\varphi_{A_a}\overset{!}{\underset{a}{:\approx}}0$ be a set of $M_a$ number of functionally independent Lagrangian constraints in the $a$-th stage, with $A_{a}=1,2,\ldots,M_a$.
These constraints are relations between the generalized coordinates $Q^A$ and velocities $\dot{Q}^A$ of the field theory under consideration.
They define the so-called $a$-th constraint surface
\begin{align}
T\mathcal{C}_a := \{ (Q^A, \dot{Q}^A) \hspace{1 mm} \rvert  \hspace{1mm}\varphi_{A_a}\underset{a-1}{\approx} 0 \} \subseteq T \mathcal{C}_{a-1} \subset T \mathcal{C}_{a-2}
\subset \dots T\mathcal{C}_{1}\subseteq T \mathcal{C}_{0},
\end{align}
where $T \mathcal{C}_{0}$ is the moduli space of the theory, defined in \eqref{eq:TC_1,0}.
In order to ensure the preservation of the said constraints under time evolution, we demand
\begin{align}
\label{eq:ELeqsa}
E_{A_a}:=\dot{\varphi}_{A_a}\overset{!}{\underset{a}{\approx}}0.
\end{align}
We refer to $E_{A_a}$ as the $(a+1)$-th stage Euler-Lagrange equations.
Next, we will explicitly write $E_{A_a}$.
But to do so, we must first define the following objects.

Let $W_{A_{a} A_{b}}$ denote the $(a+1)$-th stage Hessian.
This is a square matrix of dimension $M_{a}$
that allows us to define $M_{a+1}:=\textrm{dim}(W_{A_{a} A_{b}})-\textrm{rank}(W_{A_{a} A_{b}})$.
We refer to the $M_{a+1}$ number of linearly independent null vectors associated to $W_{A_{a} A_{b}}$ as $\gamma_{A_{a+1}}$.
Explicitly, $(\gamma_{A_{a+1}})^{A_a} W_{A_{a} A_{b}}=0$, with $A_{a+1}=1,2,\ldots,M_{{a+1}}$.
We require them to fulfil the normalization condition
\begin{align}
(\gamma_{A_{a+1}})^{A_a}(\gamma^{A_{b+1}})_{A_a}=\delta_{A_{a+1}}{}^{A_{b+1}}, \qquad \textrm{with } \gamma^{A_{a+1}}:=(\gamma_{A_{a+1}})^T,
\end{align}
so that they form a basis in the kernel of $W_{A_{a} A_{b}}$.
Here, $T$ stands for the transpose operation.
With the help of the above null vectors, the $(a+1)$-th stage Hessian can be expressed in terms of
the the functionally independent $a$-th stage Lagrangian constraints as follows:
\begin{align}
\label{eq:Hessa}
W_{A_{a} A_{b}}=(\gamma_{A_a})^{A_{a-1}}(\gamma_{A_{a-1}})^{A_{a-2}}\ldots (\gamma_{A_1})^{A}\partial_{\dot{A}}\varphi_{A_b},
\qquad \textrm{with } A=1,2,\ldots,N=\textrm{dim}(\mathcal{C}).
\end{align}
Finally, we introduce the auxiliary matrix $M^{A_{a} A_{b}}$ (the Moore-Penrose pseudo-inverse of $W_{A_{a} A_{b}}$),
which always exists and is uniquely determined from the relations
\begin{align}
M^{A_{a} A_{b}}W_{A_{b} A_{c}}-\delta^{A_a}{}_{A_c}+(\gamma^{A_{a+1}})_{A_c}(\gamma_{A_{a+1}})^{A_a}=0, \qquad
M^{A_{a} A_{b}}(\gamma^{A_{a+1}})_{A_b}=0.
\end{align}

Using the above, the $(a+1)$-th stage Euler-Lagrange equations in (\ref{eq:ELeqsa}) can be written as
\begin{align}
\label{eq:nthDtphi}
E_{A_b}=\ddot{Q}^A(\gamma^{A_1})_{A}(\gamma^{A_2})_{A_1}\ldots (\gamma^{A_{a}})_{A_{a-1}} W_{A_aA_b}+\alpha_{A_b}
\overset{!}{\underset{a}{\approx}}0,
\end{align}
where the expression (\ref{eq:Hessa}) is to be employed for $W_{A_aA_b}$ and where we have (recursively) defined
\begin{align}
\displaystyle
\begin{array}{lll}
\alpha_{A_b}:=\Big[&\hspace*{-0.3cm}
-\alpha_{A_{a-1}}M^{A_{a-1}A_{b-1}}(\gamma_{A_{b-1}})^{A_{a-2}}(\gamma_{A_{a-2}})^{A_{a-3}}\ldots (\gamma_{A_1})^{A}\partial_{\dot{A}}\\
&\hspace*{-0.3cm}
-\alpha_{A_{a-2}}M^{A_{a-2}A_{b-2}}(\gamma_{A_{b-2}})^{A_{a-3}}(\gamma_{A_{a-3}})^{A_{a-4}}\ldots (\gamma_{A_1})^{A}\partial_{\dot{A}}\\
&\hspace*{-0.3cm}
-\ldots
-\alpha_{A}M^{AB}\partial_{\dot{B}}+\dot{Q}^A\partial_A+(\partial_i\dot{Q}^A)\partial^i_A\Big]\varphi_{A_b},
\end{array}
\end{align}
with $\alpha_A$ as given in (\ref{eq:alphaB}).
To obtain the presented $\alpha_{A_b}$, the previous $a$-th stage Euler-Lagrange equations are employed.
These in turn depend on the $(a-1)$-th stage Euler-Lagrange equations and so on.
This is the origin of the noted recursion.
Notice we therefore explicitly employ the primary Euler-Lagrange equations and so the expression (\ref{eq:nthDtphi}) is an on shell statement.

In order to reproduce the results in section \ref{sec:onshell} from the above discussion,
the reader only needs to do the index replacements $(A_0\equiv A,B,\ldots)$, $(A_1\rightarrow I,J,\ldots)$, $(A_2\rightarrow R,S,\ldots)$,
etc., as well as take footnote \ref{fn:not} into account.

%%%%%%%%%%%%%%%%%%%%%%%%%%%%%%%%%%%%%%%%%%%%%%%%%%%%%%%%%%%%%%%%%%%%%%%%%%%%%%%%%%%%%%%%%

%%%%%%%%%%%%%%%%%%%%%%%%%%%%%%%%%%%%%%%%%%%%%%%%%%%%%%%%%%%%%%%%%%%%%%%%%%%%%%%%%%%%%%%%%

\end{document}